\documentclass[preprint,12pt]{elsarticle}
\usepackage{tikz} 
\usepackage{pgfplots} \pgfplotsset{compat=newest}
\usepackage{hyperref}
\usepackage{float} 
\usepackage{graphicx}	
\usepackage{amsmath}	
\usepackage{amssymb}	
\usepackage{multicol}   
\usepackage{bm}		    
\usepackage{pdflscape}	
\usepackage{soul}       
\usepackage[normalem]{ulem} 
\usepackage{xcolor}     
\usepackage{array}      
\usepackage{caption}    
\usepackage{cleveref}   
\usepackage{float}

\usepackage[T1]{fontenc}
\usepackage{comment}

\begin{document}

\begin{frontmatter}

\title{$w_{dm}$-$w_{de}$ cosmological model with new data samples of cosmological observations}

\author[1,2]{Dorian Araya%
}

\author[1]{Cristian Castillo%
}

\author[3,4]{Genly Leon%
}

\author[1]{Juan Maga\~na \corref{cor1}%
}

\author[1]{Angie Barr Domínguez %
}

\author[5]{Miguel A. Garc\'ia-Aspeitia%
}

\cortext[cor1]{juan.magana@ucentral.cl}

\affiliation[1]{organization={Facultad de Ingenier\'ia y Arquitectura, Universidad Central de Chile},
addressline={Avenida Francisco de Aguirre 0405},
postcode={171-0164},
city={La Serena, Coquimbo},
country={Chile}}

\affiliation[2]{organization={Instituto de Física, Pontificia Universidad Católica de Valparaíso},
addressline={Avenida Brasil 2950},
postcode={2362852},
city={Valparaíso},
country={Chile}}

\affiliation[3]{organization={Departamento de Matem\'{a}ticas, Universidad Cat\'{o}lica del Norte},
addressline={Avenida
Angamos 0610, Casilla 1280}, 
postcode={1270709},
city={Antofagasta},
country={Chile}}

\affiliation[4]{organization={Department of Mathematics, Faculty of Applied Sciences, Durban University of Technology},
addressline={P.O. Box 1334}, 
postcode={4000},
city={Durban},
country={South Africa}}

\affiliation[5]{organization={Depto. de F\'isica y Matem\'aticas, Universidad Iberoamericana},
addressline={Ciudad de M\'exico, Prolongaci\'on Paseo de la Reforma 880}, 
postcode={01219},
city={M\'exico D. F.},
country={M\'exico.}}

\begin{abstract}
We revisit a cosmological model where dark matter (DM) and dark energy (DE) follow barotropic equations of state, allowing deviations from the standard $\Lambda$CDM framework (i.e. $w_{dm} \neq 0$, $w_{de} \neq -1$), considering both flat and non-flat curvature. Using a dynamical system approach, we identify equilibrium states that govern stability, expansion, and contraction. Expansion occurs when $H>0$, while contraction is linked to $H < 0$. Accelerated expansion arises from DE dominance, whereas radiation- and matter-dominated phases lead to deceleration. Some solutions are unphysical due to density constraints, but viable cases offer insights into cosmic transitions, including the Einstein static universe, which allows for shifts between accelerating and decelerating phases. We perform a Bayesian analysis with updated datasets, including observational Hubble data, Pantheon+ Type Ia supernovae, strong lensing systems, baryon acoustic oscillations and cosmic microwave background, to constrain the parameters $w_{dm}$ and $w_{de}$. Our results from the data joint analysis show consistency with $\Lambda$CDM within $3\sigma$, but none of the cases reproduce $w_{dm} = 0$ and $w_{de} = -1$. Nevertheless, the comparison with the standard model using the Akaike and Bayesian information criteria indicates that only the non-flat scenario has the potential to be competitive. This suggests that a non-dust-like DM may impact structure formation, while DE could shift toward quintessence fluid. While $\Lambda$CDM remains a strong model, our findings indicate that alternative dark sector models with non-standard EoS could be viable and offer new insights into cosmic evolution.
\end{abstract}
\end{frontmatter}

\section{Introduction}

Based on the framework of general relativity, a standard cosmological model is supported by different cosmological observations, and it posits the existence of two mysterious and exotic components known as dark matter (DM) and dark energy (DE). \cite{1933Zwicky} first suggested the possible existence of DM to explain the motion of galaxies in the Coma cluster, a concept that later supported \cite{1970ApJ...159..379R}, examining the rotation curves of spiral galaxies. Subsequently, the advancement of N-body simulations, cosmological perturbation theory, and observational data has pointed to dark matter as the driving force behind large-scale structure formation in the Universe \cite{Arbey:2021gdg}.
The initial evidence for accelerated expansion in the Universe came from observations of high-redshift Type Ia supernovae (SNIa) conducted by \cite{1998RIESS, Perlmutter:1999}, which identified the cosmological constant as the source of this cosmic acceleration. Thus, a standard cosmological model $\Lambda$CDM model (cold dark matter with a cosmological constant $\Lambda$) has been established by consensus \cite{Perivolaropoulos:2021jda}. It proposes that dark matter consists of collisionless massive particles that are non-relativistic at the time of decoupling (i.e., they are "cold"). In contrast, DE is represented as a cosmological constant linked with quantum vacuum energy.

Multiple cosmological observations across a broad range of redshifts, including cosmic microwave background (CMB) radiation \cite{Hinshaw_2013}, baryon acoustic oscillations (BAO) \cite{eBOSS:2020yzd}, observational Hubble data (OHD), and SNIa, support the $\Lambda$CDM model. These observations indicate that dark matter constitutes approximately $27\%$ of the total content of the Universe. In comparison, the cosmological constant $\Lambda$ accounts for about $68\%$, meaning that these two enigmatic components together constitute roughly $95\%$ of the total content of the Universe \cite{Planck:2020}. Despite their significant contribution, the fundamental nature and properties of DM and DE remain some of the greatest mysteries in cosmology and fundamental physics.

Although the $\Lambda$CDM model has proven to be a remarkable framework for explaining the dynamics of the Universe, it still faces significant theoretical and observational challenges that need to be addressed \cite{DiValentino:2022fjm_challenges}. 
At galactic scales, a discrepancy exists between the predictions from cold dark matter $\Lambda$CDM simulations and astronomical observations of mass distributions in dwarf galaxies and other low-mass systems. While $\Lambda$CDM simulations predict that DM halos have a density that increases sharply toward the center, forming a high-density "cusp" \cite{Navarro1997}, the observations of the kinematics of dwarf and low surface brightness galaxies show constant or nearly constant central density profiles \cite{dwarfcuspycoreAmorisco_2011}; this is known as the cusp-core problem.
Another problem is that the Hubble diagram of a sample of 1598 quasars up to $z = 5.5$ showed a discrepancy of approximately $4\sigma$ with the $\Lambda$CDM prediction \cite{quasarhubbleproblem1}. Other concerns regarding the cosmological constant include the coincidence problem, which explains why DE dominates the Universe at late times, and the fine-tuning issue related to the discrepancy between the $\Lambda$ value from quantum field theory and cosmological data \cite{Weinberg:1988cp, Copeland:2006wr}.
One of the most critical problems is the well-known Hubble tension, which refers to a discrepancy of $\sim 5.5\sigma$ in the $H_0$ estimate from early universe measurements, such as the CMB, and late universe observations, such as SNIa data. Several reviews of this tension are discussed in \cite{Freedman_2019,Freedman_2020,H0_review_Shah_2021,Di_Valentino_2021,H0_review_verde,H0_review_Tully2024,H0_review_Khalife_2024}.
In general, there are two approaches: either to propose DM and DE components with specific properties distinct from those of CDM or a cosmological constant, or to modify Einstein gravity, with or without dark fluids. For the former, understanding the equation of state (EoS) of DM and DE and consequently the evolution and fate of our Universe is crucial. A standard description of the cosmic components is through a barotropic EoS as $p = w \rho$ where $p$ is the pressure of the component, and $\rho$ is its density. For instance, the EoS for baryonic or dark matter is $w=0$ (dust), whereas for radiation, it is $w=1/3$. A EoS $w=-1$ corresponds to a cosmological constant, while $-1<w <-1/3$ characterizes quintessence,  and $w<-1$ describes a phantom field \cite{comparinphquitessce}.
\\
Several studies have explored EoS for DM and DE, which differ from the standard model. Many of these studies consider time-varying EoS through various parametrizations, as discussed in \cite{Chevallier:2000qy, Linder:2002et, Nojiri_EDE:2021, Escamilla:2023oce, diValentino:2020, RoyChoudhury:2024wri}. In this context, some authors claim that some DE models with a dynamical EoS could solve the coincidence or the Hubble tension problems \cite{Yang_IDE:2018}. Although large-scale N-body simulations suggest that DM is cold \cite{Angulo:2021kes}, its true nature remains unknown. Several studies have tested the consistency of the non-coldness condition of dark matter by relaxing the condition that $w=0$ \cite{Muller:2004yb,avelino2012testing, Armendariz-Picon:2013jej,Barranco:2015,Kopp:2018zxp,Non-zeroEOS_LI,Ghose:2021EPJC, abedin2025darkmatterheatsup}. \cite{avelino2012testing} investigated a cosmological model in which both the DM and DE EoS differ from those of cold dark matter and the cosmological constant. Using a statistical analysis, the authors constrained the model parameters with data from SNIa (Union 2.1 sample), observational Hubble parameter, BAO, and CMB (WMAP-7) data. Their results suggest the presence of warm dark matter $(w_{dm}\sim 0.006)$ and phantom dark energy $(w_{de}\sim -1.1)$. 

In this work, we revisit this cosmological model to constrain the EoS of DM and DE using recent observational datasets, including SNIa (Pantheon+), OHD, strong lensing systems (SLS), BAO, and their joint analysis. A dynamical system analysis is conducted to identify equilibrium points and explore the universe's evolutionary behavior. 
\\
The manuscript is organized as follows: Section \ref{COSMOMODEL} shows the mathematical framework of the cosmological model. Section \ref{dynsyst} presents a detailed dynamical system analysis to find the critical points for the model parameters and determine the dynamical behavior of the universe.
Section \ref{constraintsdatasamples} describes the cosmological data, the merit functions to constrain the cosmological parameters, and the statistical tools to compare the $\Lambda$CDM and $w_{dm}-w_{de}$ models with different priors and data.
Section \ref{results} shows the 
results for the EoS for DM and DE using the different data samples. Finally, in Section \ref{sec:conclusiones}, we discuss the results obtained and, with this, gain insight into the cosmological model, concluding how well we can describe the universe. 

\section{The cosmological model} \label{COSMOMODEL}

We begin our analysis by presenting the standard Einstein equations as
\begin{align} \label{EFE}
G_{\mu\nu}  &= 8\pi G T_{\mu\nu},
\end{align}
where $ G_{\mu\nu} $ is the Einstein tensor which is given by  $R_{\mu\nu} - \frac{1}{2} g_{\mu\nu} R$, $R_{\mu\nu}$ is the Ricci tensor, $R$ is the Ricci scalar,
$ G $ is the gravitational constant, and $T_{\mu\nu} $ is the energy-momentum tensor. We consider the Friedmann-Lema\^{i}tre-Robertson-Walker (FLRW) metric given by

\begin{equation}\label{dsmetric}
ds^2 = -dt^2 + a(t)^2 \left(\frac{dr^2}{1 - kr^2} + r^2(d\theta^2 + \sin^2\theta \, d\phi^2)\right),
\end{equation}
where $a(t)$ is the scale factor and $k$ is the spatial curvature.

Resolving the Einstein equations for the FLRW metric, we obtain the Friedman equation as 
\begin{equation}\label{Friedmanneq}
    H^2 \equiv \left(\frac{\dot{a}}{a}\right)^2 = \frac{8\pi G}{3}\sum_i\rho_i - \frac{k}{a^2},
\end{equation}
and the Raychaudhuri equation is
\begin{equation}
    \dot{H} - \frac{k}{a^2}=-4 \pi  G \sum_i(p_i+ \rho_i),
\end{equation}
where $H$ is the Hubble parameter, $\rho_i$ is the density of the universe components, $p_i$ is the pressure, being $i=dm, de$, dark matter and dark energy respectively, and the dot is the derivative with respect to time. Considering that the Universe is composed of radiation, baryonic matter, dark matter, and dark energy, with the dark components described by a barotropic equation of state (EoS), the Friedmann equation \eqref{Friedmanneq}, coupled with the continuity equations, can be expressed as:

\begin{equation}\label{eq4}
H^2+\frac{k}{a^2} = \frac{8 \pi G}{3}\left(\rho_r + \rho_b + \rho_{de} + \rho_{dm}\right),
\end{equation}
\begin{equation}\label{eq5}
\dot{\rho}_r + 4H\rho_r=0,
\end{equation}
\begin{equation}\label{eq6}
\dot{\rho}_b + 3H\rho_b=0,
\end{equation}
\begin{equation}\label{eq7}
\dot{\rho}_{i} + 3H\rho_{i}(1 + w_{i})=0,
\end{equation}
and the Raychaudhuri equation 
\begin{equation}
 \dot{H}= \frac{k}{a^2}+\frac{4\pi  G }{3} [-4 \rho_r-3 (\rho_{b}+ \rho_{de}+ \rho_{dm}+w_{de}\, \rho_{de}+ w_{dm}\, \rho_{dm})],
\end{equation}
where the constants $w_i$ represent the EoS parameter for each fluid. The conservation equations \eqref{eq5}-\eqref{eq7} have the respective solutions in terms of the scale factor: $\rho_r(a) = \rho_{r0}\,a^{-4}$, $\rho_b(a) = \rho_{b0}\,a^{-3}$, $\rho_{i}(a) = \rho_{0i}\,a^{-3(1+w_{i})}$, where the subscript zero at $\rho_{i0}$ indicates the present-day values of the respective matter-energy densities and $i=dm,de$. We assume a spatially flat FLRW cosmology ($k=0$). However, we reinsert $k$ and investigate the effect of curvature in Sect. \ref{Sect:curvature}. Using these equations on the Friedmann equation \eqref{eq4}, and dividing by the Hubble
parameter evaluated today $H_0$ is obtained
\begin{equation}\label{eq11}
\begin{split}
E^2(a) \equiv \frac{H^2(a)}{H_0^2} = \frac{8\pi G}{3H_0^2}\left(\frac{\rho_{r0}}{a^4} + \frac{\rho_{b0}}{a^3} +  \sum_{i=dm,de} \frac{\rho_{i0}}{a^{3(1+w_{{i}})}}\right),
\end{split}
\end{equation}
where we assume $w_{dm}$ and $w_{de}$ are constant. 
From eq. \eqref{eq11} we obtain
\begin{equation}\label{eq13}
\begin{split}
E^2(z) &= \Omega_{r0}\,(1+z)^{4} +  \Omega_{b0}\,(1+z)^{3}  + \sum_{i=dm,de} \Omega_{i0}\,(1+z)^{3(1+w_{i})}. 
\end{split}  
\end{equation}
where the relation between the scale factor $a$ and the redshift $z$ is given by $a = (1 + z)^{-1}$ and the dimensionless parameter for the densities of the parameters is defined as $\Omega_{i0} = \rho_{i0}/\rho_{0\text{crit}}$, where $\rho_{0\text{crit}}$ is the critical density evaluated today defined as $\rho_{0\text{crit}} = 3H_0^2/8\pi G$.
Setting $E(z = 0) = 1$, we have the flatness condition
\begin{flalign}\label{OmegaDE}
\hspace{1cm} \Omega_{de0} = 1 - (\Omega_{r0} + \Omega_{b0} + \Omega_{dm0}). \hspace{2cm}
\end{flalign}
We can define the deceleration parameter $q(z)$ using equation \eqref{eq13} and \eqref{OmegaDE}
\begin{equation}\label{eq15}
q(z) = \frac{(z+1)}{{E(z)}}\frac{dE(z)}{dz} - 1.
\end{equation}
Using equation \eqref{eq13} and \eqref{OmegaDE}, we can define the time scale of the universe in function of redshift as
\begin{equation}\label{eq16}
 t(z)=\frac{1}{H_0} \int_z^{\infty} \frac{1}{(1 + z)\,E(z)}  dz.
\end{equation} 
Here the time $t(z)$ is in [Gyr]. 

\section{Dynamical system analysis}
\label{dynsyst}

Recent studies apply dynamical systems theory to various cosmological models, including interacting dark energy, non-canonical scalar fields, and modified gravity \cite{Copeland:1997et, Barreiro:1999zs, Copeland:2006wr}. These analyses refine theoretical frameworks by identifying viable scenarios and ruling out others \cite{Pourtsidou:2013nha, Bahamonde:2017ize}. Standard mathematical tools, as outlined in reference works \cite{wainwright_ellis_1997, Coley:2003mj, Leon:2012ccj}, combined with observational data, provide a structured method for evaluating the physical behavior of these models \cite{Hernandez-Almada:2020uyr, Leon:2021wyx, Hernandez-Almada:2021rjs, Hernandez-Almada:2021aiw}.

\subsection{Flat FLRW model}
\label{sect:3.1}
For these kinds of models, it is customary to define the dimensionless variables: 
\begin{equation}\label{eqOmegas11}
\begin{split}
 \left(\Omega_{r}, \Omega_{b}, \Omega_{dm}, \Omega_{de}\right)\equiv \frac{8\pi G}{3H^2}\left(\rho_{r} , \rho_{b} ,  \rho_{dm}, \rho_{de}\right),
\end{split}
\end{equation}
which satisfy
\begin{equation}\label{Fried_2}
 \Omega_{b}+\Omega_{de}+\Omega_{dm}+\Omega_{r}=1,
\end{equation}
and the logarithmic time variable 
\begin{equation}
    \tau = \ln a, \quad f^{\prime}=\frac{d f}{d \tau}= \frac{1}{H} \dot{f}. 
\end{equation}

We assume $w_{dm}=0$ and $\Omega_b=0$ as a first step for simplicity.
The field equations become 
\begin{subequations}
\label{ds1}
\begin{align}
& \Omega_r^{\prime}= 2 (q-1)\,\Omega_r, \\ 
& \Omega_{dm}^{\prime}= (2 q-1)\,\Omega_{dm},\\
& \Omega_{de}^{\prime}= \Omega_{de}\,(2 q-3 w_{de}-1),
\end{align}
\end{subequations}
together with the Raychaudhuri equation
\begin{equation}
    \dot{H}= -\frac{3}{2}H^2\left(\Omega_{dm}+\frac{4 \Omega_r}{3}+(w_{de}+1) \Omega_{de}\right),
\end{equation}
that is used to define the deceleration parameter 
\begin{equation}
    q\equiv -1- \frac{\dot{H}}{H^2}= \frac{3}{2} (w_{de}+1)\Omega_{de}+\frac{3 \Omega _{dm}}{2}+2
   \Omega_{r}-1. 
\end{equation}
Using \eqref{Fried_2} to reduce the system dimensionality and solving for $\Omega_{dm}$, 
\begin{equation}
\Omega_{dm}= 1-\Omega_{de}-\Omega_{r},
\end{equation} 
replacing in the definition of $q$, we have
\begin{equation}\label{decc-param}
    q= \frac{1}{2} (3 w_{de}\,\Omega_{de}+\Omega_{r}+1).
\end{equation}
Substituting in the dynamical equations, we obtain the reduced dynamical system 
\begin{subequations}
\label{ds2}
\begin{align}
& \Omega_r^{\prime}= \Omega_{r} (3 w_{de} \,\Omega_{de}+\Omega_{r}-1),\\ 
& \Omega_{de}^{\prime}=\Omega_{de} (3 w_{de} (\Omega_{de}-1)+\Omega_{r}),
\end{align}
\end{subequations}
defined on the compact phase space
\begin{align}
\left\{ (\Omega_r,  \Omega_{de}) \in \mathbb{R}^2: \Omega_r\geq 0,  \Omega_{de}\geq 0,   0\leq \Omega_{de}+\Omega_{r}\leq 1\right\}. \label{COMPACT_1}
\end{align}

The free parameter $w_{de}$ is theoretically considered within the range $-1 \leq w_{de} \leq 1$. In the regime $w_{de} < -1/3$, $\Omega_{de}$, characterized by  $w_{de}$, properly corresponds to  dark energy fluid. However, we also theoretically allow the regime $w_{de} \in [0,1]$, where $\Omega_{de}$ mimics the background fluid. Therefore, when $w_{de} = 1/3$, it behaves like radiation.

\begin{table}[]
    \centering
\begin{tabular}{|ccccccc|}\hline 
Label & $\Omega_r$ & $\Omega_{dm}$ & $\Omega_{de}$  & $k_1$ & $k_2$  & $q$ \\\hline 
R & $1$ & $0$ & $0$ & $1$ &  $1-3 w_{de}$ & $1$\\\hline
DM & $0$ & $1$ & $0$ & $-1$ &  $-3 w_{de}$ & $\frac{1}{2}$ \\\hline 
DE& $0$ & $0$ & $1$  & $3 w_{de}$  & $3
   w_{de}-1$ & $\frac{1}{2} (3 w_{de}+1)$ \\\hline
\end{tabular}
    \caption{Equilibrium points of system \eqref{ds2} for a flat FLRW model with $w_{dm}=0$ and $\Omega_b=0$. The eigenvalues $k_1$ and $k_2$ of the Jacobian matrix of the system, as well as the deceleration parameter evaluated at the equilibrium points, are also shown.} We assume $w_{de}<0$.
    \label{tab:Eq-Points}
\end{table}

To assess the stability of equilibrium points in the dynamical system \eqref{ds2}, we compute the Jacobian matrix

\begin{equation}
J =
\begin{pmatrix}
3 w_{de} \Omega_{de} + 2 \Omega_r - 1 & 3 w_{de} \Omega_r \\
\Omega_{de} & 6 w_{de} \Omega_{de} - 3 w_{de} + \Omega_r
\end{pmatrix}
\end{equation}

The stability of each equilibrium point in the system \eqref{ds2} is determined by the eigenvalues $k_1$ and $k_2$ of the Jacobian matrix. Negative real parts indicate attractors, while positive values correspond to sources or saddle points. This classification, rooted in standard dynamical systems theory \cite{wainwright_ellis_1997, Coley:2003mj, Leon:2012ccj}, offers a robust framework for interpreting cosmic dynamics. The deceleration parameter $q$ complements this analysis by linking mathematical stability to physical expansion phases, radiation, matter, or dark energy dominance.

Table \ref{tab:Eq-Points} summarizes the equilibrium points of system \eqref{ds2}, along with their associated eigenvalues and deceleration parameters.

The equilibrium point, DM, corresponds to a solution dominated by dark matter. This solution resembles the behavior of the dark matter sector and exhibits deceleration, characterized by a deceleration parameter $q = \frac{1}{2}$. The points along this line are classified as saddle points when $w_{de} < 0$. 
Consequently, this matter-dominated solution cannot act as a late-time attractor in standard cosmology.

The radiation-dominated solution, denoted R, is characterized by a deceleration factor $q = 1$. This equilibrium point acts as a source for $w_{de} < \frac{1}{3}$ or as a saddle if $w_{de} > \frac{1}{3}$. However, based on our earlier discussion, the relevant physical region is $w_{de} < 0$. Therefore, assuming $w_{de} < 0$, the radiation-dominated solution behaves as a source.

\begin{figure}[H]
    \centering
\includegraphics[scale=0.6]{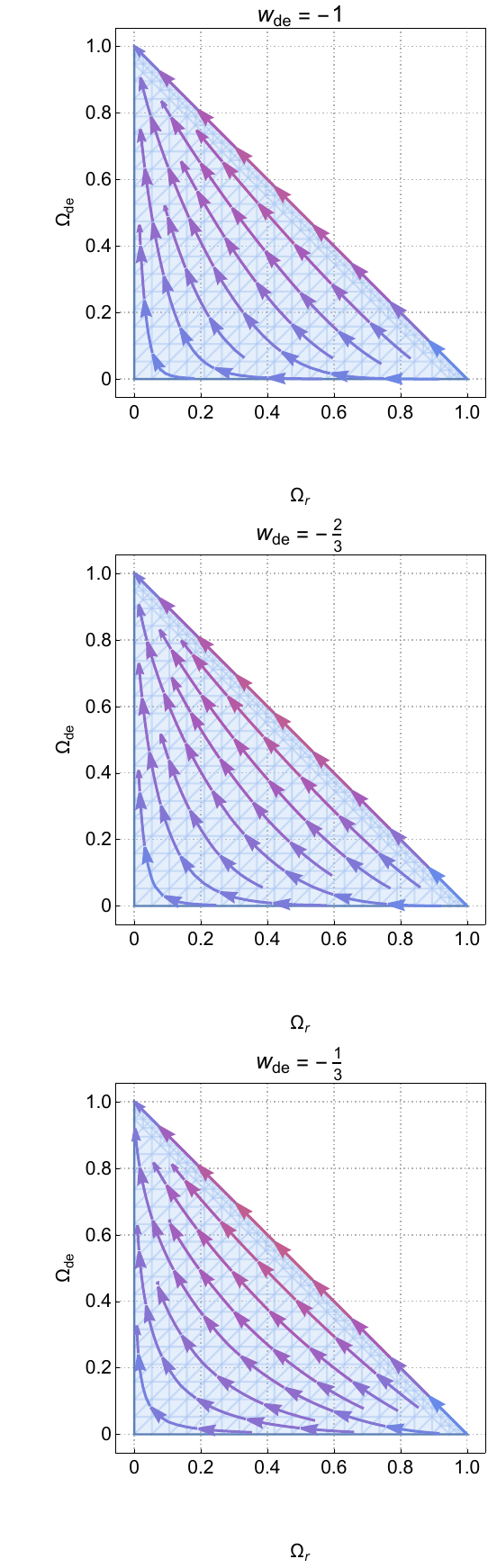}
    \caption{$\Omega_{de}$-$\Omega_{r
    }$ phase planes of system \eqref{ds2} for $w_{de}= -1, -2/3, -1/3$.}
    \label{fig:DS}
\end{figure}

Lastly, the solution dubbed DE corresponds to a state dominated by dark energy. It functions as a sink provided $w_{de} < 0$. Notably, this solution is accelerated when $q = \frac{1}{2} (3 w_{de} + 1) < 0$, which leads to the condition $w_{de} < -\frac{1}{3}$, meaning that the solution is accelerated in the quintessence regime. Additionally, this solution can represent a phantom state when $w_{de} < -1$. 
Therefore, we can represent the dynamics in two-dimensional phase planes as shown in Fig. \ref{fig:DS} for $w_{de}= -1, -2/3, -1/3$.

\subsubsection{Scaling Solutions and Their Dynamical Implications}

Scaling solutions are of particular interest in cosmology because they describe scenarios in which the energy densities of different components, such as radiation, dark matter, dark energy, and curvature, evolve proportionally over time. These solutions are especially relevant for addressing the so-called coincidence problem.

When $w_{de} = 1/3$, and assuming $\Omega_{de} + \Omega_{r} < 1$ (i.e., $\Omega_{m} > 0$), we have from \eqref{ds2} that the ratio of the derivatives of the density parameters is equal to the ratio of the parameters themselves. This implies that the ratio $\Omega_{de} / \Omega_r$ is constant with respect to the evolution variable (often the number of e-folds $N = \ln a$, or redshift $z$)
\begin{equation*}
\frac{\Omega_{de}^{\prime}}{\Omega_r^{\prime}} = \frac{\Omega_{de}}{\Omega_{r}} \implies \frac{d}{dN} \left( \frac{\Omega_{de}}{\Omega_r} \right) = 0 \Rightarrow \frac{\Omega_{de}}{\Omega_r} = \text{const.}
\end{equation*}
This result indicates that the dark energy density parameter and the radiation density parameter evolve in proportion, maintaining a constant ratio throughout cosmic evolution in the considered regime. That is, we have a dark energy–radiation scaling solution. This regime is absent if we choose $w_{de} \neq 1/3$.

Similar statements follow if, for instance, $w_{de} = 0$, when $\Omega_{de}$ mimics dust. We start from the relation
\begin{equation*}
\frac{\Omega_{de}^{\prime}}{\Omega_r^{\prime}} = \frac{\Omega_{de}}{\Omega_{r}-1}.
\end{equation*}

By solving this equation, it is obtained
\begin{equation*}
|\Omega_{de}| = A |1 - \Omega_r|,
\end{equation*}
with $A = e^C$ a positive constant.

Assuming $\Omega_{de} > 0$ and $\Omega_r < 1$, we drop the absolute values
\begin{equation*}
\Omega_{de} = A (1 - \Omega_r) \propto \Omega_{m}.
\end{equation*}
That is, we have a dark energy–dark matter scaling solution. This regime is absent if we choose $w_{de} \neq 0$.

From physical considerations, we regard the regime $w_{de} < -1/3$ as the physically relevant one. This regime forbids scaling solutions. However, we treat $w_{de}$ as a free parameter to be constrained through Bayesian analysis. 

The combination of dynamical systems techniques and Bayesian analysis provides a robust framework for investigating the physical behavior of cosmological models~\citep{Hernandez-Almada:2020uyr, Leon:2021wyx, Hernandez-Almada:2021rjs, Hernandez-Almada:2021aiw, Garcia-Aspeitia:2022uxz}.

\subsection{Dynamics with Constant Equation-of-State Parameters and Baryonic Matter}
After this initial exploration, let us consider the constant EoS as 
$w_{dm}\neq 0$ and $w_{de}$ and assume $\Omega_{b}\geq 0$. In this case,
\begin{equation}
\label{decc_2}
 q= \frac{1}{2} (3 w_{de} \Omega_{de}+3 w_{dm} \Omega_{dm}+\Omega_r+1),    
\end{equation}
and the matter equation is not decoupled: 
\begin{subequations}\label{systA}
\begin{align}
   & \Omega_r^{\prime} = \Omega_r (3 w_{de} \Omega_{de}+3 w_{dm} \Omega_{dm}+\Omega_r-1),\\
   & \Omega_{dm}^{\prime} = \Omega_{dm} (3 w_{de} \Omega_{de}+3 w_{dm} (\Omega_{dm}-1)+\Omega_r),\\
   & \Omega_{de}^{\prime} = \Omega_{de} (3
   w_{de} (\Omega_{de}-1)+3 w_{dm} \Omega_{dm}+\Omega_r),
\end{align}
\end{subequations}
defined on the compact phase-space 
\begin{align}
\left\{ (\Omega_r, \Omega_{dm}, \Omega_{de}) \in \mathbb{R}^3: \Omega_r\geq 0, \Omega_{dm}\geq 0, \Omega_{de}\geq 0,  \right. \nonumber \\
\left. 0\leq \Omega_{de}+\Omega_{dm}+\Omega_{r}\leq 1\right\}. \label{COMPACT_1B}
\end{align}
\begin{table*}[]
    \centering
    \begin{tabular}{|cccccccc|}\hline
    Label & $\Omega_{r}$ & $\Omega_{dm}$ & $\Omega_{de}$  & $k_1$ & $k_2$ & $k_3$ & $q$ \\\hline
   B & $0$ & $0$ & $0$  & $-1$ & $-3 w_{de}$ & $-3 w_{dm}$ & $\frac{1}{2}$  \\\hline
   R &  $1$ & $0$ & $0$  & $1$ & $1-3 w_{de}$ & $1-3 w_{dm}$ & $1$  \\\hline
   DM & $0$ & $1$ & $0$  & $-3\,(w_{de}-w_{dm})$ & $3 w_{dm}$
   & $3 w_{dm}-1$ & $\frac{1}{2} (3 w_{dm}+1)$  \\\hline
   DE & $0$ & $0$ &$1$  & $3 w_{de}$ & $3 w_{de}-1 $& $3
   (w_{de}-w_{dm})$ & $\frac{1}{2} (3 w_{de}+1)$  \\\hline
    \end{tabular}
    \caption{Equilibrium points of system \eqref{systA} for a flat FLRW model that includes radiation, baryons, DM, and DE. The eigenvalues $k_1$, $k_2$ and $k_3$ of the Jacobian matrix of the system, as well as the deceleration parameter evaluated at the equilibrium points, are also shown.}
    \label{tab:systA}
\end{table*}

As in Section~\ref{sect:3.1}, we calculate the Jacobian matrix, determine the eigenvalues $k_1$, $k_2$, and $k_3$, and evaluate the deceleration parameter at the equilibrium points for the system \eqref{systA}.

Define the vector field $\mathbf{F} = (f_1, f_2, f_3)^T$ corresponding to the right-hand sides of the system. The Jacobian matrix $J$ is given by:
\begin{equation*}
J = \begin{pmatrix}
\frac{\partial f_1}{\partial \Omega_r} & \frac{\partial f_1}{\partial \Omega_{dm}} & \frac{\partial f_1}{\partial \Omega_{de}} \\
\frac{\partial f_2}{\partial \Omega_r} & \frac{\partial f_2}{\partial \Omega_{dm}} & \frac{\partial f_2}{\partial \Omega_{de}} \\
\frac{\partial f_3}{\partial \Omega_r} & \frac{\partial f_3}{\partial \Omega_{dm}} & \frac{\partial f_3}{\partial \Omega_{de}}
\end{pmatrix}
\end{equation*}
where
\begin{align*}
\frac{\partial f_1}{\partial \Omega_r} &= 3 w_{de} \Omega_{de} + 3 w_{dm} \Omega_{dm} + 2\Omega_r - 1, \quad 
\frac{\partial f_1}{\partial \Omega_{dm}} = 3 w_{dm} \Omega_r, \quad 
\frac{\partial f_1}{\partial \Omega_{de}} = 3 w_{de} \Omega_r, \\
\frac{\partial f_2}{\partial \Omega_r} &= \Omega_{dm}, \quad 
\frac{\partial f_2}{\partial \Omega_{dm}} = 3 w_{de} \Omega_{de} + 6 w_{dm} \Omega_{dm} - 3 w_{dm} + \Omega_r, \quad
\frac{\partial f_2}{\partial \Omega_{de}} = 3 w_{de} \Omega_{dm}, \\
\frac{\partial f_3}{\partial \Omega_r} &= \Omega_{de}, \quad
\frac{\partial f_3}{\partial \Omega_{dm}} = 3 w_{dm} \Omega_{de}, \quad
\frac{\partial f_3}{\partial \Omega_{de}} = 6 w_{de} \Omega_{de} - 3 w_{de} + 3 w_{dm} \Omega_{dm} + \Omega_r.
\end{align*}
Equilibrium points satisfy $\Omega_r' = \Omega_{dm}' = \Omega_{de}' = 0$.
At each equilibrium point, we evaluate the Jacobian matrix and solve the characteristic equation
$
\det(J - k I) = 0
$
to obtain the eigenvalues $k_1$, $k_2$, and $k_3$. Further analysis of the eigenvalues will determine the stability of each equilibrium point.
Evaluating $q$ \eqref{decc_2} at each equilibrium point, we obtain characteristic cosmological solutions for radiation, matter and dark energy dominated epochs.

Table \ref{tab:systA} gives the 
equilibrium points of system \eqref{systA}. The corresponding eigenvalues ($k_1$, $k_2$ and $k_3$) and the deceleration parameter ($q$) related to these equilibrium points are also presented. Point B corresponds to a Baryon-dominated solution. The deceleration parameter $q=\frac{1}{2}$ corresponds to a decelerated pressureless solution. It is a sink only if $w_{de}>0$ and $w_{dm}>0$, i.e., when the equation of state of DE mimics the matter EoS. Otherwise, it is a saddle. However, $w_{de}<-1/3$ is required to obtain acceleration, which is the physical situation. Hence, in this regime, B is a saddle, as expected. 

\begin{table*}[H]
    \centering
    \begin{tabular}{|cccccccc|}\hline
    Label & $\Omega_{r}$ & $\Omega_{dm}$ & $\Omega_{de}$  & $k_1$ & $k_2$ & $k_3$ & $q$ \\\hline
   B & $0$ & $0$ & $0$  & $-1$ & $-3 w_{de}$ & $-3 w_{dm}$ & $\frac{1}{2}$  \\\hline
   R &  $1$ & $0$ & $0$  & $1$ & $1-3 w_{de}$ & $1-3 w_{dm}$ & $1$  \\\hline
   DM & $0$ & $1$ & $0$  & $-3\,(w_{de}-w_{dm})$ & $3 w_{dm}$
   & $3 w_{dm}-1$ & $\frac{1}{2} (3 w_{dm}+1)$  \\\hline
   DE & $0$ & $0$ &$1$  & $3 w_{de}$ & $3 w_{de}-1 $& $3
   (w_{de}-w_{dm})$ & $\frac{1}{2} (3 w_{de}+1)$  \\\hline
    \end{tabular}
    \caption{Equilibrium points of system \eqref{systA} for a flat FLRW model that includes radiation, baryons, DM, and DE. The corresponding eigenvalues and deceleration parameters are also presented.}
    \label{tab:systA}
\end{table*}

Point R represents a Radiation-dominated solution with deceleration parameter $q=1$. It is a source for $w_{de}<1/3$ and $w_{dm}<1/3$ (the physical situation), or a saddle for either $w_{de}>1/3$ and $w_{dm}>1/3$. 
Point DM is a matter-dominated solution with deceleration parameter $q=\frac{1}{2} (3 w_{dm}+1)$. It is a source for $w_{de}<w_{dm}$ and $w_{dm}>0$, or a saddle for either $w_{de}>w_{dm}$ or $w_{dm}>0$. Cosmic deceleration occurs when $w_{dm}>-1/3$. The physical regime for DM is defined by $w_{dm}>0$. 
Point DE is a DE-dominated solution with deceleration parameter $q=\frac{1}{2} (3 w_{de}+1)$. It is a sink for $w_{de}<0$, and  $w_{de}<1/3$ and  $w_{de}<w_{dm}$.  As expected, cosmic acceleration occurs when $w_{de}<-1/3$.

\begin{figure}[H]
    \centering
\includegraphics[scale=0.9]{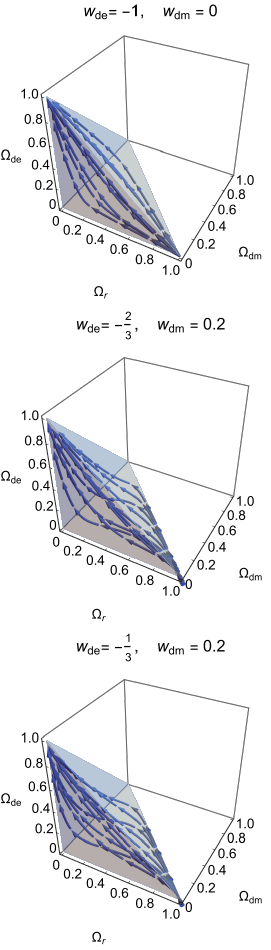}
    \caption{Phase space of system \eqref{systA} for  $(w_{de},w_{dm})= (-1,0),  (-2/3, 0.2), (-1/3, 0.2)$.}
    \label{fig:DS2}
\end{figure}

To finish this section, we can represent the dynamics in three-dimensional phase space as shown in Fig.  \ref{fig:DS2} for the values of the parameters $(w_{de},w_{dm})= (-1,0),  (-2/3, 0.2), (-1/3, 0.2)$. 

Cosmological observations will give the final parameter values, and we will discuss them in detail in section \ref{constraintsdatasamples}.

\subsubsection{Scaling Solutions and Their Dynamical Implications}

In the context of system~\eqref{systA}, a scaling solution corresponds to an equilibrium point where the ratios $\Omega_{dm} / \Omega_{de}$ and $\Omega_r / \Omega_{de}$ remain constant. Mathematically, such solutions arise when the dynamical system admits a stable equilibrium point with non-zero contributions from multiple components, i.e., $\Omega_{dm} \neq 0$, $\Omega_{de} \neq 0$, and $\Omega_r \neq 0$. These points satisfy:
\begin{equation*}
\Omega_r' = 0, \quad \Omega_{dm}' = 0, \quad \Omega_{de}' = 0,
\end{equation*}
with all $\Omega_i$ bounded and non-zero. The stability of these points, determined by the eigenvalues of the Jacobian matrix, indicates whether the system naturally evolves toward a scaling regime. If the real parts of all eigenvalues are negative, the scaling solution is stable and may act as an attractor in cosmic evolution.

The existence and nature of scaling solutions depend sensitively on the equation-of-state parameters $w_{dm}$ and $w_{de}$. For instance, choices such as $w_{de} < -1/3$ can allow for accelerated expansion while maintaining a non-negligible dark matter component, leading to viable scaling behavior.

To explore these dynamics, we consider the autonomous system governing the evolution of the density parameters $\Omega_r$, $\Omega_{dm}$, and $\Omega_{de}$, corresponding to radiation, dark matter, and dark energy, respectively.

\subsubsection{Bifurcation Analysis}

The equilibrium points of the system occur when $\Omega_i' = 0$ for all $i$. A notable equilibrium point is DE:
\begin{equation*}
\Omega_r = 0, \quad \Omega_{dm} = 0, \quad \Omega_{de} = 1,
\end{equation*}
representing a dark energy-dominated universe. Linearizing around this point yields:
\begin{align}
\Omega_r' &\approx \Omega_r (3 w_{de} - 1), \\
\Omega_{dm}' &\approx \Omega_{dm} (3 w_{de} - 3 w_{dm}), \\
\Omega_{de}' &\approx \Omega_{de} (3 w_{de}).
\end{align}
The eigenvalues of the linearized system are:
\begin{equation*}
\lambda_r = 3 w_{de} - 1, \quad
\lambda_{dm} = 3 w_{de} - 3 w_{dm}, \quad
\lambda_{de} = 3 w_{de}.
\end{equation*}

Bifurcations occur when these eigenvalues cross zero: i) $w_{de} = \frac{1}{3}$, radiation perturbations change stability; ii) $w_{de} = w_{dm}$, dark matter perturbations change stability; iii) $w_{de} = 0$, dark energy perturbations change stability. These transitions mark qualitative changes in the system's dynamics, such as the onset of acceleration or the dominance of a particular component.

We can obtain an analytical integration for the following cases.

Case 1: $w_{de} = 0$, $w_{dm} = 0$ ($\Lambda$CDM-like).  
The system simplifies to:
\begin{align}
\Omega_r' &= -\Omega_r (1-\Omega_r ), \\
\Omega_{dm}' &= \Omega_{dm} \Omega_r, \\
\Omega_{de}' &= \Omega_{de} \Omega_r.
\end{align}
Solving the first equation yields:
\begin{equation}
\Omega_r(N) = \frac{1}{1 + C e^{N}},
\end{equation}
where $C$ is an integration constant. The other components scale as
\begin{eqnarray}
\Omega_{de} &=& \frac{c_2 e^N}{C e^N+1}, \\ \Omega_{dm} &=&
   \frac{c_3 e^N}{C e^N+1}.
\end{eqnarray}
Using the constriction $\Omega_{de} + \Omega_{dm}+ \Omega_{r}=1$ we obtain the restriction for the parameters $C=c_2+c_3$.
Where the connection with the current energy density values is through 
\begin{equation*}C=\frac{1}{\Omega_{r0}}-1,\quad c_2=
   \frac{\Omega_{de0}}{\Omega_{r0}}, \quad c_3=
   \frac{\Omega_{dm0}}{\Omega_{r0}}.
\end{equation*}

Case 2: $w_{de} = \frac{1}{3}$, $w_{dm} = 0$. 
The system becomes:
\begin{align}
\Omega_r' &= \Omega_r (\Omega_{de} + \Omega_r - 1), \\
\Omega_{dm}' &= \Omega_{dm} (\Omega_{de} + \Omega_r), \\
\Omega_{de}' &= \Omega_{de} (\Omega_{de} + \Omega_r -1).
\end{align}
Integrating the system, we obtain 
\begin{align}
    \Omega_{r} & = \frac{\Omega_{r0}}{\Omega_{de0}+\Omega_{r0}+e^N (1-\Omega_{de0}-\Omega_{r0})},\\
   \Omega_{de} & = 
   \frac{\Omega_{de0}}{\Omega_{de0}+\Omega_{r0}+e^N (1-\Omega_{de0}-\Omega_{r0})},\\
   \Omega_{dm} & =  \frac{e^N \Omega_{dm0}}{\Omega_{de0}+\Omega_{r0}+e^N (1-\Omega_{de0}-\Omega_{r0})}.
\end{align}
Case 3: $w_{de} = w_{dm} = w$.
Define $\Omega_x = \Omega_{dm} + \Omega_{de}$. The system reduces to
\begin{align}
\Omega_r' &= \Omega_r (3w \Omega_x + \Omega_r - 1), \label{eq:omegar}\\
\Omega_x' &= \Omega_x (3w (\Omega_x - 1) + \Omega_r). \label{eq:omegax}
\end{align}

Although this two-dimensional system is not analytically solvable, its qualitative behavior can be explored through phase plane analysis. To address broader implications, we highlight the role of center manifold techniques in such contexts.

When the eigenvalues vanish, we apply Center Manifold Analysis to study the stability of the fixed point $(\Omega_x, \Omega_r) = (1, 0)$. To simplify the analysis, we shift coordinates:
\[
u = \Omega_x - 1, \quad v = \Omega_r,
\]
so that the fixed point becomes $(u, v) = (0, 0)$. The system transforms into
\begin{align}
u' &= (u + 1)(3w u + v), \label{eq:vprime}\\
v' &= v(3w u + 3w + v - 1). \label{eq:uprime}
\end{align}

The Jacobian matrix at $(u, v) = (0, 0)$ is
\[
J = \begin{pmatrix}
\partial_u u' & \partial_v u' \\
\partial_u v' & \partial_v v'
\end{pmatrix}_{(0,0)} =
\begin{pmatrix}
3w & 1 \\
0 & 3w - 1
\end{pmatrix}.
\]

The eigenvalues are
\[
3w, \quad 3w - 1.
\]

One of the eigenvalues is zero when $w = 0$ or $w = \frac{1}{3}$.

For $w = 0$, the system becomes
\begin{align}
u' &= (u + 1)v, \label{eq:vprime0}\\
v' &= v(v - 1). \label{eq:uprime0}
\end{align}

We assume the center manifold is of the form
\[
u = -v + h(v),
\]
where the function $h(v)$ satisfies
\[
h(0) = 0, \quad h'(0) = 0.
\]

Differentiating $u = -v + h(v)$ with respect to time and substituting from the system yields
\[
h'(v)(1 - v) + h(v) = 0.
\]

This is a first-order linear ODE. Solving it with the initial conditions $h(0) = 0$, $h'(0) = 0$, we find
\[
h(v) \equiv 0.
\]

Substituting $u = -v$ into the system, the dynamics reduce to
\[
v' = v(v - 1) = -\nabla V(v), \quad \text{where } V(v) = \frac{v^2}{2} - \frac{v^3}{3}.
\]

The center manifold at the origin is
\[
\{(u,v) \in \mathbb{R}^2 : u = -v\},
\]
and the dynamics on it are governed by a gradient flow with a local minimum at $v = 0$. Therefore, the origin is stable.

For $w = \frac{1}{3}$, the system becomes
\begin{align}
u' &= (u + 1)(u + v), \label{eq:vprime1/3}\\
v' &= v(u + v). \label{eq:uprime1/3}
\end{align}

We analyze the stability of the origin using Center Manifold Theory. The center manifold is assumed to be of the form
\[
u = -v + h(v),
\]
where the function $h(v)$ satisfies the regularity conditions
\[
h(0) = 0, \quad h'(0) = 0.
\]

Differentiating the identity $u = -v + h(v)$ with respect to time and substituting from the system yields
\[
(h'(v) - 1)v(u + v) = (u + 1)(u + v).
\]

Substituting $u = -v + h(v)$ into both sides and simplifying, we obtain the identity
\[
\left[h'(v)v - h(v) - 1\right] h(v) = 0,
\]
{which must hold for all sufficiently small $v$.

This equation implies that either $h(v) = 0$ or the bracketed term vanishes. Substituting into the latter, we obtain the differential equation
\[
h'(v)v - h(v) - 1 = 0,
\]
which is a first-order linear ODE. Solving it yields
\[
h(v) = A v - 1,
\]
for some constant $A$. Applying the initial condition $h(0) = 0$ leads to $-1 = 0$, a contradiction. Therefore, the only consistent solution is
\[
h(v) \equiv 0.
\]

Hence, the center manifold is given by $u = -v$, and the dynamics restricted to it reduce to
\[
v' = v(u + v) = v(-v + v) = 0.
\]

The dynamics on the center manifold are thus trivial, and the origin is stable but not asymptotically stable.

\subsubsection{Scaling Behavior}

In general, the energy densities scale as:
\begin{equation}
\rho_i \propto a^{-3(1 + w_i)} \quad \Rightarrow \quad \Omega_i \propto e^{-3(1 + w_i)N},
\end{equation}
Therefore, the scaling behavior depends on the EoS values. In the following we analyze several cases.

Case 1: $w_{de} = 0$, $w_{dm} = 0$}  $\Omega_r \sim \frac{1}{1 + C e^{N}}$: decays exponentially. $\Omega_{dm}, \Omega_{de} \sim 1/(C+ e^{-N})$: grow then saturate. Indeed, considering the function:
\begin{equation}
\Omega_{de,dm}(N) = \frac{c_{2,3}}{C + e^{-N}}, 
\end{equation}
it presents an early and late-time behaviors. The Early-time behavior is given by ($ N \to -\infty $):
\begin{equation*}
e^{-N} \to \infty \quad \Rightarrow \quad \Omega_{de,dm}(N) \approx \frac{1}{e^{-N}} = e^{N} \to 0,
\end{equation*}
and the late-time behavior ($ N \to +\infty $):
\begin{equation*}
e^{-N} \to 0 \quad \Rightarrow \quad \Omega_{de,dm}(N) \to \frac{c_{2,3}}{C}.
\end{equation*}

Thus, $ \Omega_{de,dm} $ grows from near zero and saturates to a constant value $ c_{2,3}/C $ as $ N \to \infty $.
It matches the standard $\Lambda$CDM model with radiation, matter and dark energy domination epochs.

Case 2: $w_{de} = \frac{1}{3}$, $w_{dm} = 0$, 
$\rho_{de} \propto a^{-4}$ scales same as radiation. $\Omega_{de}$ and $\Omega_r$ scale similarly. $\Omega_{dm} \propto a^{-3}$: decays slower. Dark energy does not drive late-time acceleration; it is an unstable scenario.

Case 3: $w_{de} = w_{dm} = w$, unified dark sector: $\rho_x \propto a^{-3(1 + w)}$. If $w > 0$, the dark sector dilutes faster. If $w < 0$, the dark sector dilutes more slowly, and we have a possible acceleration. If $w = -1$, it behaves like a cosmological constant. Thus, there are scaling solutions with a constant $\Omega_{dm} / \Omega_{de}$ ratio.

Next, we will examine the influence of the curvature term on the system's dynamics.

\subsection{Effect of curvature }
\label{Sect:curvature}

To investigate the effect of curvature, we introduce the normalized curvature variable $\Omega_{k}= 1/a^2 H^2$. The new variables satisfy
\begin{equation}\label{Fried_2-curv}
 \Omega_{b}+\Omega_{de}+\Omega_{dm}+\Omega_{r} - k\,\Omega_k=1.
\end{equation}
Let us consider the constant EoS as 
$w_{dm}\neq 0$ and $w_{de}$. 
Then, the deceleration parameter is given by 
\begin{equation}
    q= \frac{1}{2} (k\,\Omega_k+3 w_{de}\,\Omega_{de}+3 w_{dm} \,\Omega_{dm}+\Omega_r+1).
\end{equation}
Then, the dynamical system is given by 
\begin{subequations}\label{Curvature-A}
\begin{align}
& \Omega_r^{\prime}=   \Omega_r (k \Omega_k+3 w_{de} \Omega_{de}+3 w_{dm} \Omega_{dm}+\Omega_r-1),\\
& \Omega_{dm}^{\prime}= \Omega_{dm} [k \Omega_k+3 w_{de} \Omega_{de}+3 w_{dm} (\Omega_{dm}-1)+\Omega_r],\\
& \Omega_{de}^{\prime}= \Omega_{de} [k\, \Omega_k+3 w_{de} (\Omega_{de}-1)+3 w_{dm}
   \Omega_{dm}+\Omega_r],\\
& \Omega_k^{\prime}= \Omega_k (k\, \Omega_k+3 w_{de}\, \Omega_{de}+3 w_{dm} \Omega_{dm}+\Omega_r+1). 
\end{align}
\end{subequations}
defined on phase space
\begin{align}
 \left\{   (\Omega_r, \Omega_{dm}, \Omega_{de}, \Omega_{k})\in \mathbb{R}^3: \Omega_r\geq 0,  \Omega_{dm}\geq 0,   \Omega_{de}\geq 0, \right. \nonumber \\
\left. \Omega_k\geq 0,  0\leq\Omega_{r}+ \Omega_{dm} + \Omega_{de} - k \Omega_k\leq 1\right\}. \label{region3}
\end{align}
Table \ref{tab:Curv} presents the equilibrium points of the system \eqref{Curvature-A}. The corresponding eigenvalues and deceleration parameters are also shown.
Point B is a Baryon-dominated solution. Because two eigenvalues have opposite signs, it is a saddle. The deceleration parameter $q=\frac{1}{2}$ corresponds to a decelerated pressureless solution. 

The point R is a Radiation-dominated solution. It is a source for $w_{de}<1/3$ and $w_{dm}<1/3$ or a saddle otherwise. The deceleration parameter $q=1$ corresponds to a decelerated solution.  

The point labeled DM is a dark-matter-dominated solution. It is a sink for $w_{dm}<w_{de}\leq -\frac{1}{3}$, or $w_{de}>-\frac{1}{3}> w_{dm}$, whence DM mimics a quintessence field. In both cases, it is an accelerated solution. It is a source for $w_{de}\leq \frac{1}{3}<w_{dm}$ or $w_{dm}>w_{de}>\frac{1}{3}$. The first case is when dark matter dominates before radiation, and the second is when DM and DE dominate before radiation. However, these are nonphysical solutions.  It is a saddle when $w_{de}<-\frac{1}{3}, w_{de}<w_{dm}<\frac{1}{3}$ or
   $-\frac{1}{3}<w_{de}<\frac{1}{3},  -\frac{1}{3}<w_{dm}<\frac{1}{3}$ or
   $w_{de}>\frac{1}{3},  -\frac{1}{3}<w_{dm}<w_{de}$. The first choice is the physical one when the DE-dominated solution is accelerated. 

The point DE is a dark energy-dominated solution. The eigenvalues are $3 w_{de}, 3
   w_{de}-1, 3 w_{de}+1, 3 (w_{de}-w_{dm})$ and $q=\frac{1}{2} (3 w_{de}+1)$. It is a sink for $w_{de}<w_{dm} \leq -\frac{1}{3}$ or $w_{dm}>-\frac{1}{3}> w_{de}$, which gives accelerated expansion. 

The point called C is a Milne solution ($k=1$) with  $\Omega_k= -1$  or a curvature ($k=-1$)-dominated solution ($\Omega_k= 1$). It is a sink for $w_{de}>-\frac{1}{3}, w_{dm}>-\frac{1}{3}$ or a saddle for $w_{de}<-\frac{1}{3}$ or $w_{dm}<-\frac{1}{3}$. It is a zero acceleration solution ($q=0$). 

\begin{table}[]
\centering
\resizebox{.95\linewidth}{!}{%
\begin{tabular}{|cccccccccc|}\hline
Label & $\Omega_r$ & $\Omega_{dm}$ & $\Omega_{de}$ & $\Omega_k$ & $k_1$ & $k_2$ & $k_3$ & $k_4$ &  $q$ \\\hline
B & $0$ & $0$ & $0$ & $0$  & $-1$ & $1$ & $-3
   w_{de}$ & $-3 w_{dm}$ & $\frac{1}{2}$ \\\hline
R & $1$ & $0$ & $0$ & $0$  & $2$ & $1$ & $1-3 w_{de}$ & $1-3 w_{dm}$ & $1$ \\\hline 
DM & $0$ & $1$ & $0$ & $0$ &  $-3
   (w_{de}-w_{dm})$ & $3 w_{dm}$ & $3 w_{dm}-1$ & $3 w_{dm}+1$ & $\frac{1}{2} (3 w_{dm}+1)$ \\\hline
DE & $0$ & $0$ & $1$ & $0$ & $3 w_{de}$ & $3
   w_{de}-1$ & $3 w_{de}+1$ & $3 (w_{de}-w_{dm})$ & $\frac{1}{2} (3 w_{de}+1)$ \\\hline
C & $0$ & $0$ & $0 $& $ -\frac{1}{k}$ & $-2$ &
  $ -1$ & $-3 w_{de}-1$ &$ -3 w_{dm}-1$ & $0$ \\\hline
 \end{tabular}}
    \caption{Equilibrium points of system \eqref{Curvature-A}. Eigenvalues and deceleration parameters.}
    \label{tab:Curv}
\end{table}

\subsubsection{Scaling solutions and their dynamical implications with curvature}

In the context of the extended system~\eqref{Curvature-A}, a scaling solution corresponds to an equilibrium point where the ratios $\Omega_{dm} / \Omega_{de}$, $\Omega_r / \Omega_{de}$, and $\Omega_k / \Omega_{de}$ remain constant. Mathematically, such solutions arise when the dynamical system admits a stable equilibrium point with non-zero contributions from multiple components, i.e., $\Omega_{dm}$, $\Omega_{de}$, $\Omega_r$, $\Omega_k \neq 0$. These points satisfy:
\begin{equation*}
\Omega_r' = 0, \quad \Omega_{dm}' = 0, \quad \Omega_{de}' = 0, \quad \Omega_k' = 0,
\end{equation*}
with all $\Omega_i$ bounded and non-zero. The stability of these points, determined by the eigenvalues of the Jacobian matrix, indicates whether the system naturally evolves toward a scaling regime. If the real parts of all eigenvalues are negative, the scaling solution is stable and may act as an attractor in cosmic evolution.

The existence and nature of scaling solutions depend sensitively on the equation-of-state parameters $w_{dm}$ and $w_{de}$, as well as the curvature coupling parameter $k$. For instance, choices such as $w_{de} < -1/3$ can allow for accelerated expansion while maintaining a non-negligible dark matter and curvature component, leading to viable scaling behavior.

The dynamical system is defined by \eqref{Curvature-A}, and the corresponding phase space is constrained by \eqref{region3}.

\subsubsection{Bifurcation Analysis}

Equilibrium points occur when all derivatives vanish. A notable equilibrium point is DE:
\begin{equation*}
\Omega_r = 0, \quad \Omega_{dm} = 0, \quad \Omega_{de} = 1, \quad \Omega_k = 0,
\end{equation*}
representing a flat, dark energy-dominated universe.

Linearizing around this point yields:
\begin{align}
\Omega_r' &\approx \Omega_r (3 w_{de} - 1), \\
\Omega_{dm}' &\approx \Omega_{dm} (3 w_{de} - 3 w_{dm}), \\
\Omega_{de}' &\approx \Omega_{de} (3 w_{de}), \\
\Omega_k' &\approx \Omega_k (3 w_{de} + 1).
\end{align}

The eigenvalues of the linearized system are:
\begin{equation}
\lambda_r = 3 w_{de} - 1, \quad
\lambda_{dm} = 3 w_{de} - 3 w_{dm}, \quad
\lambda_{de} = 3 w_{de}, \quad
\lambda_k = 3 w_{de} + 1.
\end{equation}

Bifurcations occur when these eigenvalues cross the zero value. If $w_{de} = \frac{1}{3}$, radiation perturbations change stability. If $w_{de} = w_{dm}$, dark matter perturbations change stability. If $w_{de} = 0$, dark energy perturbations change stability. If $w_{de} = -\frac{1}{3}$, curvature perturbations change stability.

These transitions mark qualitative changes in the system's dynamics, such as the onset of acceleration, curvature dominance, or the emergence of scaling regimes.

\subsubsection{Scaling Behavior}

In general, the energy densities scale as:
\begin{equation}
\rho_i \propto a^{-3(1 + w_i)} \quad \Rightarrow \quad \Omega_i \propto e^{-3(1 + w_i)N}.
\end{equation}
We analyze the following cases.
{Case 1: $w_{de} = 0$, $w_{dm} = 0$, $k = 0$, $\Omega_r \sim \frac{1}{1 + C e^{N}}$ decays exponentially. $\Omega_{dm}, \Omega_{de} \sim 1/(C+ e^{-N})$ grow then saturate. 
It matches the standard flat $\Lambda$CDM model.

Case 2: $w_{de} = \frac{1}{3}$, $w_{dm} = 0$, $k = 1$,  $\rho_{de} \propto a^{-4}$, it is the same as radiation. $\Omega_{de}$ and $\Omega_r$ scale similarly. $\Omega_{dm} \propto a^{-3}$, it decays slower.
$\Omega_k$ grows due to a positive eigenvalue.
 Dark energy does not drive late-time acceleration; unstable scenario.

Case 3: $w_{de} = w_{dm} = w$, arbitrary $k$: 
We define a unified dark sector variable:
\begin{align}
\Omega_x := \Omega_{dm} + \Omega_{de},
\end{align}
which combines dark matter and dark energy into a single effective component. Under this assumption, the system reduces to the following autonomous set of equations:
\begin{subequations}
  \label{Omega_x}
\begin{align}
\Omega_r' &= \Omega_r \left(3w \Omega_x + k \Omega_k + \Omega_r - 1\right), \\
\Omega_x' &= \Omega_x \left(3w (\Omega_x - 1) + k \Omega_k + \Omega_r\right), \\
\Omega_k' &= \Omega_k \left(3w \Omega_x + k \Omega_k + \Omega_r + 1\right).
\end{align}
\end{subequations}

This formulation allows for a range of cosmological behaviors.
Unified dark sector: the total dark sector energy density evolves as $ \rho_x \propto a^{-3(1 + w)} $.
If $ w > 0 $ we have a dilution behavior and the dark sector dilutes faster than radiation or matter. If $ w < 0 $, it dilutes more slowly, potentially driving cosmic acceleration. For $ w = -1 $, the dark sector mimics a cosmological constant. The system also admits scaling solutions with a constant ratio $ \Omega_{dm} / \Omega_{de} $, enabling tracking behavior.
Curvature dynamics: the evolution of $ \Omega_k $ depends on both $ w $ and $ k $; curvature may grow or decay accordingly.

The three-dimensional system \eqref{Omega_x} is not integrable in closed form, but its qualitative behavior can be studied using phase space techniques. 

Table~\ref{tab:fixedpoints_stability} presents the equilibrium points of the dynamical system \eqref{Omega_x}, along with their corresponding eigenvalues and stability conditions for selected values of the curvature parameter $ k $. The behavior of each equilibrium point is determined by the equation-of-state parameter $ w $, which governs the nature of the dominant energy component.

\begin{table}[H]
\centering
\begin{tabular}{|p{5cm}|p{4cm}|p{3.5cm}|}
\hline
Equilibrium point & Eigenvalues & Stability \\
\hline
$\Omega_r = 0,\ \Omega_x = 1,\ \Omega_k = 0$ & $\{3w,\ 3w - 1,\ 3w + 1\}$ & 
Stable if $w < -\frac{1}{3}$; Saddle if $w = -\frac{1}{3}$; Unstable if $w > -\frac{1}{3}$ \\
\hline
$\Omega_r = 0,\ \Omega_x = 0,\ \Omega_k = -\frac{1}{k}$ & $\{-2,\ -1,\ -3w - 1\}$ & 
Stable if $w > -\frac{1}{3}$; Saddle if $w = -\frac{1}{3}$; Unstable if $w < -\frac{1}{3}$ \\
\hline
$\Omega_r = 1,\ \Omega_x = 0,\ \Omega_k = 0$ & $\{2,\ 1,\ 1 - 3w\}$ & 
Always unstable (2 positive eigenvalues); Saddle if $w > \frac{1}{3}$ \\
\hline
$\Omega_r = 0,\ \Omega_x = 0,\ \Omega_k = 0$ & $\{-1,\ 1,\ -3w\}$ & 
Saddle for all $w$ \\
\hline
\end{tabular}
\caption{equilibrium points of system \eqref{Omega_x}, eigenvalues, and stability for $k = 0, \pm 1$.}
\label{tab:fixedpoints_stability}
\end{table}

   Dark energy-dominated point:
  The equilibrium point $ (\Omega_r = 0,\ \Omega_x = 1,\ \Omega_k = 0) $ is characterized by eigenvalues $ \{3w,\ 3w - 1,\ 3w + 1\} $. It is stable when $ w < -\frac{1}{3} $, corresponding to accelerated expansion. At the critical value $ w = -\frac{1}{3} $, the point becomes a saddle, and for $ w > -\frac{1}{3} $, it is unstable, indicating a departure from dark energy dominance.
Curvature-dominated point:  
  The equilibrium point $ (\Omega_r = 0,\ \Omega_x = 0,\ \Omega_k = -\frac{1}{k}) $ has eigenvalues $ \{-2,\ -1,\ -3w - 1\} $. This point is stable for $ w > -\frac{1}{3} $, saddle at $ w = -\frac{1}{3} $, and unstable for $ w < -\frac{1}{3} $. The dependence on $ k $ reflects the geometric contribution to the dynamics.
  Radiation-dominated point:
  The equilibrium point $ (\Omega_r = 1,\ \Omega_x = 0,\ \Omega_k = 0) $ yields eigenvalues $ \{2,\ 1,\ 1 - 3w\} $. It is unstable due to the presence of two positive eigenvalues, regardless of $ w $. However, it may behave as a saddle when $ w > \frac{1}{3} $, allowing for transient radiation dominance.
Empty universe point:  
  The origin $ (\Omega_r = 0,\ \Omega_x = 0,\ \Omega_k = 0) $ has eigenvalues $ \{-1,\ 1,\ -3w\} $ and consistently acts as a saddle for all values of $ w $. This reflects its sensitivity to perturbations and its role as a transitional state in the phase space.

In summary, the stability of each equilibrium point depends critically on the value of $w$, with transitions between stable, saddle, and unstable behavior marking distinct cosmological phases. Table~\ref{tab:fixedpoints_stability} encapsulates these dynamics and provides a framework for understanding the evolution of the universe under various energy conditions.

\subsubsection{Discussion: Phase Space Consistency}

To ensure the physical viability of the dynamical system \eqref{Omega_x}, it is essential to impose a constraint on the phase space that reflects the generalized Friedmann condition. Specifically, the system must evolve within the domain:
\begin{align}
\left\{ (\Omega_r, \Omega_x, \Omega_k) \in \mathbb{R}^3 : \Omega_r \geq 0,\ \Omega_x \geq 0,\ \Omega_k \geq 0,\ 
0 \leq \Omega_x - k \Omega_k \leq 1 \right\},
\end{align}
which guarantees that the total effective energy density remains bounded between 0 and 1, while accounting for curvature contributions through the parameter $ k $.

We now examine the admissibility of each equilibrium point within this constrained phase space:
 Dark energy-dominated point:  
  \begin{equation*}
  (\Omega_r = 0,\ \Omega_x = 1,\ \Omega_k = 0).
  \end{equation*}
  All variables satisfy the non-negativity conditions, and the constraint evaluates to $ \Omega_x - k \Omega_k = 1 $, which is within the allowed range.  
  This point is therefore admissible for all values of $ k $.
Curvature-dominated point:  
  \begin{equation*}
  (\Omega_r = 0,\ \Omega_x = 0,\ \Omega_k = -\tfrac{1}{k}).
  \end{equation*}
  The curvature component $ \Omega_k $ is negative unless $ k < 0 $, in which case $ -\tfrac{1}{k} > 0 $. The constraint becomes $ \Omega_x - k \Omega_k = 1 $, which is valid.  
  Hence, this point is only physically consistent in open universes ($ k < 0 $), and excluded for flat or closed geometries ($ k \geq 0 $).
 Radiation-dominated point:  
  \begin{equation*}
  (\Omega_r = 1,\ \Omega_x = 0,\ \Omega_k = 0).
  \end{equation*}
  All components are non-negative, and the constraint yields $ \Omega_x - k \Omega_k = 0 $, which satisfies the condition.  
  This point is admissible for all values of $ k $.
 Empty universe point:  
  \begin{equation*}
  (\Omega_r = 0,\ \Omega_x = 0,\ \Omega_k = 0).
  \end{equation*}
  This trivial equilibrium point satisfies all non-negativity conditions, and the constraint evaluates to zero.  
  It is therefore admissible for all values of $ k $.

Among the four equilibrium points analyzed:
 The dark energy-dominated, radiation-dominated, and empty universe points are always consistent with the phase space constraint.
The curvature-dominated point is only physically meaningful in open universes ($ k < 0 $), and must be excluded from the analysis in flat ($ k = 0 $) or closed ($ k > 0 $) geometries.

This evaluation confirms that the dynamical system respects the physical bounds imposed by the generalized Friedmann constraint, and highlights the geometric dependence of certain cosmological solutions.

\subsubsection{Presureless dark matter}
 Assuming $w_{dm}=0$ and $\Omega_b\geq 0$,  and using similar arguments as before, we can reduce the system's dimensionality to the lowest possible, considering the dynamical system
\begin{subequations}
\label{Phase-Space3Dnegativek}
\begin{align}
  & \Omega_r^{\prime}=  \Omega_r (k\,\Omega_k+3 w_{de} \,\Omega_{de}+\Omega_r-1),\\
  & \Omega_{de}^{\prime}=\Omega_{de} (k\,
   \Omega_k+3 w_{de} (\Omega_{de}-1)+\Omega_r),\\
   & \Omega_k^{\prime}=\Omega_k (k \,\Omega_k+3
   w_{de}\, \Omega_{de}+\Omega_r+1),
\end{align}
\end{subequations}
defined on phase space
\begin{align}
 \left\{   (\Omega_r, \Omega_{de}, \Omega_{k})\in \mathbb{R}^3: \Omega_r\geq 0,  \Omega_{de}\geq 0, \Omega_k\geq 0,  \right. \nonumber \\
\left. 0\leq\Omega_{r}+ \Omega_{de} - k \Omega_k\leq 1\right\}.
\end{align}
We have the auxiliary matter equation
\begin{equation}
  \Omega_{dm}^{\prime}= \Omega_{dm} (k \Omega_k+3 w_{de} \Omega_{de}+\Omega_r). 
\end{equation}
The deceleration parameter reduces to
\begin{equation}
    q= \frac{1}{2} (k \Omega_k+3 w_{de} \Omega_{de}+\Omega_r+1).
\end{equation}
In table \ref{tab:systC2} are presented the 
equilibrium points of system \eqref{Phase-Space3Dnegativek}. 

\begin{table*}[]
    \centering
    \begin{tabular}{|cccccccc|}\hline
    Label & $\Omega_{r}$ & $\Omega_{de}$ & $\Omega_{k}$ & $k_1$ & $k_2$ & $k_3$ & $q$ \\\hline
B & $0$ & $0$ & $0$ & $-1$ & $1$ & $-3 w_{de}$& $\frac{1}{2}$ \\\hline
R & $1$ & $0$ & $0$ & $2$ & $1$ & $1-3 w_{de}$ & $1$ \\\hline
DE & $0$ & $1$ & $0$ & $3 w_{de}$ & $3 w_{de}-1$ & $3
   w_{de}+1$ & $\frac{1}{2} (3 w_{de}+1) $\\\hline
   C & $0$ & $0$ & $-\frac{1}{k}$ & $-2$ & $-1$ & $-3 w_{de}-1$ & $0$ \\\hline
    \end{tabular}
    \caption{Equilibrium points of system \eqref{Phase-Space3Dnegativek} for a flat FLRW model that includes radiation, baryons, DM (dust), DE, and a curvature term. The corresponding eigenvalues and deceleration parameters are also presented. }
    \label{tab:systC2}
\end{table*}

Point B is a Baryon-dominated solution. Due to two eigenvalues with opposite signs, it is a saddle. The deceleration parameter $q=\frac{1}{2}$ corresponds to a decelerated pressureless solution.

Point R is a Radiation-dominated solution. It is a source for $w_{de}<1/3$ or a saddle otherwise. The deceleration parameter $q=1$ corresponds to a decelerated solution. 

Point DE is a dark energy-dominated solution. It is a sink for $w_{de}\leq -\frac{1}{3}$. The deceleration parameter $q=\frac{1}{2} (3 w_{de}+1)$ gives accelerated expansion for $w_{de}\leq -\frac{1}{3}$.
   
For $k=-1$, the point C is a curvature ($k=-1$)-dominated solution ($\Omega_k= 1$). It is a sink for $w_{de}>-\frac{1}{3}$ or a saddle for $w_{de}<-\frac{1}{3}$. It is a zero acceleration solution ($q=0$). 

For $k=1$, the stability conditions of $C$ remain the same, but physically, it corresponds to a Milne solution. However, because the phase space is not compact, we will follow the procedure implemented in section \ref{positive-curv-section}.

\subsubsection{Negative curvature}
In this section, we numerically study the stability of the equilibrium points of system \eqref{Phase-Space3Dnegativek} for $k=-1$, which yields a compact phase space. 

In Figure \ref{fig:Phase-Space3Dnegativek} it is shown the Phase space of system \eqref{Phase-Space3Dnegativek} for $k=-1$ and  $w_{de}= -1, -2/3, -1/3$. These confirm the previous analytical results.

\begin{figure}[H]
    \centering    \includegraphics[scale=1]{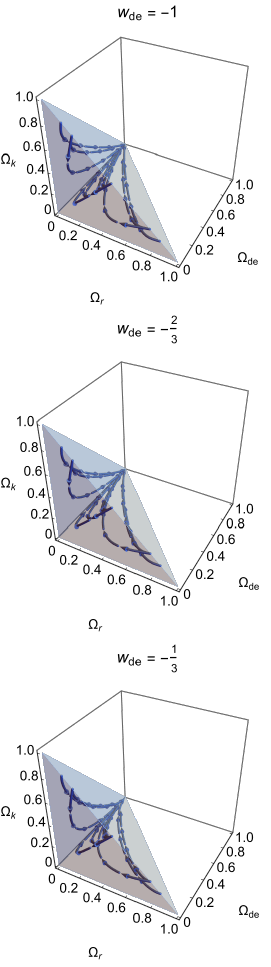}
    \caption{Phase space of system \eqref{Phase-Space3Dnegativek} for $k=-1$ and  $w_{de}= -1, -2/3, -1/3$.}
    \label{fig:Phase-Space3Dnegativek}
\end{figure}

\subsubsection{Bifurcation Analysis and Scaling Solutions for \texorpdfstring{$k = -1$}{k = -1}}

We analyze the cosmological dynamics governed by the system:
\begin{subequations}
\label{Phase-Space3DkMinus1}
\begin{align}
  & \Omega_r^{\prime}=  \Omega_r (-\Omega_k + 3 w_{de} \,\Omega_{de} + \Omega_r - 1),\\
  & \Omega_{de}^{\prime}= \Omega_{de} (-\Omega_k + 3 w_{de} (\Omega_{de} - 1) + \Omega_r),\\
  & \Omega_k^{\prime}= \Omega_k (-\Omega_k + 3 w_{de} \,\Omega_{de} + \Omega_r + 1),
\end{align}
\end{subequations}
with phase space constrained by:
\begin{equation*}
\Omega_r \geq 0, \quad \Omega_{de} \geq 0, \quad \Omega_k \geq 0, \quad 0 \leq \Omega_r + \Omega_{de} + \Omega_k \leq 1.
\end{equation*}

A notable equilibrium point is DE:
\begin{equation*}
\Omega_r = 0, \quad \Omega_{de} = 1, \quad \Omega_k = 0,
\end{equation*}
representing a flat, dark energy-dominated universe.
Linearizing the system~\eqref{Phase-Space3DkMinus1} around this point yields:
\begin{align}
\Omega_r' &\approx \Omega_r (3 w_{de} - 1), \\
\Omega_{de}' &\approx \Omega_{de} (3 w_{de}), \\
\Omega_k' &\approx \Omega_k (3 w_{de} + 1).
\end{align}

The eigenvalues of the linearized system are:
\begin{equation*}
\lambda_r = 3 w_{de} - 1, \quad
\lambda_{de} = 3 w_{de}, \quad
\lambda_k = 3 w_{de} + 1.
\end{equation*}

Bifurcations occur when the eigenvalues cross zero:
$ w_{de} = \frac{1}{3} $: radiation perturbations change stability.
 $ w_{de} = 0 $: dark energy perturbations change stability.
$ w_{de} = -\frac{1}{3} $: curvature perturbations change stability.
These bifurcation values mark transitions in the qualitative behavior of the system, such as the onset of curvature growth, radiation dominance, or dark energy acceleration.

In general, the energy densities scale as:
\begin{equation*}
\rho_i \propto a^{-3(1 + w_i)} \quad \Rightarrow \quad \Omega_i \propto e^{-3(1 + w_i)N}.
\end{equation*}

We analyze the scaling behavior for representative values of $ w_{de}$.
Case 1: $ w_{de} = 0 $,  
 $ \rho_{de} \propto a^{-3} $, it behaves like pressureless matter.
$ \lambda_r = -1 $, $ \lambda_{de} = 0 $, $ \lambda_k = 1 $, curvature grows.
 No late-time acceleration; curvature eventually dominates.\\
Case 2: $ w_{de} = \frac{1}{3} $,  
 $ \rho_{de} \propto a^{-4} $, same scaling as radiation.
 $ \lambda_r = 0 $, $ \lambda_{de} = 1 $, $ \lambda_k = 2 $, all components grow.
 System unstable; curvature dominates at late times.
Case 3: $ w_{de} = -1 $  
  $ \rho_{de} = \text{const} $, cosmological constant behavior.
$ \lambda_r = -4 $, $ \lambda_{de} = -3 $, $ \lambda_k = -2 $, all components decay.
   Stable attractor; dark energy dominates and curvature decays.
  Represents viable late-time accelerated expansion.

Setting $ k = -1 $ introduces negative spatial curvature, which competes dynamically with radiation and dark energy. The curvature term grows unless suppressed by sufficiently negative $ w_{de} $. Scaling solutions with constant ratios among components are only viable when curvature is either negligible or balanced by dark energy. The bifurcation structure reveals critical values for stability and cosmic acceleration.

\subsubsection{Positive curvature}
\label{positive-curv-section}
For $k=+1$, it is more convenient to define the dimensionless variables: 
\begin{align}
&\left(\hat{\Omega}_{r}, \hat{\Omega}_{b}, \hat{\Omega}_{dm}, \hat{\Omega}_{de} 
\right)\equiv \frac{8\pi G}{3 \left(H^2+a^{-2}\right)}\left(\rho_{r} , \rho_{b} ,  \rho_{dm}, \rho_{de}\right), \label{eqOmegas_curv}\\
& Q= \frac{H}{\sqrt{H^2+a^{-2}}},
\end{align}
which satisfy
\begin{equation}
    \hat{\Omega }_b+\hat{\Omega }_{de}+\hat{\Omega }_{dm}+\hat{\Omega }_r=1
\end{equation}
and the time derivative
\begin{equation}
 f^{\prime}=\frac{d f}{d \eta}= \frac{1}{\sqrt{H^2 + a^{-2}}} \dot{f}, 
\end{equation}
where $a$ is the scale factor. 
Hence, we obtain the reduced dynamical system
\begin{subequations}
\label{system_positive_curvature}
\begin{align}
& \hat{\Omega }_r^{\prime}= Q\,\hat{\Omega }_r \left(3 w_{de} \hat{\Omega }_{de}+3 w_{dm} \hat{\Omega }_{dm}+\hat{\Omega }_r-1\right),\\
& \hat{\Omega }_{dm}^{\prime}= Q \hat{\Omega }_{dm} \left(3 w_{de}
   \hat{\Omega }_{de}+3 w_{dm} \left(\hat{\Omega }_{dm}-1\right)+\hat{\Omega }_r\right),\\
& \hat{\Omega }_{de}^{\prime}=Q \hat{\Omega }_{de} \left(3 w_{de} \hat{\Omega }_{de}+3 w_{dm}
   \hat{\Omega }_{dm}+\hat{\Omega }_r-3 w_{de}\right),\\
& Q^{\prime}= -\frac{1}{2} \left(1-Q^2\right) \left(3 w_{de} \hat{\Omega }_{de}+3 w_{dm} \hat{\Omega }_{dm}+\hat{\Omega
   }_r+1\right), \label{EQ.(40d)}
\end{align}
\end{subequations}
defined on phase space
\begin{align}
 \left\{   (\hat{\Omega }_r, \hat{\Omega }_{dm}, \hat{\Omega }_{de}, Q)\in \mathbb{R}^3: \hat{\Omega }_r\geq 0,  \hat{\Omega }_{dm}\geq 0,   \hat{\Omega }_{de}\geq 0, \right. \nonumber \\
\left. -1\leq Q\leq 1,  0\leq\hat{\Omega }_{r}+ \hat{\Omega }_{dm} + \hat{\Omega }_{de} \leq 1\right\}.
\end{align}

The region $Q<0$ corresponds to contraction, and $Q>0$ corresponds to the expanded branch. 

Moreover, the deceleration parameter $q$ is related to the rest of the variables through
\begin{equation}
  q  Q^2=3 w_{de} \hat{\Omega }_{de}+3 w_{dm} \hat{\Omega }_{dm}+\hat{\Omega }_r+1.
\end{equation} 

\begin{figure}[H]
    \centering
    \includegraphics[width=0.9\linewidth]{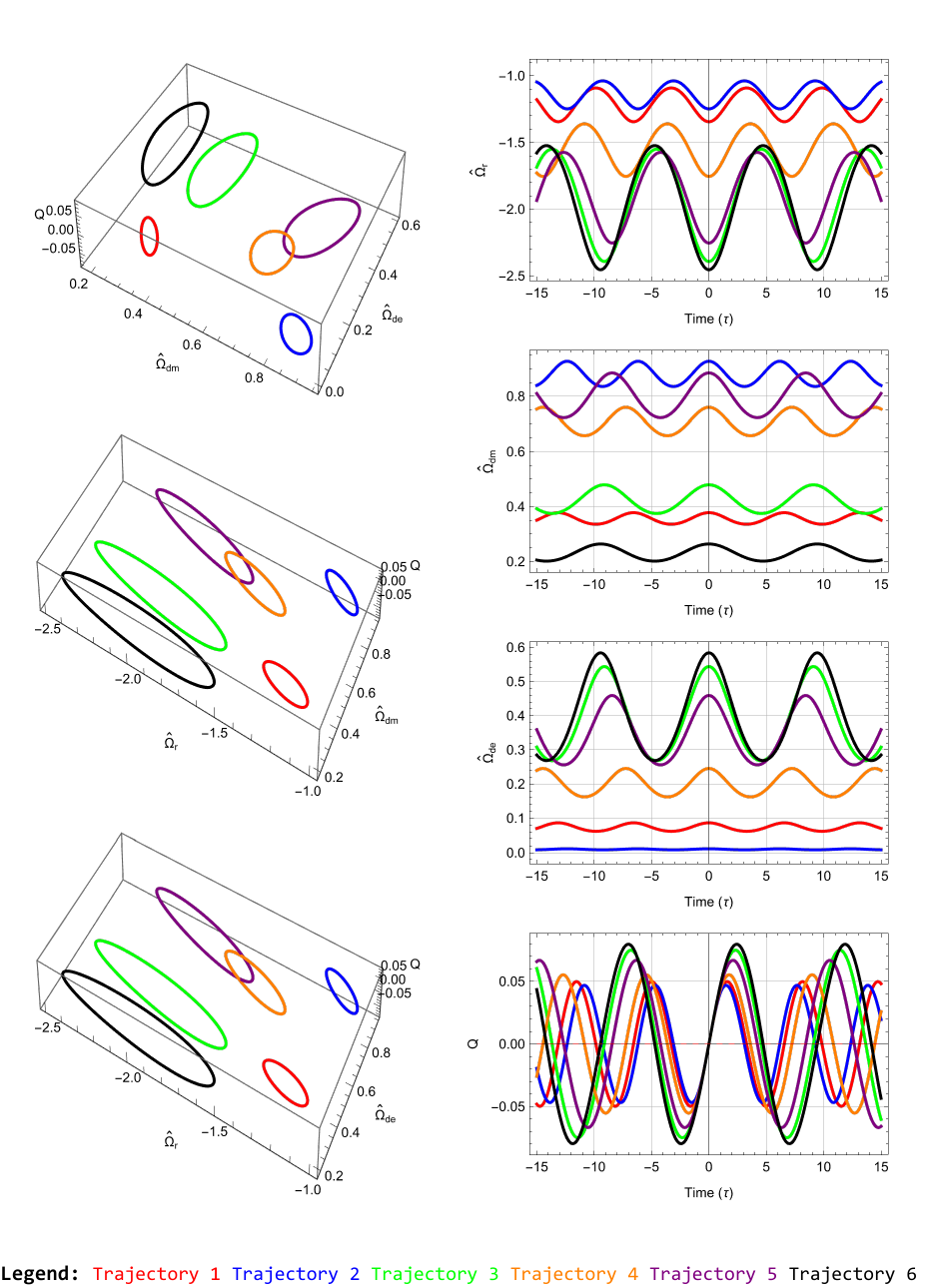}
    \caption{Some trajectories of system \eqref{system_positive_curvature} using initial conditions from table \ref{tab:ini2} and setting $w_{de}=0.753217$ and $w_{dm}=0.0439285$. It shows the crossing through $H=0$ and the center behavior of $S_1$. This behavior, corresponding to cyclic cosmology, is nonphysical since it requires  $\hat{\Omega }_{r}<0$.}
    \label{fig:centerpositive-curvature}
\end{figure}
In the figure \ref{fig:centerpositive-curvature}, we solve the system numerically using the initial conditions displayed in table \ref{tab:ini2}.

\begin{table*}
\centering
\begin{tabular}{|c|c|c|c|c|}
\hline
Legend& $\hat{\Omega}_{de}(0)$ & $\hat{\Omega}_{dm}(0)$ & $\hat{\Omega}_{r}(0)$ & $Q(0)$ \\
\hline
\textcolor{red}{Trajectory 1}& $0.0862234$ & $0.377913$ & $-1.34464$ & $0$ \\
\textcolor{blue}{Trajectory 2}& $0.0116446$ & $0.927266$ & $-1.24851$ &$0$ \\
\textcolor{green}{Trajectory 3}& $0.543757$ & $0.479332$ & $-2.39187$ & $0$ \\
\textcolor{orange}{Trajectory 4}& $0.245349$ & $0.759896$ & $-1.75455$ & $0$ \\
\textcolor{purple}{Trajectory 5}& $0.459017$ & $0.884729$ & $-2.25381$ & $0$ \\
\textcolor{black}{Trajectory 6}& $0.583854$ & $0.263973$ & $-2.45409$ & $0$ \\
\hline
\end{tabular}
    \caption{Initial conditions to solve the system \eqref{system_positive_curvature} for $Q=0$. The plots show that the curves $S_1$ are centered.}
    \label{tab:ini2}
\end{table*}

\begin{table*}[]
    \centering
    \resizebox{.95\linewidth}{!}{%
 \begin{tabular}{|cccccccccc|}\hline
Scenario & $\hat{\Omega }_r$ & $\hat{\Omega }_{dm}$ & $\hat{\Omega }_{de}$  & $Q$ & $k_1$ & $k_1$ & $k_3$ & $k_4$ & $q$ \\\hline
$S_1$ & $\hat{\Omega }_{\text{r}}$ & $\hat{\Omega }_{dm}$ & $\hat{\Omega }_{de}$ & $0$ &$\hat{k}_1$ & $\hat{k}_2$ & $0$ & $0$  & -\\\hline
$S_{2,3}$ & $1$ & $0$ & $0$ & $\pm1$ & $\pm2$ & $\pm1$ & $\pm(1-3 w_{de})$ & $\pm(1-3 w_{dm})$ &$ 2$ \\\hline
$S_{4,5}$ & $0$ & $0$ & $0$ & $\pm1$ & $\mp1$ & $\pm1$ & $\mp3 w_{de}$ & $\mp3 w_{dm}$ & $1 $\\\hline
$S_{6,7}$ & $0$ & $1$ & $0$ & $\pm1$ & $\pm3 (w_{dm}-w_{de})$ & $\pm3 w_{dm}$ & $\pm(3 w_{dm}-1)$ & $\pm(3 w_{dm}+1)$ & $3 w_{dm}+1$ \\\hline
$S_{8,9}$ & $0$ & $0$ & $1 $& $\pm1$ & $\pm3 w_{de}$ & $\pm(3 w_{de}-1)$ & $\pm(3 w_{de}+1 )$& $\pm 3 (w_{de}-w_{dm})$ & $3 w_{de}+1$ \\\hline
\end{tabular}}
    \caption{Equilibrium points of the system \eqref{system_positive_curvature}. For $S_1$ we have $3 w_{de} \hat{\Omega }_{de}+3 w_{dm} \hat{\Omega }_{dm}+\hat{\Omega }_r+1=0$, $\hat{k}_1= -\frac{\sqrt{3 w_{de} (3
   w_{de}-1) \hat{\Omega }_{de}+3 w_{dm} (3 w_{dm}-1) \hat{\Omega }_{dm}-2}}{\sqrt{2}}$, $\hat{k}_2=\frac{\sqrt{3 w_{de} (3 w_{de}-1) \hat{\Omega }_{de}+3
   w_{dm} (3 w_{dm}-1) \hat{\Omega }_{dm}-2}}{\sqrt{2}}$.}
    \label{tab:positive-curvature}
\end{table*}

Nine scenarios are listed in the table \ref{tab:positive-curvature}. Say, 
\begin{itemize}
\item[$S_1$]: The equilibrium condition $3 w_{de} \hat{\Omega}_{de} + 3 w_{dm} \hat{\Omega}_{dm} + \hat{\Omega}_r + 1 = 0$ with $Q = 0$ forms a normally-hyperbolic surface, where stability depends on $w_{de}$ and $w_{dm}$. The equilibrium points behave as a center or saddle based on whether $\hat{k}_1$ and $\hat{k}_2$ given by $\hat{k}_1= -\frac{\sqrt{3 w_{de} (3
   w_{de}-1) \hat{\Omega }_{de}+3 w_{dm} (3 w_{dm}-1) \hat{\Omega }_{dm}-2}}{\sqrt{2}}$ and $\hat{k}_2=\frac{\sqrt{3 w_{de} (3 w_{de}-1) \hat{\Omega }_{de}+3
   w_{dm} (3 w_{dm}-1) \hat{\Omega }_{dm}-2}}{\sqrt{2}}$ are real or imaginary. If $3 w_{de} (3 w_{de} - 1) \hat{\Omega}_{de} + 3 w_{dm} (3 w_{dm} - 1) \hat{\Omega}_{dm} - 2$ is negative, the system forms a center (see Figure \ref{fig:centerpositive-curvature} and Table \ref{tab:ini2}) that is nonphysical because $\hat{\Omega }_r<0$; otherwise, it forms a saddle. Since $ qQ^2 = 0$, acceleration is indeterminate. Negative $\hat{\Omega}_r$ is unlikely, and the case $\hat{\Omega}_{de} = 0$ represents an absence of dark energy. This set includes the Einstein static universe, allowing transitions between decelerated and accelerated regimes.
\item[$S_2$:] $(\hat{\Omega }_r, \hat{\Omega }_{dm}, \hat{\Omega }_{de}, Q) = (1,0,0,1)$ has eigenvalues $2, 1$, \newline $1 - 3w_{de}, 1 - 3w_{dm}$, acting as a source for $w_{de}, w_{dm} < 1/3$, becoming nonhyperbolic for $w_{de}=1/3$ or $w_{dm}=1/3$, and a saddle otherwise. The deceleration parameter, $q = 2$, represents a radiation-dominated, decelerating solution, likely unstable but expanding for $Q > 0$.
\item[$S_3$:] $(\hat{\Omega }_r, \hat{\Omega }_{dm}, \hat{\Omega }_{de}, Q) = (1,0,0,-1)$ has eigenvalues $-1, -2$, \newline $1, -(1-3 w_{de}), -(1-3 w_{dm})$ is nonhyperbolic for $w_{de}=1/3$ or $w_{dm}=1/3$, and a saddle otherwise; also has $q = 2$ but is contracting due to negative $Q$, making both cases relevant in the early radiation-dominated universe.
\item[$S_4$:] $(\hat{\Omega }_r, \hat{\Omega }_{dm}, \hat{\Omega }_{de}, Q) = (0,0,0,1)$ with eigenvalues $-1, 1$, \newline $ -3w_{de}, -3w_{dm}$ is nonhyperbolic for $w_{de}=0$ or 
$w_{dm}=0$ or a saddle otherwise. The deceleration parameter $q=1$ indicates that the universe is curvature-dominated ($k=+1$). It is an unstable saddle. 
\item[$S_5$:] $(\hat{\Omega }_r, \hat{\Omega }_{dm}, \hat{\Omega }_{de}, Q) = (0,0,0,-1)$ with eigenvalues \newline $1, -1, 3w_{de}, 3w_{dm}$ is nonhyperbolic for $w_{de}=0$ or 
$w_{dm}=0$ or a saddle otherwise. Its behavior is similar to $S_4$. It is an unstable saddle and curvature-dominated ($k=+1$). However, it contracts due to the negative $Q$, making both cases relevant to the curvature-dominated universe.
\item[$S_6$:] $(\hat{\Omega }_r, \hat{\Omega }_{dm}, \hat{\Omega }_{de}, Q) = (0,1,0,1)$ has eigenvalues \newline $3w_{dm} - 3w_{de}, 3w_{dm}, 3w_{dm} - 1$ and deceleration parameter $q = 3w_{dm} + 1$. It is stable for $w_{dm} < w_{de}, w_{dm}<0$. 
It is expanding ($Q>0$) decelerated ($q > 0$) for $w_{dm}>-1/3$. It is expanding ($Q>0$)  and accelerated ($q < 0$) for $w_{dm}<-1/3$. Hence, it is an accelerated sink if $w_{dm} < w_{de} \leq -\frac{1}{3}$ or $w_{de} > -\frac{1}{3}> w_{dm}$, where dark matter behaves like a quintessence field. It is a source if $w_{de} \leq \frac{1}{3}<w_{dm}$ or $w_{dm} > w_{de} > \frac{1}{3}$, corresponding to nonphysical scenarios.  It is nonhyperbolic for $w_{dm} = w_{de}$ or $w_{dm}=0$ or $w_{dm}=1/3$. Otherwise, it is a saddle, remaining physical if $w_{dm}$ produces a positive deceleration parameter, representing a universe dominated by dark matter.
\item[$S_7$:] $(\hat{\Omega }_r$, $\hat{\Omega }_{dm}$, $\hat{\Omega }_{de}, Q) = (0,1,0,-1)$ has eigenvalues \newline $-(3w_{dm} - 3w_{de}), -3w_{dm}, -(3w_{dm} - 1)$ and deceleration parameter $q = 3w_{dm} + 1$. 
It is a source for $w_{dm} < w_{de}$ and $ w_{dm}<0$. 
It is contracting ($Q<0$) decelerated ($q > 0$) for $w_{dm}>-1/3$. It is contracting ($Q<0$)  and accelerated ($q < 0$) for $w_{dm}<-1/3$. Hence, it is an accelerated source if $w_{dm} < w_{de} \leq -\frac{1}{3}$ or $w_{de} > -\frac{1}{3}> w_{dm}$, where dark matter behaves like a quintessence field. It is a sink if $w_{de} \leq \frac{1}{3}<w_{dm}$ or $w_{dm} > w_{de} > \frac{1}{3}$, corresponding to nonphysical scenarios.  It is nonhyperbolic for $w_{dm} = w_{de}$ or $w_{dm}=0$ or $w_{dm}=1/3$. Otherwise, it is a saddle, remaining physical if $w_{dm}$ produces a positive deceleration parameter, representing a contracting dark matter-dominated universe.
\item[$S_8$:] $(\hat{\Omega }_r, \hat{\Omega }_{dm}, \hat{\Omega }_{de}, Q) = (0,0,1,1)$ has eigenvalues \newline $3w_{de}, 3w_{de} - 1, 3w_{de} + 1, 3(w_{de} - w_{dm})$ and deceleration parameter $q = 3w_{de} + 1$. Stability depends on $w_{de}$, requiring $w_{de} > -\frac{1}{3}$ for deceleration ($q > 0$) and $w_{de} < -\frac{1}{3}$ for acceleration ($q < 0$). It represents a dark energy-dominated universe and is physical if $w_{de}$ ensures a negative deceleration parameter. Reducing the parameter space to  $w_{de} < -\frac{1}{3}$, it is a sink for $w_{de} < w_{dm} \leq -\frac{1}{3}$ or $w_{dm} > -\frac{1}{3}>w_{de}$, providing an accelerated expansion.
\item[$S_9$:] $(\hat{\Omega }_r, \hat{\Omega }_{dm}, \hat{\Omega }_{de}, Q) = (0,0,1,-1)$ has eigenvalues \newline $-3w_{de}, -(3w_{de} - 1), -(3w_{de} + 1), -3(w_{de} - w_{dm})$ and deceleration parameter $q = 3w_{de} + 1$. It is a contracting solution because $ Q = 1$. Deceleration ($q>0$) requires $w_{de} > -\frac{1}{3}$ for decelerated contraction, and accelerated ($q < 0$) contraction ($Q<0$) requires $w_{de} <-\frac{1}{3}$. Representing a dark energy-dominated universe, it remains physical if $w_{de}$ ensures a negative deceleration parameter. Under the assumption $w_{de} <-\frac{1}{3}$, it is a source for $ w_{de} < w_{dm} \leq -\frac{1}{3}$ or $w_{dm} > -\frac{1}{3}>w_{de}$, leading to accelerated contraction.
\end{itemize}

From equation \eqref{EQ.(40d)}, we determine the invariant set of solutions: expanding ($Q=+1$) and contracting ($Q=-1$). The system follows a discrete time-reversal symmetry:
\begin{equation}
(\eta, \hat{\Omega }_r, \hat{\Omega }_{dm}, \hat{\Omega }_{de}, Q) \mapsto (-\eta, \hat{\Omega }_r, \hat{\Omega }_{dm}, \hat{\Omega }_{de}, -Q)
\end{equation}
Thus, the behavior in one branch mirrors the other. We focus on characterizing attractors for the positive branch numerically, while the negative branch’s dynamics follow the above transformation.

\begin{figure}[H]
    \centering
    \includegraphics[width=0.9\linewidth]{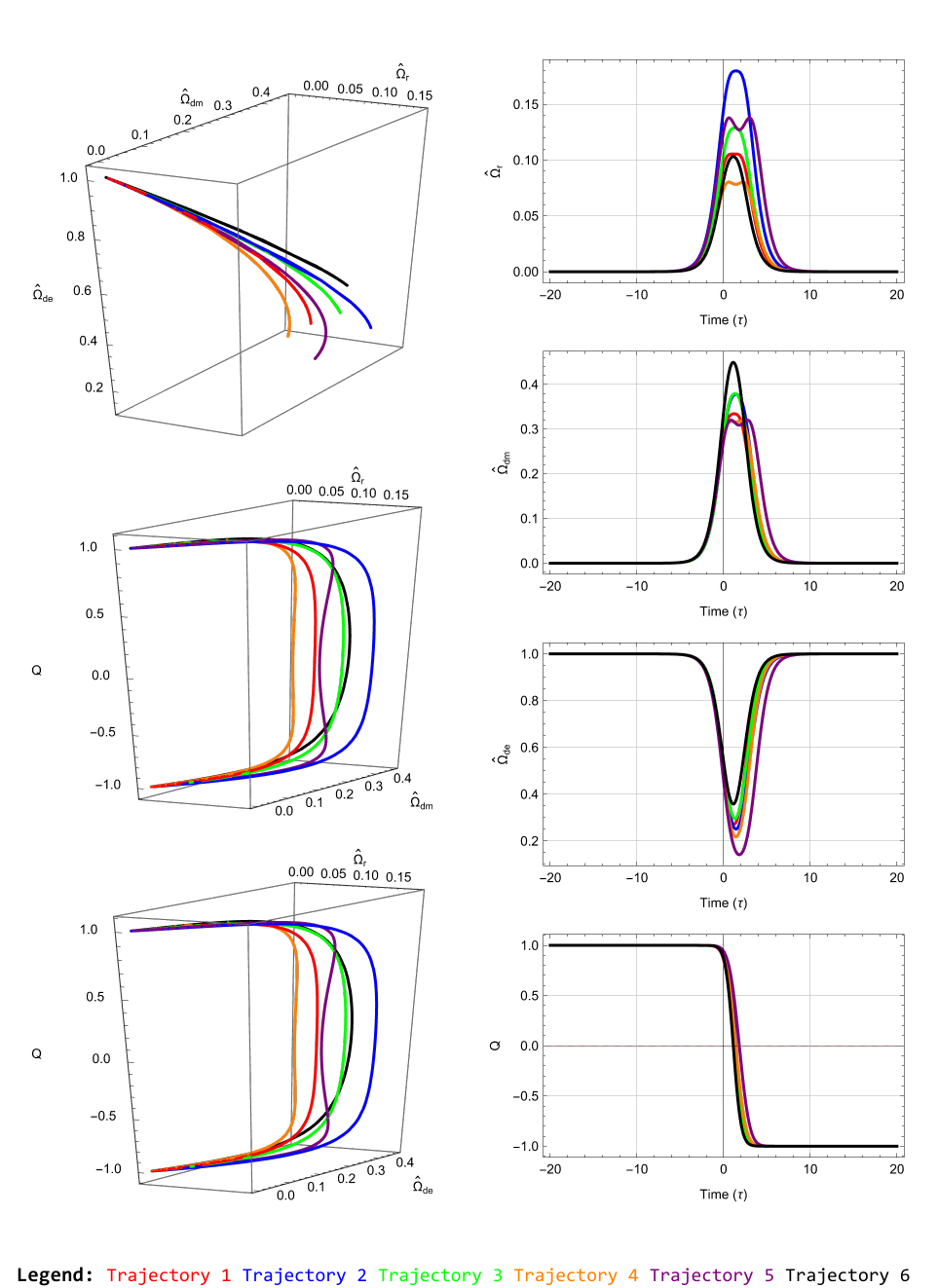}
    \caption{Some trajectories of system \eqref{system_positive_curvature} using initial conditions from table \ref{tab:ini} and setting $w_{de}=0.7$ and $w_{dm}=0.3$. It shows the crossing through $H=0$.}
    \label{fig:positive-curvature}
\end{figure}
In the figure \ref{fig:positive-curvature}, we solve the system numerically using the initial conditions displayed in table \ref{tab:ini}. 

 \begin{table*}[]
    \centering
   \begin{tabular}{|c|c|c|c|c|}
   \hline
Legend& $\Omega_{r}(0)$ & $\Omega_{dm}(0)$ & $\Omega_{de}(0)$ & Q(0) \\ \hline
\textcolor{red}{Trajectory 1}& 0.0734429 & 0.318689 & 0.532257 & 0.942885 \\
\textcolor{blue}{Trajectory 2}& 0.0740592 & 0.338936 & 0.475942 & 0.946468 \\
\textcolor{green}{Trajectory 3}& 0.101068 & 0.28298 & 0.538756 & 0.84803 \\
\textcolor{orange}{Trajectory 4}& 0.115268 & 0.324484 & 0.476893 & 0.989655 \\
\textcolor{purple}{Trajectory 5}& 0.0762049 & 0.337009 & 0.515713 & 0.927947 \\
\textcolor{black}{Trajectory 6}& 0.0830375 & 0.296938 & 0.46441 & 0.801824 \\
\hline
\end{tabular}
    \caption{Initial conditions to solve the system \eqref{system_positive_curvature}.}
    \label{tab:ini}
\end{table*}

The dynamical system exhibits different equilibrium states, which determine whether it is stable, expanding, or contracting. Stability depends on the parameters $w_{de}$ and $w_{dm}$, influencing whether the system acts as a sink, source, or saddle. Radiation-dominated and curvature-dominated universes appear under different conditions, with expansion occurring when $Q>0$ and contraction when $Q < 0$. Accelerated expansion happens when dark energy dominates, while deceleration is typical in radiation- and matter-dominated phases. Some solutions are nonphysical due to density constraints, but others provide insight into transitions between cosmic phases, including the Einstein static universe, which allows shifts between acceleration and deceleration.
Three key factors define the system: stability, acceleration, and physical viability. Stability relies on the interactions between $w_{de}$, $w_{dm}$, and density parameters $\Omega_r$, $\Omega_{dm}$, and $\Omega_{de}$. Accelerated expansion (negative $q$) is linked to dark energy, while deceleration occurs in radiation- and matter-driven phases. For a physically viable solution, density parameters must be non-negative ($\Omega\geq 0$), and their sum should not exceed $1$. These findings underscore the influence of cosmic parameters on the universe's evolution, particularly the interactions between dark matter and dark energy.

We consider the dynamical system \eqref{system_positive_curvature} derived for a positively curved universe ($ k = +1 $), using the compactified variables $ \hat{\Omega}_i $ and the normalized Hubble parameter $ Q $. This formulation is particularly suited for analyzing the interplay between curvature and cosmic components.

Equilibrium points are obtained by setting all derivatives in system~\eqref{system_positive_curvature} to zero.

Notable equilibrium points include a dark energy dominated universe:  
  $ \hat{\Omega}_{de} = 1 $, $ \hat{\Omega}_{dm} = \hat{\Omega}_r = 0 $, $ Q = \pm 1 $.  
  This represents a universe dominated by dark energy. The evolution of $ Q $ is governed by:
  \begin{equation}
  Q' = -\frac{1}{2}(1 - Q^2)(3w_{de} + 1).
  \end{equation}
  For $ w_{de} < -\frac{1}{3} $, the expanding branch ($ Q > 0 $) is stable, indicating late-time acceleration.
  
   Matter Dominated Universe:  
  $ \hat{\Omega}_{dm} = 1 $, others zero, $ Q = \pm 1 $.  
  For cold dark matter ($ w_{dm} = 0 $), we find:
  \begin{equation}
   Q' = -\frac{1}{2}(1 - Q^2),
  \end{equation}
  implying deceleration in the expanding branch.
  
Radiation Dominated Universe:  
  $ \hat{\Omega}_r = 1 $, others zero, $ Q = \pm 1 $.  
  Then: 
  \begin{equation}
  Q' = -(1 - Q^2),
  \end{equation}
  which is always negative, confirming that radiation leads to deceleration.

  Mixed Scaling Solutions:  
  These occur when the energy components maintain constant ratios, i.e., $ \hat{\Omega}_r : \hat{\Omega}_{dm}: \hat{\Omega}_{de} = \text{const} $.  
  Such solutions require $ Q' = 0 $, which imposes a constraint on the effective equation of state:
  \begin{equation}
  3 w_{de} \hat{\Omega}_{de} + 3 w_{dm} \hat{\Omega}_{dm} + \hat{\Omega}_r + 1 = 0.
   \end{equation}
These solutions are rare and typically require fine-tuning of parameters.

Bifurcations occur when the stability of equilibrium points changes due to variations in $ w_{de} $ or $ w_{dm} $. Key bifurcation values include:
$ w_{de} = -\frac{1}{3} $, marks the onset of accelerated expansion.
  $ w_{dm} = -\frac{1}{3} $, unphysical for matter, but mathematically relevant for curvature dynamics.
 $ Q = 0 $: corresponds to a bounce or turnaround between contraction and expansion.
At these values, the sign of $ Q' $ changes, altering the cosmic trajectory and stability of the universe.

Scaling solutions imply that each component evolves proportionally, maintaining fixed ratios. These are characterized by:
\begin{equation*}
\rho_i \propto a^{-3(1 + w_i)} \quad \Rightarrow \quad \hat{\Omega}_i = \text{const}.
\end{equation*}
Representative cases include Case 1: $ w_{de} = -1 $, $ w_{dm} = 0 $,  
  dark energy behaves as a cosmological constant, and matter is pressureless.  
  The system evolves toward $ \hat{\Omega}_{de} \to 1 $, $ Q \to 1 $, representing a stable, accelerating universe.
   Case 2: $ w_{de} = 0 $, $ w_{dm} = 0 $, 
  no acceleration; curvature eventually dominates.  
  $ Q' < 0 $, indicating deceleration and possible recollapse.
   Case 3: $ w_{de} = \frac{1}{3} $,  
  dark energy mimics radiation.  
  All components dilute rapidly, and curvature dominates the system.  
  No stable attractor exists in this regime.
The normalized Hubble parameter $ Q $ compactifies the dynamics and encodes the curvature effects:
 $ Q = 1 $, pure expansion ($ H \gg a^{-1} $).
$ Q = 0 $, bounce or turnaround.
  $ Q = -1 $, pure contraction.
Its evolution is governed by:
\begin{equation}
Q' = -\frac{1}{2}(1 - Q^2)\left(3 w_{de} \hat{\Omega}_{de} + 3 w_{dm} \hat{\Omega}_{dm} + \hat{\Omega}_r + 1\right),
\end{equation}
which determines whether the universe accelerates, decelerates, or reverses.

The positively curved universe exhibits rich dynamics governed by the competition between radiation, matter, dark energy, and curvature. Bifurcations mark transitions in cosmic behavior, and scaling solutions reveal attractors or unstable regimes. The variable $ Q $ plays a central role in compacting the phase space and distinguishing between expansion and contraction branches.

\subsection{Summary: Dynamical Systems Analysis}

To understand how different cosmic components evolve over time, we applied a dynamical systems approach across three curvature regimes: flat, negatively curved, and positively curved universes. This method allowed us to identify equilibrium points, assess their stability, and explore how the equation-of-state parameters for dark matter ($w_{dm}$) and dark energy ($w_{de}$) influence cosmic dynamics. Each curvature scenario exhibits distinct behaviors, with specific conditions leading to radiation, matter, or dark energy dominance, as well as scaling solutions that mirror the standard $\Lambda$CDM model or reveal alternative evolutionary paths. Below, we summarize the key dynamical features observed in each curvature case.

     Flat Universe ($k = 0$): 
        Radiation-dominated solutions act as sources under physical conditions.
         Matter-dominated solutions drive deceleration when $w_{dm} > -1/3$.
       Dark energy-dominated solutions enable accelerated expansion when $w_{de} < -1/3$.
         Scaling solutions with constant $\Omega_{dm}/\Omega_{de}$ ratios are possible and consistent with $\Lambda$CDM evolution.
   Negatively Curved Universe ($k = -1$):
     Radiation and baryon-dominated solutions behave as sources or saddle points depending on $w_{de}$ and $w_{dm}$.
     Dark matter-dominated solutions vary between sinks (quintessence-like), sources (nonphysical), and saddles.
     Dark energy-dominated solutions ensure accelerated expansion under viable conditions.
     Bifurcation values such as $w_{de} = -1/3$ mark transitions in curvature growth and acceleration.
     Positively Curved Universe ($k = +1$):
     Nine equilibrium scenarios emerge, including dark energy, matter, and radiation-dominated equilibrium points.
     Mixed scaling solutions require fine-tuning and are rare.
     The normalized Hubble parameter $Q$ compactifies the phase space and distinguishes between expansion ($Q = 1$), contraction ($Q = -1$), and bounce ($Q = 0$).
     Stability and acceleration depend on $w_{dm}$, $w_{de}$, and the evolution of $Q$.

\subsubsection{Scaling Solutions and Stability}

Scaling solutions correspond to equilibrium points where the ratios among energy densities remain constant. These arise when all derivatives vanish and each component contributes non-trivially. Stability is determined by the eigenvalues of the linearized system; negative real parts indicate attractor behavior in cosmic evolution.
For all curvature cases, scaling behavior follows:
  \begin{equation*}
  \rho_i \propto a^{-3(1 + w_i)} \quad \Rightarrow \quad \Omega_i \propto e^{-3(1 + w_i)N}.
  \end{equation*}
 Viable scaling requires a balance between curvature and dark energy, especially when $w_{de} < -1/3$, enabling late-time acceleration.

\subsubsection{Bifurcation Analysis Across Curvature Regimes}

Flat Universe ($k = 0$)  
    System defined by dynamical equations with curvature set to zero.
      Equilibrium point: $\Omega_r = 0$, $\Omega_{dm} = 0$, $\Omega_{de} = 1$, $\Omega_k = 0$.
  Eigenvalues:
      \begin{equation*}
      \lambda_r = 3w_{de} - 1, \quad
      \lambda_{dm} = 3w_{de} - 3w_{dm}, \quad
      \lambda_{de} = 3w_{de}, \quad
      \lambda_k = 3w_{de} + 1.
      \end{equation*}
     Bifurcation values:  
        $w_{de} = \frac{1}{3}$ (radiation),  
        $w_{de} = w_{dm}$ (dark matter),  
        $w_{de} = 0$ (dark energy),  
        $w_{de} = -\frac{1}{3}$ (curvature).
     Representative case:  
        $w_{de} = 0$, $w_{dm} = 0$ reproduces standard $\Lambda$CDM evolution.  
        Scaling solutions with constant $\Omega_{dm}/\Omega_{de}$ are possible.  
        Curvature remains zero throughout.

   Negatively Curved Universe ($k = -1$). System includes curvature coupling in $\Omega_r'$, $\Omega_{de}'$, and $\Omega_k'$. The equilibrium point $\Omega_r = 0$, $\Omega_{de} = 1$, $\Omega_k = 0$ has the eigenvalues:
      \begin{equation*}
      \lambda_r = 3w_{de} - 1, \quad
      \lambda_{de} = 3w_{de}, \quad
      \lambda_k = 3w_{de} + 1.
      \end{equation*}
   Bifurcation values:  
        $w_{de} = \frac{1}{3}$ (radiation),  
        $w_{de} = 0$ (dark energy),  
        $w_{de} = -\frac{1}{3}$ (curvature).
     Representative cases:  
        $w_{de} = 0$, curvature grows, no acceleration.  
        $w_{de} = \frac{1}{3}$, unstable growth of all components.  
        $w_{de} = -1$, stable attractor, curvature decays.
 Positively Curved Universe ($k = +1$).
   
    System uses compactified variables $\hat{\Omega}_i$ and normalized Hubble parameter $Q$.
      Equilibrium points include radiation, matter, and dark energy dominated states, as well as mixed scaling solutions.
   Bifurcation values:  
        $w_{de} = -\frac{1}{3}$ (acceleration onset),  
        $w_{dm} = -\frac{1}{3}$ (curvature dynamics),  
        $Q = 0$ (bounce or turnaround).
   Representative cases:  
        $w_{de} = -1$, stable acceleration, $\hat{\Omega}_{de} \to 1$, $Q \to 1$.  
        $w_{de} = 0$, deceleration, curvature dominates.  
        $w_{de} = \frac{1}{3}$, rapid dilution, unstable.
     Role of $Q$:  
        $Q = 1$ (expansion),  
        $Q = 0$ (bounce),  
        $Q = -1$ (contraction).  
        Evolution governed by:
        \begin{equation}
        Q' = -\frac{1}{2}(1 - Q^2)\left(3w_{de} \hat{\Omega}_{de} + 3w_{dm} \hat{\Omega}_{dm} + \hat{\Omega}_r + 1\right).
        \end{equation}
    Unified dark sector case:  
       when $w_{de} = w_{dm} = w$, the system reduces to a 3D phase space with $\Omega_x = \Omega_{dm} + \Omega_{de}$.
       Behavior depends on $w$:  
        $w > 0$, fast dilution.  
        $w < 0$, slow dilution, possible acceleration.  
        $w = -1$, mimics cosmological constant.
        
\section{Cosmological constraints with data samples} \label{constraintsdatasamples}
\normalfont


In this section, we present the cosmological data sets employed in the Bayesian analysis to constrain the model parameters in both scenarios: the flat and non-flat universe.
 \subsection{Observational Hubble Data}
Galaxies evolving passively over a time much longer than their age difference allow us
to measure the Hubble parameter $H(z)$ using the differential age method within the framework of general relativity \cite{Jimenez_2002}, this consists of observing galaxies with the same age of formation (observing their 4,000 \text{\AA} break in the passively evolving spectra) and with two different but very close redshifts we can use these galaxies as cosmic clocks using the formula given by

\begin{equation}
    H(z)=- \frac{1}{1+z} \frac{dz}{dt},
\end{equation}
where $dt$  is the differential time evolution
of the universe at redshift interval $dz$.
This can be applied to any model assuming an FLRW cosmology and using OHD samples \cite{DATAHZ}. Additionally, we will utilize new points for the OHD data \citep{HZnewpoint1, hznewpoint2, hznewpoint3}. With this data, we can use the chi-square function given by

\begin{equation}
\chi^2_{\text{OHD}} = \sum\limits_{i=1}^{N} \left( \frac{H_{\text{th}}(z_i) - H_{\text{obs}}(z_i)}{\sigma_{i,\text{obs}}} \right)^2,
\end{equation}
where $H_{\text{th}}(z_i)$ is the theoretical Hubble parameter  $H_{\text{obs}}(z_i) \pm \sigma_{i,\text{obs}}$ is the observational Hubble parameter  (from the differential age method) with its uncertainty at the redshift $z_i$, and $N$ is the number of points used. Where $N=33$ and the redshift range is $0.07<z<1.965$

\subsection{Type Ia Supernovae}

White dwarfs are compact remnants of low- to intermediate-mass stars, supported against gravitational collapse by electron degeneracy pressure. In close binary systems, a white dwarf can accrete from its companion, usually hydrogen or helium \cite{whitedwarcompo}. As its mass increases and approaches the Chandrasekhar limit ($\sim 1.44,M_{\odot}$, \cite{chandrasekharlimit}), the conditions at its core become extreme enough to trigger the runaway thermonuclear fusion of carbon (C) and oxygen (O) \cite{biografiaSNIa}.
This results in a catastrophic explosion, characteristic of a Type Ia supernova, which releases a large amount of energy and can be observed in galaxies at high redshift.

Because their peak luminosity can be standardized, SNIa serve as essential cosmological distance indicators. Observations of these events provide measurements of the distance modulus, $\mu$, which can be compared with the theoretical predictions of cosmological models. In this work, we utilized the most recent compilation of SNIa, known as Pantheon +, which comprises 1701 data points spanning the redshift range $0.001 < z < 2.26$ \cite{DATASN}. The $\chi^2$ function for these data is given as

\begin{equation}\label{eq:chi_snia}
 \chi^2_{\text{SNIa}} = \alpha + \log\left(\frac{\gamma}{2\pi}\right) - \frac{\beta^2}{\gamma}, 
\end{equation} 
 where
 
\begin{equation*}
\alpha = (\Delta{\mu})^T \cdot \text{Cov}_P^{-1} \cdot \Delta{\mu},
\end{equation*}
\begin{equation*}
\beta = (\Delta{\mu})^T \cdot \text{Cov}_P^{-1} \cdot \Delta1,
\end{equation*} 
\begin{equation*}
\gamma = \Delta1^T \cdot \text{Cov}_P^{-1} \cdot \Delta1,
\end{equation*} 
and  $\Delta\mu $ is the vector of residuals between the theoretical distance modulus and the observed one, $ \Delta1 = (1, 1, \ldots, 1)^T $, $ \text{Cov}_P $ is the covariance matrix formed by adding the systematic and statistical uncertainties, i.e., $ \text{Cov}_P = \text{Cov}_{P,\text{sys}} + \text{Cov}_{P,\text{stat}} $. The superscript $ T$ on the above expressions denotes the transpose of the vectors.
The theoretical distance modulus is estimated by
\begin{equation}
m_{\text{th}} = M + 5 \log_{10} \left[\frac{D_L(z)}{10 \text{pc}}\right],
\end{equation}
\noindent
where $ M $ is a nuisance parameter which has been marginalized in \eqref{eq:chi_snia}
and the luminosity distance ($d_{L}$) is given by

\begin{equation}
D_L(z) = (1+z) \frac{c}{H_0} d(z),
\end{equation}
where $d(z)$ is a function that depends on the curvature parameter as
\begin{align}
    d(z) =
    \begin{cases}
        \frac{1}{\sqrt{\Omega_k}} \sinh\left( \sqrt{\Omega_k} \, \chi(z) \right) & \text{if } \Omega_k > 0, \\
        \chi(z) & \text{if } \Omega_k = 0, \\
        \frac{1}{\sqrt{|\Omega_k|}} \sin\left( \sqrt{|\Omega_k|} \, \chi(z) \right) & \text{if } \Omega_k < 0,
    \end{cases}
    \label{eq:dz}
\end{align}

being $\chi(z)$ the comoving radial distance given by:

\begin{equation}
    \chi(z) = \int_0^{z} \frac{dz'}{E(z')}, \qquad
\end{equation}
which depends on the cosmological parameters through the dimensionless Friedmann equation ($E(z)$).

\subsection{Strong lensing systems}
The deflection of light paths characterizes the gravitational lensing phenomenon due to the curvature of spacetime induced by massive objects. This distortion enables the observation of objects located behind the massive object; however, it results in the apparent displacement of the source from the observer's perspective.

The deflection of light paths characterizes the gravitational lensing phenomenon, which is caused by the curvature of spacetime induced by matter. This distortion enables the observation of objects located behind the massive lensing structure; however, it results in the apparent displacement of the source from the observer’s perspective.

Measuring the universe's expansion using gravitational lenses is based on the Einstein radius ($\theta_E$), which represents the angular separation between the light source and its image as observed. By reconfiguring the relationship between angular distances ($\frac{D_{LS}}{D_S}$), we obtain:
\begin{equation}
    \frac{D_{LS}}{D_S} = \frac{c^2 \theta_E}{4\pi \sigma^2_{SIS}},
\end{equation}
where $D_S$ is the angular diameter distance to the source, and $D_{LS}$ is the angular diameter distance between the lens and the source. Here, the velocity dispersion $\sigma_{SIS}$ within a Singular Isothermal Sphere (SIS) model is employed as an observational proxy for mass, owing to its simplicity.

Several groups analyzing Strong Lensing Systems (SLS) have applied different methodologies to constrain cosmological parameters. These systems encompass lens redshifts within the range $0.06 \lesssim z_l \lesssim 1$ and source redshifts within $0.2 \lesssim z_s \lesssim 3.5$. The chi-square function for SLS is formulated as:
\begin{equation}
    \chi^2_{SLS} = \sum_{i=1}^{N_{SLS}} \left[ \frac{D_{th}(z_l^i, z_s^i; \theta) - D_{obs}^i}{\delta D_{obs}^i} \right]^2,
\end{equation}
where the observed value is computed via:
\begin{equation}
    D_{obs} = \frac{c^2 \theta_E}{4 \pi \sigma^2_{SIS}},
\end{equation}
and its uncertainty is propagated as:
\begin{equation}
    \delta D_{obs} = D_{obs} \sqrt{ \left( \frac{\delta \theta_E}{\theta_E} \right)^2 + 4 \left( \frac{\delta \sigma}{\sigma} \right)^2 }.
\end{equation}

In cosmological models where spatial curvature is non-zero, the theoretical quantity $D_{th}$ must be computed using curvature-corrected comoving distances. In such cases, we use the distance sum rule given by \cite{distancesumrule}:

\begin{equation}
    D_{th}(z_l, z_s; \theta) = \sqrt{1 + \Omega_k d_l^2} - \left( \frac{d_l}{d_s} \right) \sqrt{1 + \Omega_k d_s^2},
\end{equation}
here, $d_i(z_i)$ (with $i = l, s$) is given by \eqref{eq:dz}, where the subscript $i$ refers to the lens or the source, respectively.

The compilation of SLS data used for this work is \cite{DEMODELSSLS} alongside measurements for $ \sigma $ and $ \theta_E $ offers an avenue to estimate cosmological parameters by minimizing the chi-square function described earlier. This computation of uncertainties follows the standard method of error propagation.

\subsection{Baryonic Acoustic Oscillations}
Baryon acoustic oscillations emerged from the interplay of two fundamental forces during the universe's early stages: gravity and pressure. These forces clashed in an immensely dense region of primordial plasma composed mainly of electrons and baryons (protons and neutrons). The plasma confined photons through Thomson scattering, exerting significant outward pressure, while simultaneously, gravity pulled matter towards the dense center.

As the universe expanded and cooled during the epoch of recombination $(z\simeq 1100)$, hydrogen atoms began to form from baryonic matter. With hydrogen being electrically neutral, photons could travel more freely through space, no longer trapped by constant interactions. This allowed the photons to decouple from the plasma, leaving behind patterns of baryonic matter in the form of waves.

These patterns, or baryon acoustic oscillations, imprinted the CMB. The distance from the central dense region to the first ripple of baryonic matter, where galaxies later formed, became a standard ruler for measuring cosmic expansion \cite{bassett2009baryon}. 
The theoretical BAO angular scale ($\theta_{\text{th}}$) is estimated as

\begin{equation}
\theta_{\text{th}}(z) = \frac{r_{\text{drag}}}{(1 + z)D_{\text{A}}(z)}.
\end{equation}

The comoving sound horizon, $r_{s}(z)$, is defined as

\begin{equation}\label{eq:83}
r_{s}(z) = \frac{c}{H_{0}} \int_{0}^{z} \frac{c_{s}(z^{'})}{E(z^{'})} \, dz^{'},
\end{equation}
where the sound speed $c_{s}(z) = \frac{1}{\sqrt{3(1 + R_{\bar{b}}/(1 + z))}}$, with \newline $R_{\bar{b}} = 31500 \Omega_{b}h^{2}(T_{\text{CMB}}/2.7\,\text{K})^{-4}$, and $T_{\text{CMB}}$ is the CMB temperature. The redshift $z_{\text{drag}}$ at the baryon drag epoch is well-fitted with the formula proposed by \cite{BAOFORMULA}, given by
\begin{equation}
\begin{aligned}
z_{\text{drag}} = & \ 1291(\Omega_{m0}h^{2})^{0.251} \left[1 + 0.659 (\Omega_{m0}h^{2})^{0.828} \right. \\
& \hspace{3cm} \cdot \left. \left(1 + b_{1}\left(\frac{\Omega_{b0}h^{2}}{b_{2}}\right)\right)\right],
\end{aligned}
\end{equation}
where
\begin{align*}
b_{1} &= 0.313 (\Omega_{m0} h^{2})^{-0.419} h^{1 + 0.607(\Omega_{m0} h^{2})^{0.674}}, \\
b_{2} &= 0.238 (\Omega_{m0} h^{2}).
\end{align*}
where $\Omega_{\text{m}0}=\Omega_{dm0}+\Omega_{\text{b}0}$.
Compilation of transversal BAO measurements $\theta_{\text{BAO}}(z)$ is presented in \cite{DATABAO}.
As transversal angular BAO points are considered uncorrelated, the chi-square function is built as

\begin{equation}
\chi^2 = \sum_{i=1}^{N} \frac{(\theta_{\text{BAO}}^{i} - \theta_{\text{th}}(z_i))^2}{(\sigma_{\theta_{\text{BAO}}^{i}})^2},
\end{equation}
where $\theta_i^{\text{BAO}} \pm \sigma_{\theta_{\text{BAO}}^{i}}$ is the BAO angular scale, $N$ is the number of data and its uncertainty at 68\% measured at $z$.
In addition, we include nine new measurements from the DESI 2025 DR2 release \cite{desi2025} with redshift range of $0.295<z<2.330$, with eight values of the comoving distance ratio $D_M/r_d$, and one of the dilation scale ratio $D_V/r_d$, where $r_d = r_s(z_{\text{drag}})$. The comoving distance $D_M(z)$ and the dilation scale $D_V(z)$ \citep{DILATIONSCALE2005} are defined as follows
\begin{align}
D_M(z) &= \frac{c}{H_0} \int_0^z \frac{dz'}{E(z')}, \\
D_H(z) &= \frac{c}{H(z)}, \\
D_V(z) &= \left[z D_H(z) D_M^2(z)\right]^{1/3},
\end{align}
with normalized observables
\begin{equation}
D_{M,rd} = \frac{D_M}{r_d}, \quad D_{V,rd} = \frac{D_V}{r_d}.
\end{equation}

The contribution from DESI is encoded in the chi-square function
\begin{equation}
\chi^2_{D_V,D_M} = \sum_{i} \frac{\left(D_i^{\text{obs}} - D_i^{\text{th}}(z_i)\right)^2}{\sigma_{D_i}^2},
\end{equation}
where $D_i = \{ D_{M,rd}, D_{V,rd} \}$, and $\sigma_{D_i}$ are the reported uncertainties. The total BAO chi-square is then
\begin{equation}
\chi^2_{\text{BAO}} = \chi^2_{\theta} + \chi^2_{D_V,D_M}.
\end{equation}
\subsection{Cosmic Microwave Background}

In the early epochs of the universe, photons were tightly coupled to baryonic matter. As the universe expanded and cooled, electrons and protons recombined to form neutral hydrogen in the process of recombination, drastically reducing the number of free electrons. At that moment, at redshift $z \simeq 1100$ \cite{CMB_1968,CMB_1969}, photons decoupled from baryonic matter and began to travel freely through space, generating the Cosmic Microwave Background observed today. We can employ the compressed distance posterior derived from the acoustic peaks of the CMB, namely the acoustic scale ($\mathbf{l_A}$), the shift parameter ($R$), and the decoupling redshift ($z_\ast$) to constrain cosmological models.
The acoustic scale is defined as
\begin{equation}
l_A = \frac{\pi r(z_\ast)}{r_s(z_\ast)} ,
\end{equation}

where $r_s$ is the sound horizon defined in eq.\ref{eq:83} at the redshift of decoupling $z_\ast$ given
by Hu and Sugiyama \cite{Hu_1996},
\begin{equation}
z_\ast = 1048 \left[1 + 0.00124(\Omega_{b0}h^2)^{-0.738}\right]\left[1 + g_1(\Omega_{m0}h^2)^{g_2}\right] ,
\end{equation}

where
\begin{equation}
g_1 = \frac{0.0783(\Omega_b h^2)^{-0.238}}{1 + 39.5(\Omega_{b0}h^2)^{0.763}} , 
\quad 
g_2 = \frac{0.560}{1 + 21.1(\Omega_{b0}h^2)^{1.81}} .
\end{equation}

The shift parameter is defined as \cite{SHIFT_PARAMETER}
\begin{equation}
R = \frac{\sqrt{\Omega_{m0}} \, H_0}{c}\, r(z_\ast) ,
\end{equation}
where $\Omega_{m0}=\Omega_{dm0}+\Omega_{b0}.$

Thus, the $\mathbf{\chi^2}$ for the CMB data is given as
\begin{equation}
\chi^2_{CMB} = X^T \, {Cov}^{-1}_{CMB} \, X ,
\end{equation}

where $\mathbf{{Cov}^{-1}_{CMB}}$ is the inverse covariance matrix of the distance posteriors and
\begin{equation}
\mathbf{X} =
\begin{pmatrix}
l_A^{\rm th} - l_A^{\rm obs} \\
R^{\rm th} - R^{\rm obs} \\
z_\ast^{\rm th} - z_\ast^{\rm obs}
\end{pmatrix} ,
\end{equation}
the superscripts th and obs refer to the theoretical and observational estimations, respectively. For this work we use the distance posteriors (represented as the obs superscript) for the flat and non-flat cases given by \cite{CMB_DATA2}.

\subsection{Comparison of models} \label{comparison}

After obtaining the constraints from the different cosmological data, we compare the different DE models using the Akaike information criterion (AIC, \cite{1974ITAC...19..716A}) and Bayesian information criterion (BIC, \cite{BIC1974}) defined as:

\begin{equation}
\text{AIC} = \chi^2_{\text{min}} + 2\alpha,
\end{equation}

\begin{equation}
\text{BIC} = \chi^2_{\text{min}} + \alpha \ln N,
\end{equation}
where $\chi^2_{\text{min}}$ is the chi-square obtained from the best fit of the parameters, $\alpha$ is the number of parameters, and $N$ is the number of data points used in the fit. The methodology involves computing the difference between the value of each information criterion and that of a reference model, which is defined as the one with the minimum AIC or BIC. Specifically, $\Delta \text{AIC} = \text{AIC}i - \text{AIC}{\text{min}}$ provides insight into the relative support of the models, whilst $\Delta \text{BIC} = \text{BIC}i - \text{BIC}{\text{min}}$ quantifies the strength of evidence against a given model, as summarized in Table 1. 

\begin{table}
\centering
\begin{tabular}{|c|c|}
\hline
$\Delta$AIC & Empirical support for model $i$ \\
\hline
0 - 2 & Substantial \\
4 - 7 & Considerably less \\
$>$ 10 & Essentially none \\
\hline
$\Delta$BIC & Evidence against model $i$ \\
\hline
0 - 2 & Not worth more than a bare mention \\
2 - 6 & Positive \\
6 - 10 & Strong \\
$>$ 10 & Very strong \\
\hline
\end{tabular}
\caption{Reference values for the $\Delta$AIC and $\Delta$BIC criteria. Note that smaller values of $\Delta$AIC and $\Delta$BIC indicate a preference for the model.}
\end{table}

\section{Results and discussion}\label{results}
\subsection{Bayesian analysis results}
We performed a Bayesian analysis using the MCMC Python module \textit{emcee} \cite{mcmchammer}, setting 200 walkers near the maximum probability region and carrying out a burn-in phase until convergence was reached. After that, we performed 10000 MCMC steps. The convergence of the MCMC analysis is tested using an autocorrelation time test discussed by \cite{Sokal1996MonteCM}. Flat priors were assumed for the equations of state of dark matter and dark energy, specifically, $w_{dm}$: $[-0.3,1.0]$ and $w_{de}$: $[-3.0,0.0]$. For the remaining parameters, we considered the following cases: CI) Gaussian priors for $h=(0.674 \pm 0.005)$ \cite{Planck:2020} and $\Omega_{b0} \cdot h^2 = (0.02218 \pm 0.00055)$ \cite{DESI:2024mwx}; CII) Gaussian priors on $h=(0.7330 \pm 0.0104)$ \citep{Riess_2022} and $\Omega_{b0} \cdot h^2 = (0.02218 \pm 0.00055)$ \cite{DESI:2024mwx}; 
CIII) Gaussian priors on $\Omega_{b0} \cdot h^2 = (0.02218 \pm 0.00055)$ \cite{DESI:2024mwx} and a flat prior on $h$: [0.2,1.0]. In addition, we considered analogous scenarios including a curvature term with a flat prior $\Omega_k:[-0.5,0.5]$, denoted as CIK, CIIK, and CIIIK, respectively.

Table \ref{tab:flatparameters} presents the mean values of the constraints of the $wdm$-$wde$ model parameter in a flat universe, along with the $\chi^2$ and reduced chi-squared ($\chi_{\text{red}}^2$) values for Cases CI, CII, and CIII, respectively, obtained from the OHD, SLS, SNIa, and BAO data sets, as well as for their joint analysis.

\begin{table*}
\centering
{\renewcommand{\arraystretch}{1.2} 
\resizebox{0.9\textwidth}{!}{%
\begin{tabular}{|c|c|c|c|c|c|c|c|}
\hline
\multicolumn{8}{|c|}{Flat Universe}\\
\hline
Data set & $h$ & $\Omega_{dm0}$ & $\Omega_{b0}$ & $w_{dm}$ & $w_{de}$ & $\chi^2$ & $\chi^2_{\text{red}}$ \\
\hline
\multicolumn{8}{|c|}{CI}\\
\hline
OHD & $0.6741^{+0.0275}_{-0.0239}$ & $0.4022^{+0.2187}_{-0.1206}$ & $0.0494^{+0.0032}_{-0.0030}$ & $-0.1443^{+0.2189}_{-0.1995}$ & $-1.1974^{+0.6000}_{-0.4526}$ & $28.94$ & $1.07$ \\
\hline
BAO & $0.6769^{+0.0291}_{-0.0347}$ & $0.3782^{+0.1810}_{-0.2090}$ & $0.0489^{+0.0040}_{-0.0031}$ & $-0.2393^{+0.3112}_{-0.1139}$ & $-0.8301^{+0.5990}_{-1.3736}$ & $35.48$ & $2.36$ \\
\hline
SN Ia & $0.6748^{+0.0179}_{-0.0130}$ & $0.3475^{+0.2267}_{-0.1374}$ & $0.0492^{+0.0019}_{-0.0024}$ & $-0.2076^{+0.2263}_{-0.0661}$ & $-0.8350^{+0.1780}_{-0.4184}$ & $3219.01$ & $1.89$ \\
\hline
SLS & $0.6741^{+0.0274}_{-0.0273}$ & $0.2232^{+0.1230}_{-0.0196}$ & $0.0493^{+0.0031}_{-0.0030}$ & $-0.3049^{+0.1347}_{-0.0262}$ & $-1.7193^{+0.6890}_{-1.1541}$ & $230.56$ & $1.67$ \\
\hline
CMB & $0.6739^{+0.0153}_{-0.0151}$ & $0.3031^{+0.2935}_{-0.1028}$ & $0.04932^{+0.00236}_{-0.00223}$ & $0.00464^{+0.01596}_{-0.02181}$ & $-0.9642^{+0.3785}_{-0.0988}$ & 69.70 & 34.85 \\
\hline
Joint & $0.6730^{+0.0267}_{-0.0258}$ & $0.2748^{+0.0134}_{-0.0129}$ & $0.0495^{+0.0040}_{-0.0036}$ & $0.00089^{+0.00275}_{-0.00288}$ & $-0.9847^{+0.0884}_{-0.0891}$ & 2231.2285 & 1.18 \\
\hline

\multicolumn{8}{|c|}{CII}\\
\hline
OHD & $0.7319^{+0.0275}_{-0.0239}$ & $0.3215^{+0.2187}_{-0.1206}$ & $0.0418^{+0.0032}_{-0.0030}$ & $-0.0968^{+0.2189}_{-0.1995}$ & $-1.5435^{+0.6000}_{-0.4526}$ & 27.78 & 1.02 \\
\hline
BAO & $0.7335^{+0.0291}_{-0.0347}$ & $0.4134^{+0.1810}_{-0.2090}$ & $0.0417^{+0.0040}_{-0.0031}$ & $-0.2158^{+0.3112}_{-0.1139}$ & $-1.5441^{+0.5990}_{-1.3736}$ & 30.48 & 2.03 \\
\hline
SN Ia & $0.7312^{+0.0179}_{-0.0130}$ & $0.3376^{+0.2267}_{-0.1374}$ & $0.0418^{+0.0019}_{-0.0024}$ & $-0.2633^{+0.2263}_{-0.0661}$ & $-0.8430^{+0.1780}_{-0.4184}$ & 3846.36 & 2.26 \\
\hline
SLS & $0.7328^{+0.0274}_{-0.0273}$ & $0.2197^{+0.1230}_{-0.0196}$ & $0.0417^{+0.0031}_{-0.0030}$ & $-0.3036^{+0.1347}_{-0.0262}$ & $-1.7624^{+0.6890}_{-1.1541}$ & 229.20 & 1.66 \\
\hline
CMB & $0.7333^{+0.0300}_{-0.0302}$ & $0.3490^{+0.2487}_{-0.1484}$ & $0.04169^{+0.00362}_{-0.00329}$ & $0.01346^{+0.00987}_{-0.02115}$ & $-1.1428^{+0.4625}_{-0.1850}$ & 69.70 & 34.85 \\
\hline
Joint & $0.6983^{+0.0205}_{-0.0206}$ & $0.2661^{+0.0116}_{-0.0110}$ & $0.04606^{+0.00280}_{-0.00260}$ & $0.00295^{+0.00231}_{-0.00245}$ & $-1.0565^{+0.0713}_{-0.0743}$ & 2122.045 & 1.12 \\
\hline

\hline
\multicolumn{8}{|c|}{CIII}\\
\hline
OHD   & $0.7258^{+0.2023}_{-0.1624}$ & $0.3623^{+0.2243}_{-0.1614}$ & $0.0424^{+0.0261}_{-0.0164}$ & $-0.1746^{+0.2992}_{-0.1543}$ & $-1.8265^{+1.4705}_{-1.1615}$ & $30.57$ & $1.13$ \\
\hline
SN Ia & $0.6461^{+0.3466}_{-0.1679}$ & $0.3486^{+0.2511}_{-0.1482}$ & $0.0537^{+0.0445}_{-0.0309}$ & $-0.2184^{+0.2964}_{-0.1098}$ & $-0.8455^{+0.2267}_{-0.5379}$ & $2411.46$ & $1.42$ \\
\hline
SLS   & $0.8556^{+0.1429}_{-0.3366}$ & $0.2218^{+0.1289}_{-0.0215}$ & $0.0307^{+0.0529}_{-0.0083}$ & $-0.3016^{+0.1522}_{-0.0284}$ & $-1.7573^{+0.6827}_{-1.2128}$ & $227.67$ & $1.64$ \\
\hline
BAO   & $0.8460^{+0.1521}_{-0.3363}$ & $0.2238^{+0.1116}_{-0.0234}$ & $0.0312^{+0.0556}_{-0.0087}$ & $-0.3024^{+0.1257}_{-0.0274}$ & $-1.6982^{+0.6498}_{-1.0465}$ & $100.18$ & $6.67$ \\
\hline
CMB & $0.7372^{+0.2534}_{-0.1265}$ & $0.3043^{+0.2818}_{-0.1043}$ & $0.04120^{+0.01870}_{-0.01838}$ & $0.01122^{+0.01311}_{-0.03993}$ & $-1.1848^{+0.6808}_{-1.1241}$ & 43.53 & 21.76 \\
\hline
Joint & $0.6734^{+0.0143}_{-0.0144}$ & $0.2105^{+0.0529}_{-0.0105}$ & $0.04944^{+0.00228}_{-0.00213}$ & $0.01269^{+0.00606}_{-0.01064}$ & $-0.8442^{+0.0941}_{-0.1426}$ & 2114.07 & 1.11 \\
\hline

\hline

\hline
\end{tabular}
}}
\caption{Best-fit values and corresponding 3$\sigma$ errors for the Wdm–Wde model parameter constraints in a flat universe, for Cases CI, CII, and CIII, obtained from the OHD, SNIa, SLS, and BAO data sets, as well as their joint combination.}
\label{tab:flatparameters}
\end{table*}

Figure \ref{fig:wdmwde_flat_grid} shows the 1D posterior distributions and 2D confidence contours of the free parameters in a flat universe within 1$\sigma$, 2$\sigma$ and 3$\sigma$ (from darker to lighter color bands), obtained from OHD, SLS, SNIa, BAO data, and their joint combination for the CI (top left panel), CII (top right panel), and CIII (bottom left panel) cases.

The reconstruction of $H(z)$ for the $w_{dm}$–$w_{de}$ model, obtained using the joint constraints and compared to the $\Lambda$CDM scenario, is presented in Figure~\ref{hzcombined} for the cases CI (top left panel), CII (middle left panel) and CIII (bottom left panel). The shaded regions indicate the confidence intervals within $1\sigma$ and $3\sigma$. In addition, the left panel of this figure illustrates the reconstruction of the deceleration parameter, $q(z)$ using the joint constraints for the cases CI (top left panel), CII (middle left panel) and CIII (bottom left panel). For CI, it is obtained to $3\sigma$ values of $H(z=0)=67.31^{+1.17}_{-1.31}$, $q(z=0)=-0.43^{+0.09}_{-0.10}$, $z_{t}=0.79^{+0.118}_{-0.04}$, and $t(z=0)=13.96^{+0.41}_{-0.38}\ [Gyr]$.

For CII, to $3\sigma$ values of $H(z=0)=69.84^{+1.73}_{-1.96}$, $q(z=0)=-0.59^{+0.08}_{-0.08}$, $z_{t}=0.63^{+0.06}_{-0.06}$, and $t(z=0)=13.42^{+0.83}_{-0.76}\ [Gyr]$.

For CIII, to $3\sigma$ values of $H(z=0)=67.23^{+2.25}_{-2.43}$, $q(z=0)=-0.49^{+0.11}_{-0.09}$, $z_{t}=0.604^{+0.03}_{-0.04}$, and $t(z=0)=13.68^{+0.91}_{-0.93}\ [Gyr]$.
 
\begin{figure}[H]
    \centering

    \includegraphics[width=0.48\textwidth]{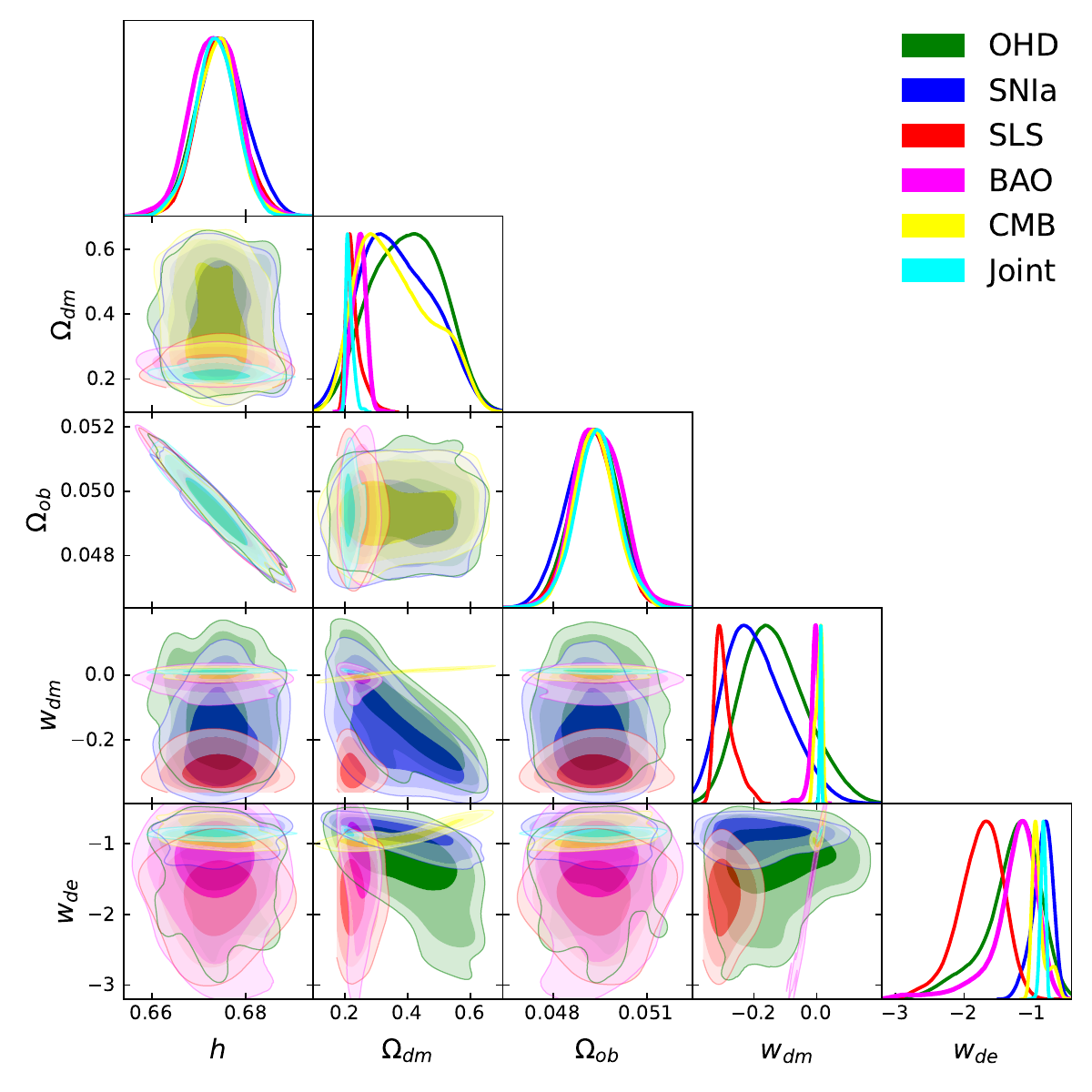}
    \hfill
    \includegraphics[width=0.48\textwidth]{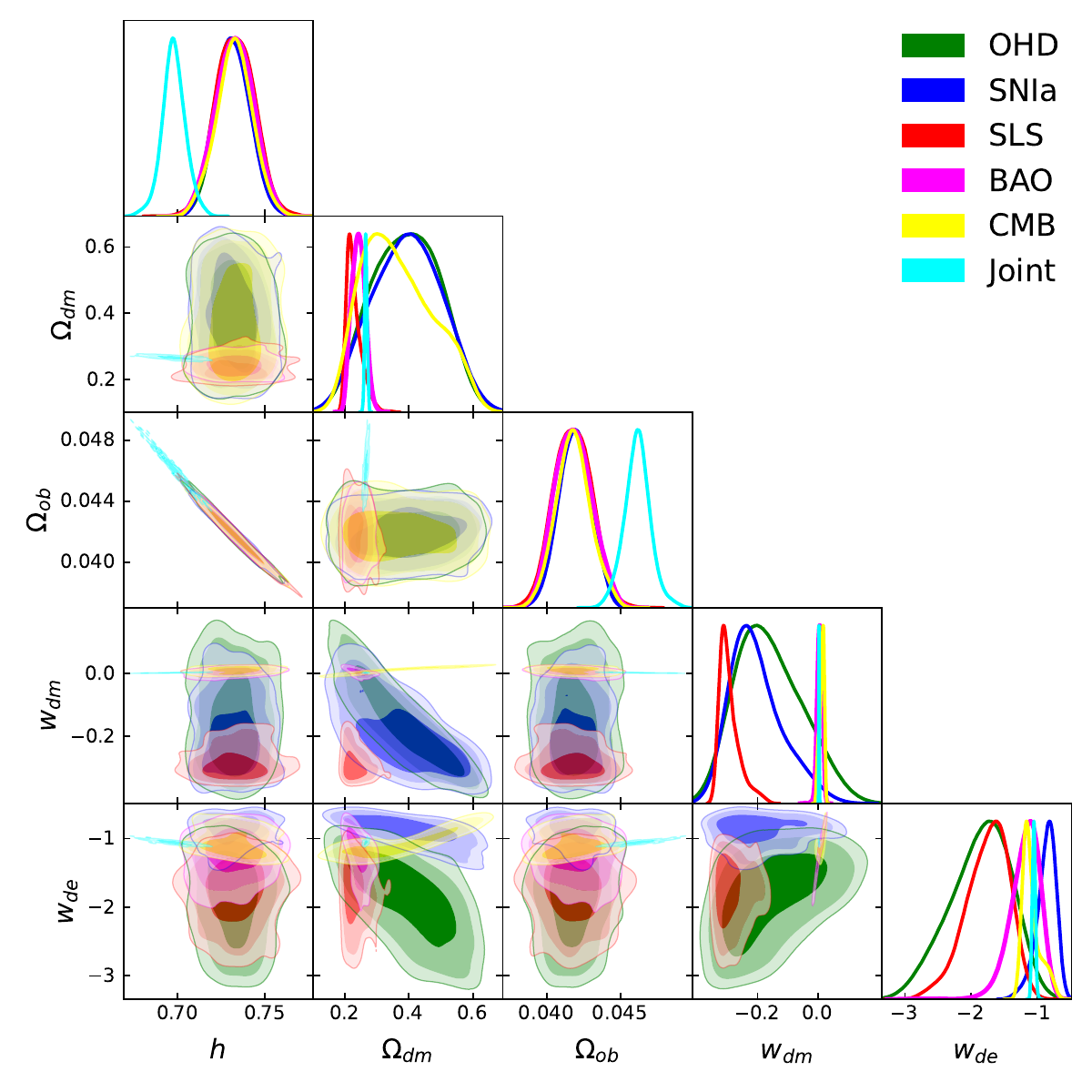}

    \vspace{0.5cm}

    \includegraphics[width=0.48\textwidth]{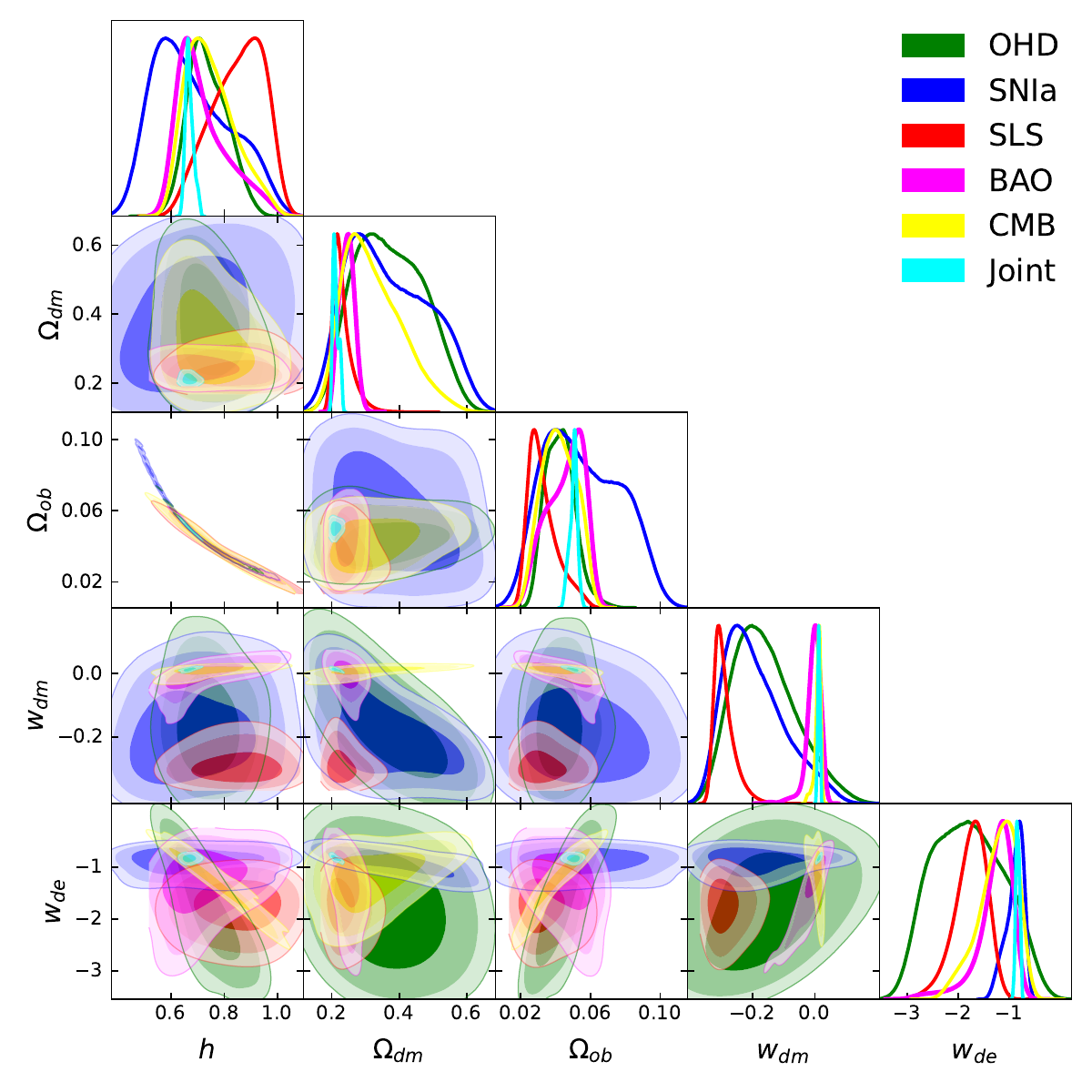}
    \hfill
    \includegraphics[width=0.48\textwidth]{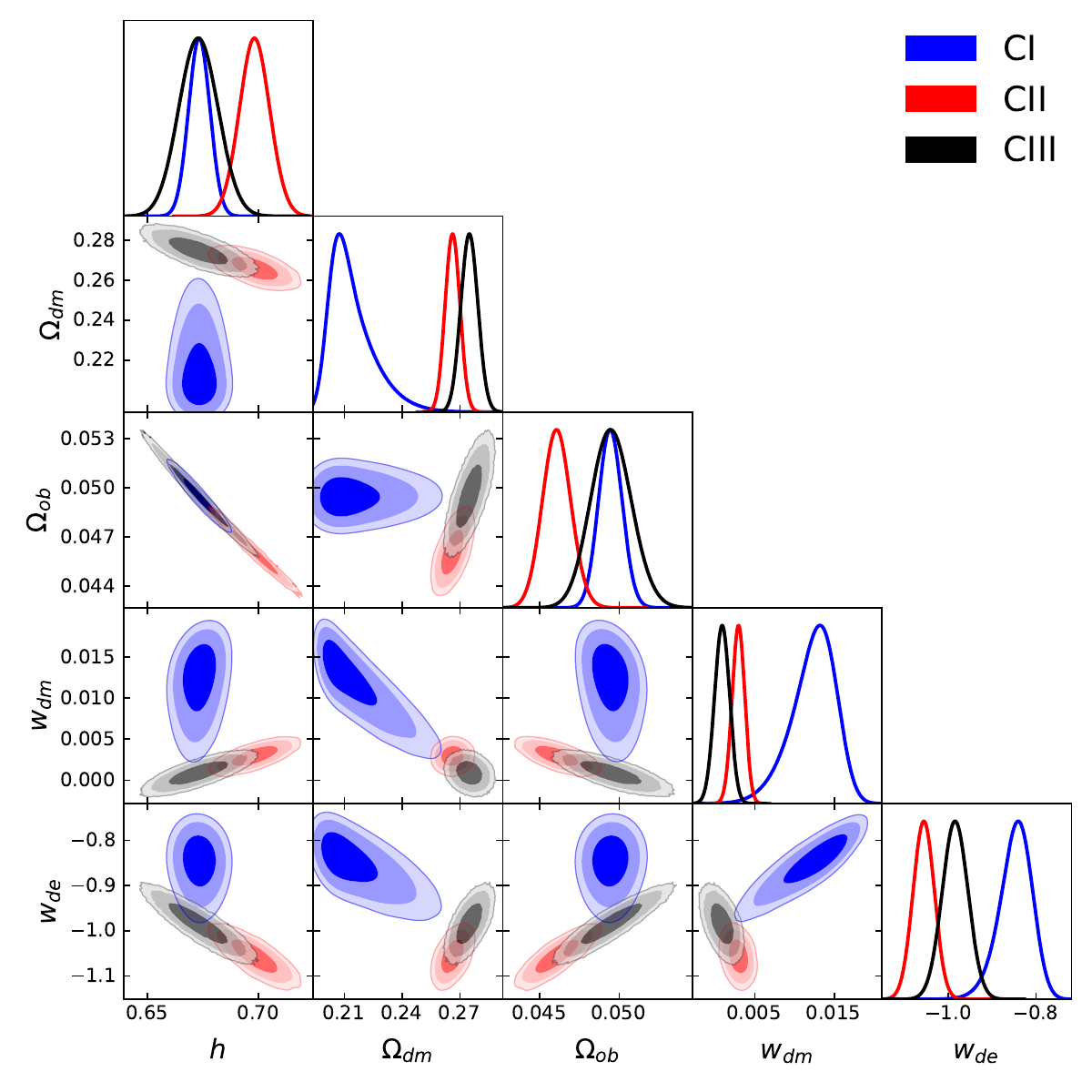}

    \caption{1D posterior distributions and 2D confidence contours for the Wdm-Wde model free parameters from different priors using OHD, SLS, SNIa, BAO, and their joint combinations. Top-left: CI (Gaussian prior on $h$ from \cite{Planck:2020} and  $\Omega_{b0} h^2$ from \cite{DESI:2024mwx}); Top-right: CII (Gaussian prior on $h$ from \cite{Riess_2022} and on $\Omega_{b0} h^2$ from \cite{DESI:2024mwx}); Bottom-left: CIII (Gaussian prior on $\Omega_{b0} h^2$ from \cite{DESI:2024mwx}; and Bottom-right: comparison of the three cases using the joints constraints. Contours represent 1$\sigma$, 2$\sigma$, and 3$\sigma$ levels (from darker to lighter color bands).}
    \label{fig:wdmwde_flat_grid}
\end{figure}

\begin{figure}[H]
\centering
\includegraphics[width=0.4\textwidth]{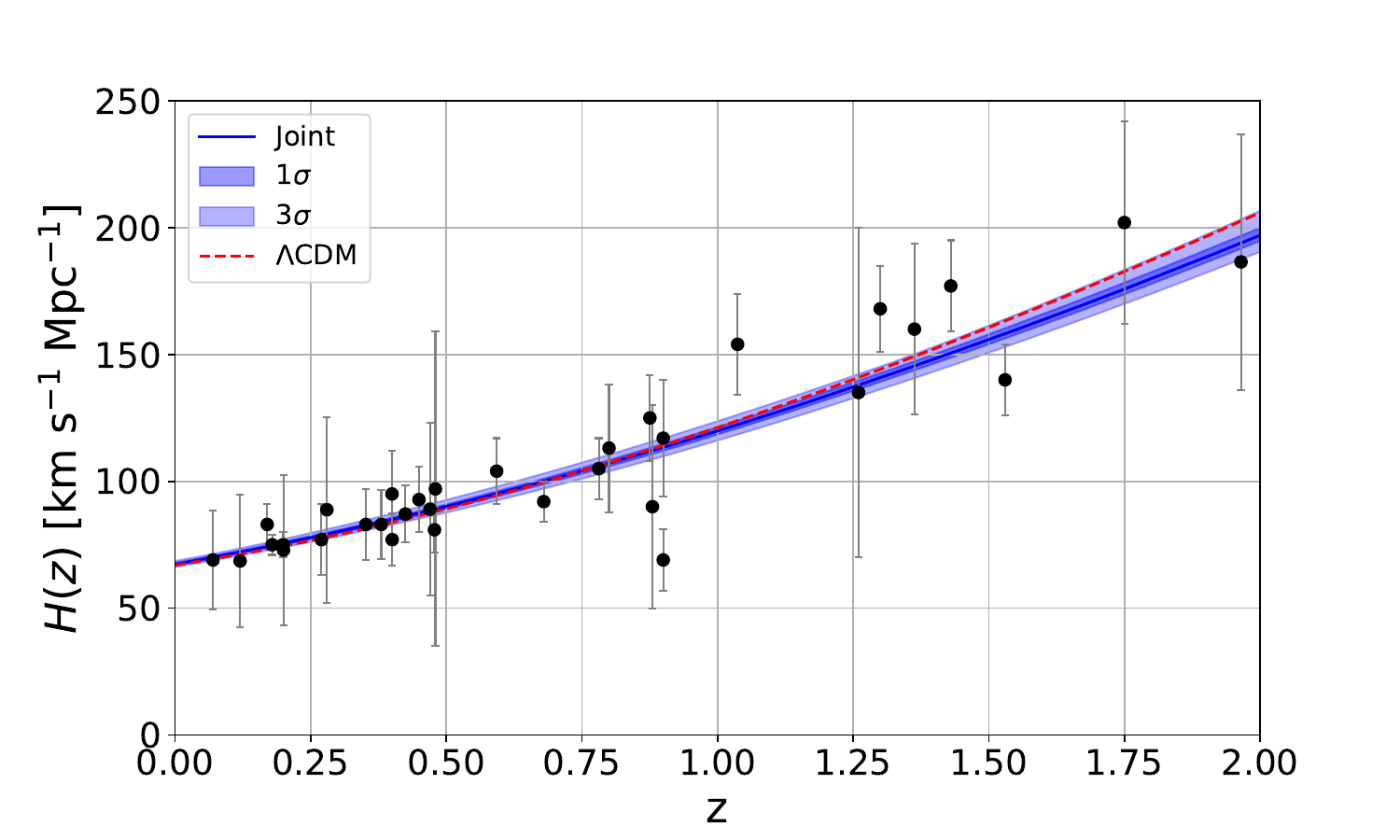}
\includegraphics[width=0.4\textwidth]{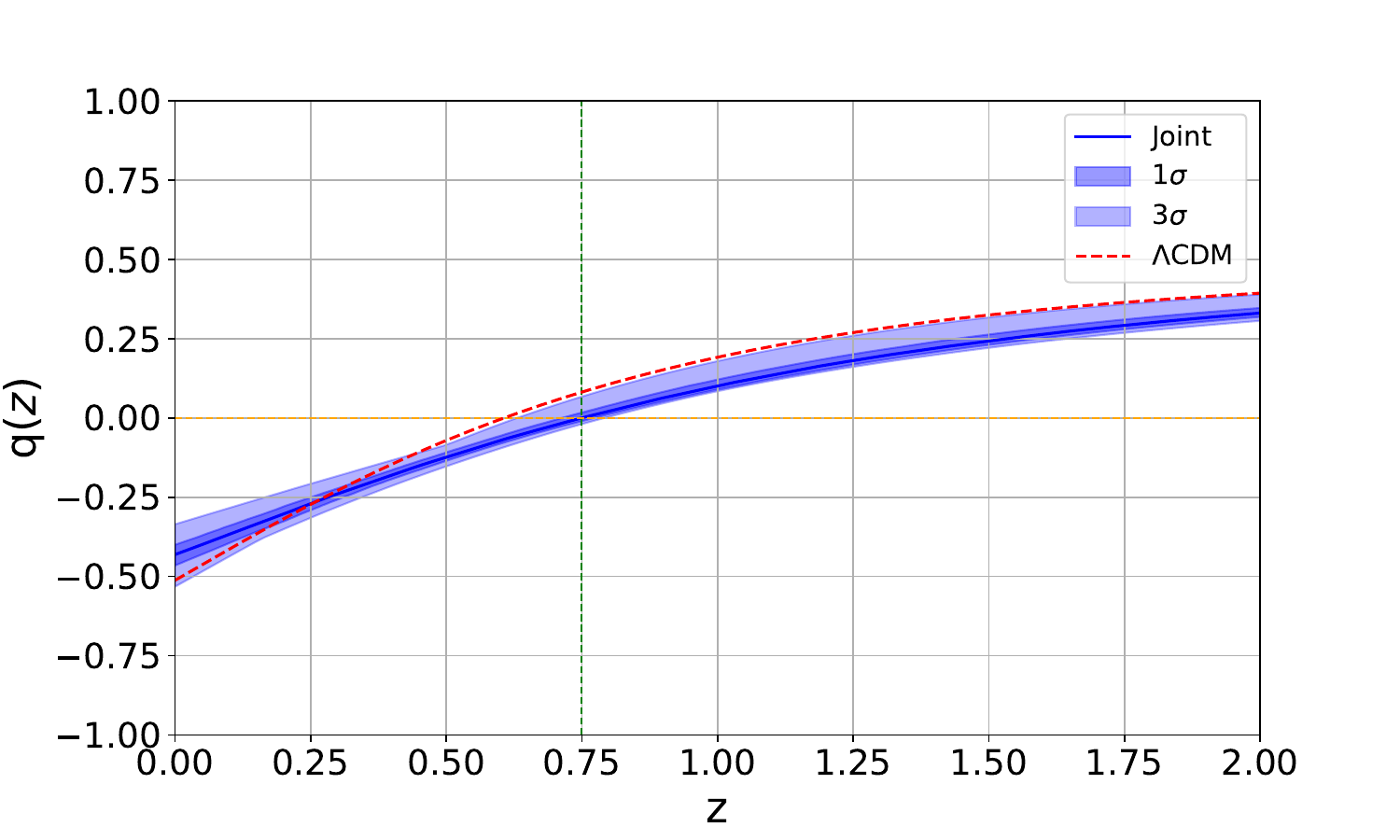} \\
\includegraphics[width=0.4\textwidth]{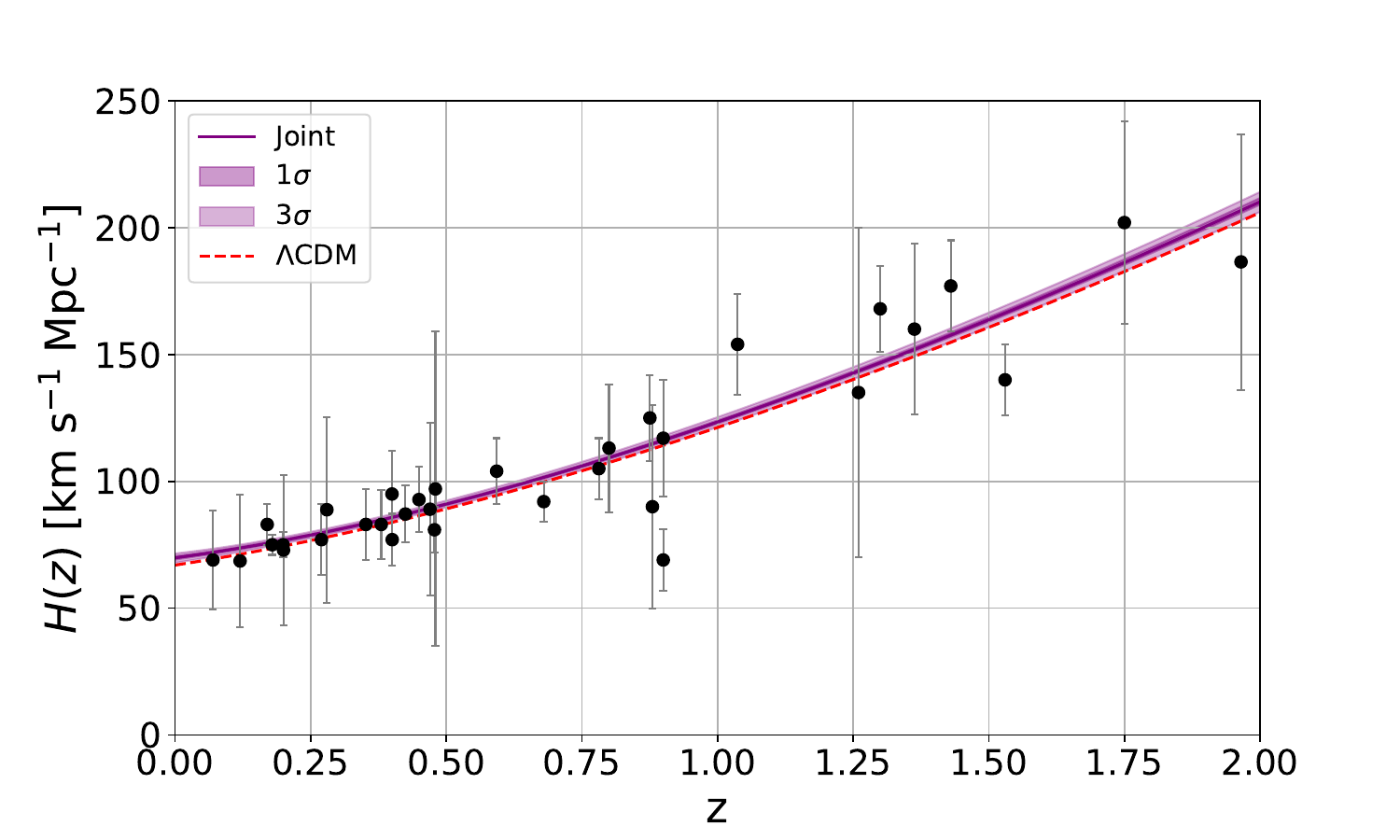}
\includegraphics[width=0.4\textwidth]{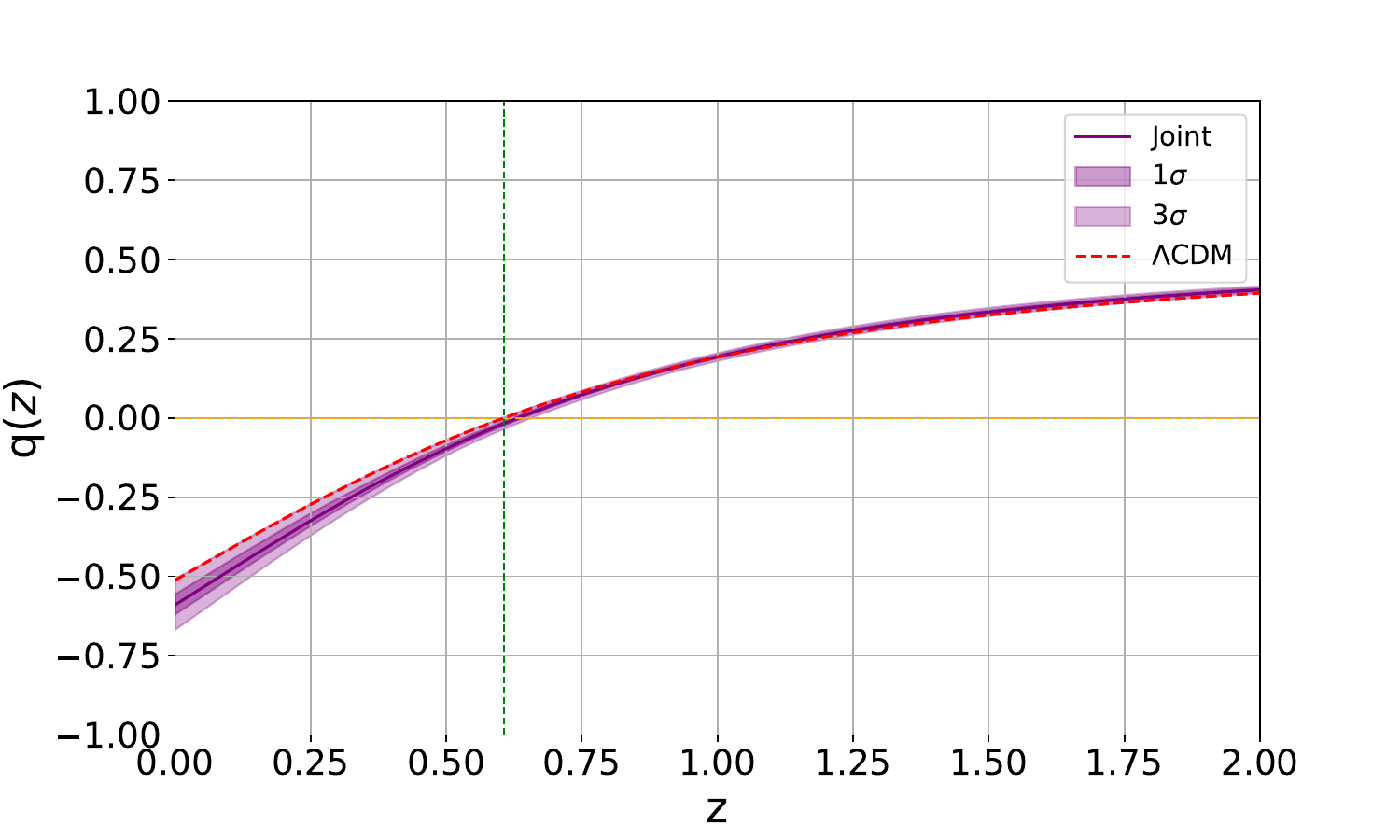}\\
\includegraphics[width=0.4\textwidth]{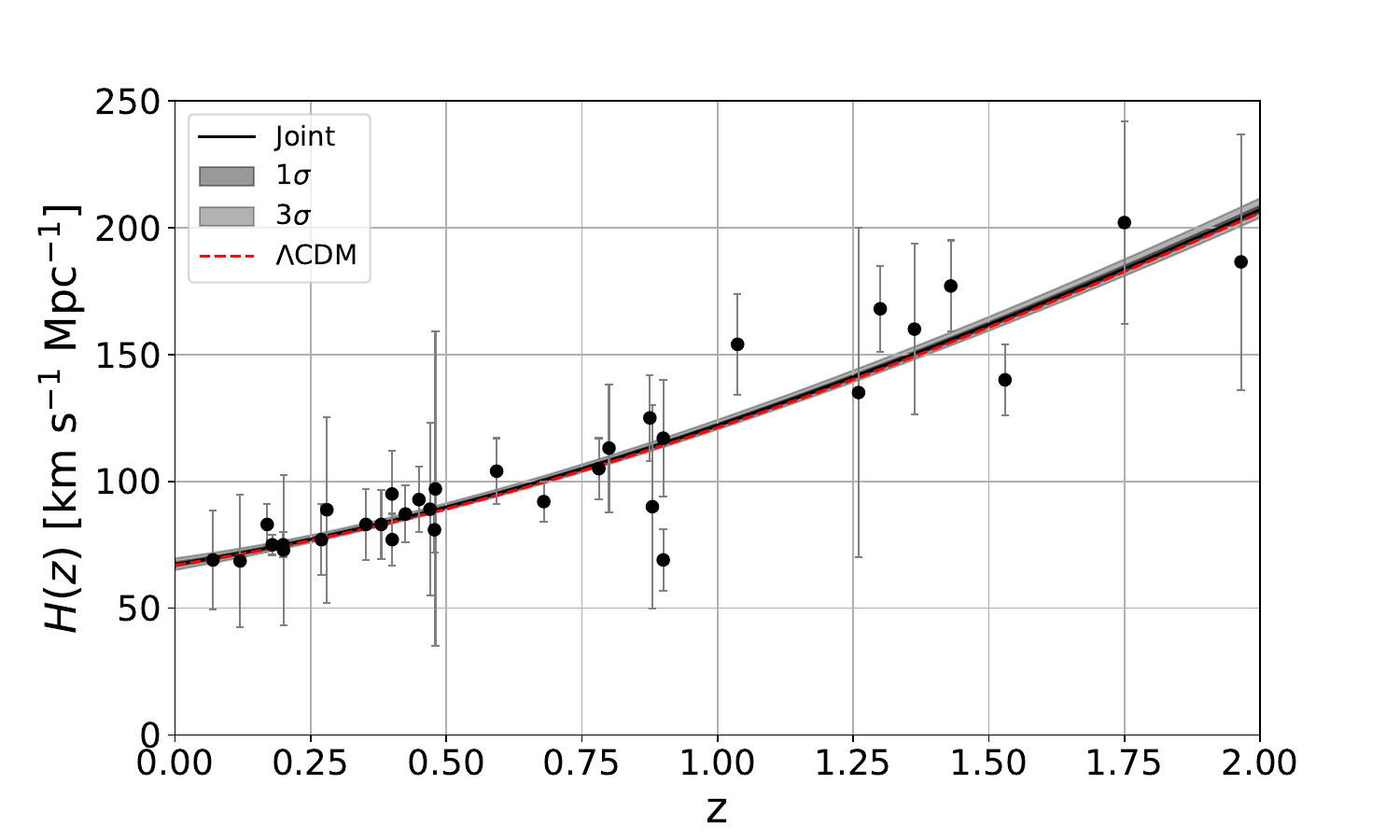}
\includegraphics[width=0.4\textwidth]{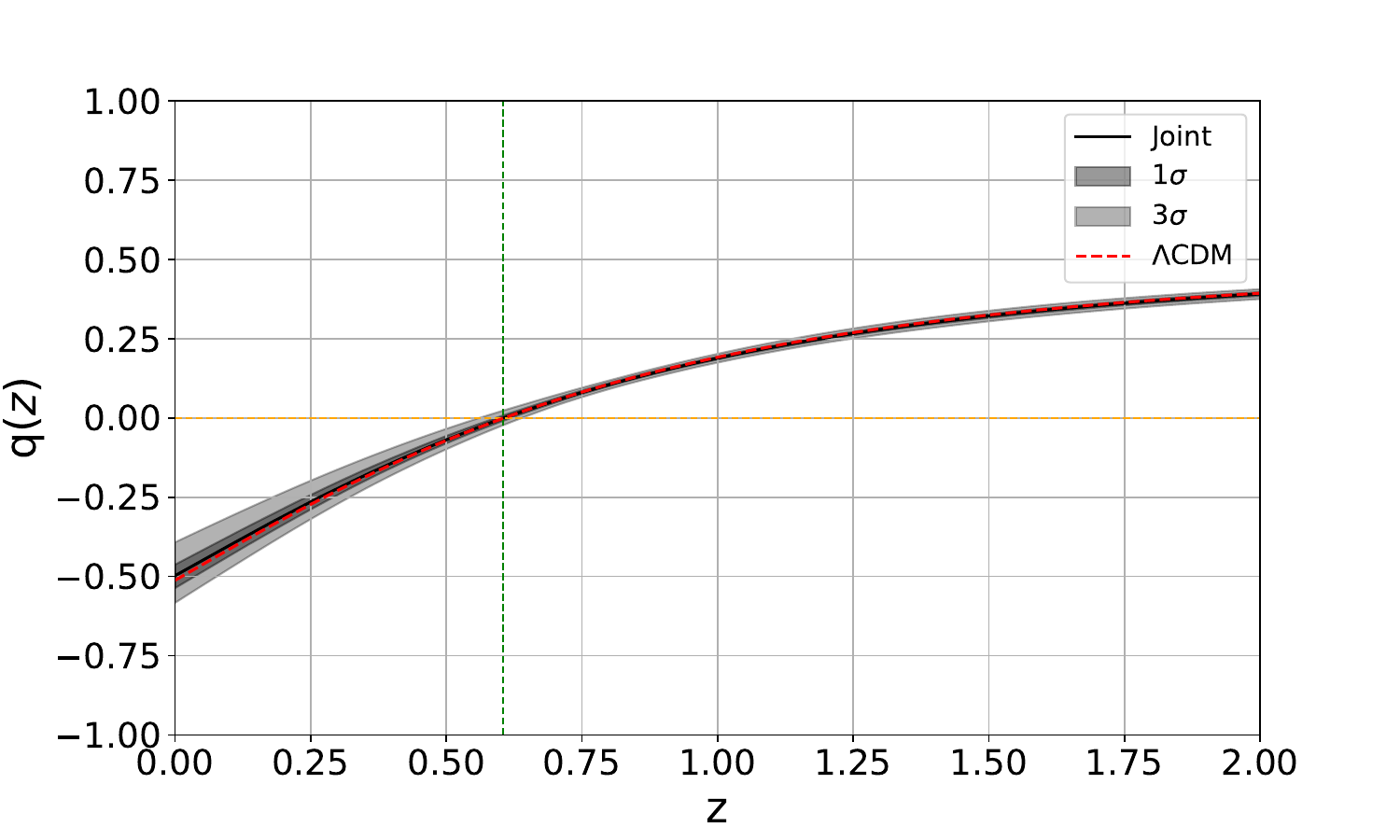}\\
\caption{Reconstruction of the Hubble parameter (left columns) and the deceleration parameter (right columns) for a flat universe using the mean values from the joint constraints for cases CI (top), CII (middle), and CIII (bottom). The shaded regions represent the $1\sigma$ (darker) and $3\sigma$ (lighter) error bands.}
    \label{hzcombined}
\end{figure}

Table \ref{tab:nonflat_parameters} presents the mean values of the constraints of the $w_{dm}-w_{de}$ model parameter in a non-flat universe, along with the $\chi^2$ and reduced chi-squared ($\chi_{\text{red}}^2$) values for Cases CI, CII, and CIII, respectively, obtained from the OHD, SLS, SNIa, and BAO data sets, as well as for their joint analysis.

Figure \ref{fig:wdmwde_curvature_all} shows the 1D posterior distributions and 2D confidence contours of the free parameters in a non-flat universe within 1$\sigma$, 2$\sigma$ and 3$\sigma$ (from darker to lighter color bands), obtained from OHD, SLS, SNIa, BAO data, and their joint combination for the CI (top left panel), CII (top right panel), and CIII (bottom left panel) cases.

Figure~\ref{hzcombined_curvature} illustrates the reconstruction of the Hubble parameter, $H(z)$, (left panels) and the deceleration parameter, $q(z)$, (right panels) for the non-flat $w_{dm}$–$w_{de}$ model employing the joint constraints, compared to the $\Lambda$CDM scenario. The results are shown for cases CIK (top), CIIK (middle), and CIIIK (bottom), with shaded regions representing the confidence levels within $1\sigma$ and $3\sigma$. For case CIK, to $3\sigma$ values of 
$H(z=0) = 67.54^{+1.42}_{-1.44}$, 
$q(z=0) = -0.46^{+0.10}_{-0.07}$, 
$z_{t} = 0.63^{+0.02}_{-0.03}$, 
and $t(z=0) = 13.63^{+0.30}_{-0.29}~\mathrm{[Gyr]}$.

For case CIIK, it is obtained to $3\sigma$ values of 
$H(z=0) = 72.12^{+2.68}_{-2.68}$, 
$q(z=0) = -0.45^{+0.14}_{-0.19}$, 
$z_{t} = 0.74^{+0.09}_{-0.11}$, 
and $t(z=0) = 13.03^{+0.39}_{-0.42}~\mathrm{[Gyr]}$.

For case CIIIK, to $3\sigma$ values of 
$H(z=0) = 68.81^{+4.80}_{-4.75}$, 
$q(z=0) = -0.46^{+0.13}_{-0.15}$, 
$z_{t} = 0.65^{+0.08}_{-0.11}$, 
and $t(z=0) = 13.41^{+1.13}_{-1.04}~\mathrm{[Gyr]}$.

\begin{table*}
\centering
{\renewcommand{\arraystretch}{1.2} 
\resizebox{0.9\textwidth}{!}{%
\begin{tabular}{|c|c|c|c|c|c|c|c|c|}
\hline
\multicolumn{9}{|c|}{Non-flat Universe}\\
\hline
Data set & $h$ & $\Omega_{dm0}$ & $\Omega_{b0}$ & $w_{dm}$ & $w_{de}$ & $\Omega_k$ & $\chi^2$ & $\chi^2_{\text{red}}$ \\
\hline
\multicolumn{9}{|c|}{CIK}\\
\hline
OHD & $0.6746^{+0.0108}_{-0.0132}$ & $0.3532^{+0.2411}_{-0.1514}$ & $0.0492^{+0.0023}_{-0.0015}$ & $-0.1510^{+0.3700}_{-0.1776}$ & $-1.1954^{+0.6097}_{-1.7370}$ & $0.0388^{+0.4175}_{-0.5365}$ & $29.47$& $1.092$\\
\hline
Sne Ia & $0.6739^{+0.0142}_{-0.0136}$ & $0.3370^{+0.2576}_{-0.1356}$ & $0.0494^{+0.0019}_{-0.0022}$ & $-0.2214^{+0.3531}_{-0.1081}$ & $-0.9700^{+0.4054}_{-1.5967}$ & $0.1135^{+0.3784}_{-0.5318}$ & $1761.55$&$1.035$ \\
\hline
SLS & $0.6748^{+0.0106}_{-0.0126}$ & $0.3069^{+0.2831}_{-0.1066}$ & $0.0492^{+0.0019}_{-0.0017}$ & $-0.2169^{+0.3795}_{-0.0825}$ & $-1.5236^{+0.6894}_{-1.4472}$ & $-0.2354^{+0.1970}_{-0.2620}$ &$226.97$ &$1.641$\\
\hline
BAO & $0.6732^{+0.0121}_{-0.0151}$ & $0.2164^{+0.0900}_{-0.0163}$ & $0.0495^{+0.0023}_{-0.0018}$ & $0.0107^{+0.0075}_{-0.0159}$ & $-0.8274^{+0.1570}_{-0.4023}$ & $-0.0101^{+0.3218}_{-0.3966}$ &$80.43$& $5.360$ \\
\hline
CMB & $0.6739^{+0.0157}_{-0.0143}$ 
             & $0.3307^{+0.2687}_{-0.1300}$ 
             & $0.0495^{+0.0023}_{-0.0021}$ 
             & $0.0077^{+0.0132}_{-0.0238}$ 
             & $-1.4027^{+0.8698}_{-1.5932}$ 
             & $0.1111^{+0.0998}_{-0.5031}$ 
             & $283.714$ 
             & $94.57$ \\
\hline
Joint & $0.6754^{+0.0142}_{-0.0144}$
      & $0.2705^{+0.0147}_{-0.0140}$
      & $0.0492^{+0.0022}_{-0.0021}$
      & $0.0007^{+0.0026}_{-0.0030}$
      & $-0.9163^{+0.1072}_{-0.1187}$
      & $-0.0281^{+0.0559}_{-0.0684}$
      & $2189.85$ & $1.349$ \\
\hline
\multicolumn{9}{|c|}{CIIK}\\
\hline
OHD & $0.7323^{+0.0284}_{-0.0268}$ & $0.3269^{+0.2697}_{-0.1243}$ & $0.0418^{+0.0029}_{-0.0032}$ & $-0.1437^{+0.3535}_{-0.1840}$ & $-1.6359^{+0.9396}_{-1.3227}$ & $-0.0256^{+0.4315}_{-0.4723}$ & $30.97$ & $1.143$ \\
\hline
Sne Ia & $0.7318^{+0.0264}_{-0.0244}$ & $0.3491^{+0.2470}_{-0.1452}$ & $0.0419^{+0.0029}_{-0.0031}$ & $-0.2125^{+0.2902}_{-0.1170}$ & $-0.9192^{+0.3859}_{-1.2194}$ & $0.0705^{+0.4236}_{-0.5412}$ & $1760.69$ & $1.037$ \\
\hline
SLS & $0.7327^{+0.0333}_{-0.0328}$ & $0.3119^{+0.2681}_{-0.1109}$ & $0.0417^{+0.0042}_{-0.0037}$ & $-0.2305^{+0.2921}_{-0.0994}$ & $-1.5433^{+0.6771}_{-1.2926}$ & $-0.1970^{+0.2039}_{-0.2752}$ & $225.70$ & $1.630$ \\
\hline
BAO & $0.7367^{+0.0254}_{-0.0199}$ & $0.3511^{+0.2476}_{-0.1490}$ & $0.0414^{+0.0022}_{-0.0029}$ & $0.0105^{+0.3116}_{-0.2472}$ & $-1.1756^{+0.3519}_{-1.3599}$ & $-0.3660^{+0.6147}_{-0.1312}$ & $53.11$ & $3.541$ \\
\hline
CMB & $0.7325^{+0.0312}_{-0.0312}$ 
             & $0.3648^{+0.2346}_{-0.1646}$ 
             & $0.0419^{+0.0039}_{-0.0034}$ 
             & $0.0146^{+0.0092}_{-0.0226}$ 
             & $-1.3990^{+0.8683}_{-1.5975}$ 
             & $0.0683^{+0.1135}_{-0.5556}$ 
             & $316.62$ 
             & $105.800$ \\
\hline
Joint & $0.7212^{+0.0268}_{-0.0268}$
      & $0.2636^{+0.0182}_{-0.0171}$
      & $0.0432^{+0.0034}_{-0.0031}$
      & $0.0047^{+0.0032}_{-0.0039}$
      & $-0.8209^{+0.1014}_{-0.1256}$
      & $-0.1418^{+0.0858}_{-0.1003}$
      & $2229.76$ & $1.176$ \\
\hline
\multicolumn{9}{|c|}{CIIIK}\\
\hline
OHD & $0.7107^{+0.2024}_{-0.1390}$ & $0.3538^{+0.2333}_{-0.1506}$ & $0.0445^{+0.0242}_{-0.0174}$ & $-0.1349^{+0.2644}_{-0.1641}$ & $-1.5639^{+1.0626}_{-1.4339}$ & $-0.0138^{+0.4941}_{-0.4803}$ & $29.35$ & $1.087$ \\
\hline
SNe Ia & $0.6684^{+0.1908}_{-0.1908}$ & $0.3144^{+0.2561}_{-0.1115}$ & $0.0502^{+0.0484}_{-0.0271}$ & $-0.2117^{+0.3573}_{-0.1165}$ & $-1.0012^{+0.4362}_{-1.3751}$ & $0.1050^{+0.3810}_{-0.5773}$ & $1767.82$ & $1.043$ \\
\hline
SLS & $0.6744^{+0.3194}_{-0.1999}$ & $0.3133^{+0.2836}_{-0.1062}$ & $0.0493^{+0.0497}_{-0.0267}$ & $-0.2377^{+0.3399}_{-0.0916}$ & $-1.5052^{+0.5131}_{-1.2831}$ & $-0.2219^{+0.2278}_{-0.2697}$ & $226.42$ & $1.641$ \\
\hline
BAO & $0.7831^{+0.1193}_{-0.0906}$ & $0.3499^{+0.2448}_{-0.1492}$ & $0.0366^{+0.0102}_{-0.0092}$ & $0.0288^{+0.2996}_{-0.3181}$ & $-1.4772^{+0.6702}_{-1.4790}$ & $-0.3279^{+0.4717}_{-0.1695}$ & $46.02$ & $3.068$ \\
\hline
CMB & $0.6750^{+0.1627}_{-0.0632}$ 
             & $0.3227^{+0.2750}_{-0.1219}$ 
             & $0.0493^{+0.0107}_{-0.0174}$ 
             & $0.0084^{+0.0135}_{-0.0340}$ 
             & $-1.2799^{+0.7813}_{-1.7148}$ 
             & $0.0754^{+0.1677}_{-0.4246}$ 
             & $260.41$ 
             & $86.805$ \\
\hline
Joint & $0.6881^{+0.0488}_{-0.0475}$
      & $0.2707^{+0.0147}_{-0.0142}$
      & $0.0474^{+0.0073}_{-0.0061}$
      & $0.0022^{+0.0052}_{-0.0069}$
      & $-0.9101^{+0.1025}_{-0.1281}$
      & $-0.0483^{+0.1011}_{-0.0880}$
      & $2187.162$ & $1.1536$ \\
\hline
\end{tabular}
}}
\caption{Best-fit values and corresponding 3$\sigma$ errors for the $w_{dm}$–$w_{de}$ model parameter constraints in a non-flat universe, for Cases CI, CII, and CIII, obtained from the OHD, SNIa, SLS, and BAO data sets, as well as their joint combination. }
\label{tab:nonflat_parameters}
\end{table*}

Furthermore, to investigate whether the $w_{dm}$–$w{de}$ model is competitive with the $\Lambda$CDM scenario, we perform a MCMC analysis for this model using the joint of same data sets, applying only a Gaussian prior on the baryonic matter density for the flat universe, $\Omega_{b0} \cdot h^2 = 0.02218 \pm 0.00055$ \cite{DESI:2024mwx}, and the same curvature prior for the non-flat case. This yields constraints with $\chi^2 = 2241.23$ for the flat universe, a Hubble constant of $H(z=0) = 67.01^{+0.58}_{-0.55}$, and a deceleration parameter of $q(z=0) = -0.51^{+0.02}_{-0.02}$, with a transition redshift of $z_t = 0.67^{+0.03}_{-0.02}$ and $t(z=0)=13.76^{+0.13}_{-0.12}\ [Gyr]$ at the level $3\sigma$. For the non-flat universe, we obtain 
$\chi^2 = 2197.12$, a Hubble constant of $H(z=0) = 67.94^{+1.27}_{-1.41}$, and a deceleration parameter of $q(z=0) = -0.54^{+0.03}_{-0.03}$, with a transition redshift of $z_t = 0.64^{+0.04}_{-0.05}$ and $t(z=0)=13.72^{+0.11}_{-0.09}~\mathrm{[Gyr]}$ at the $3\sigma$ level. 

\begin{figure}[H]
    \centering

    \includegraphics[width=0.48\textwidth]{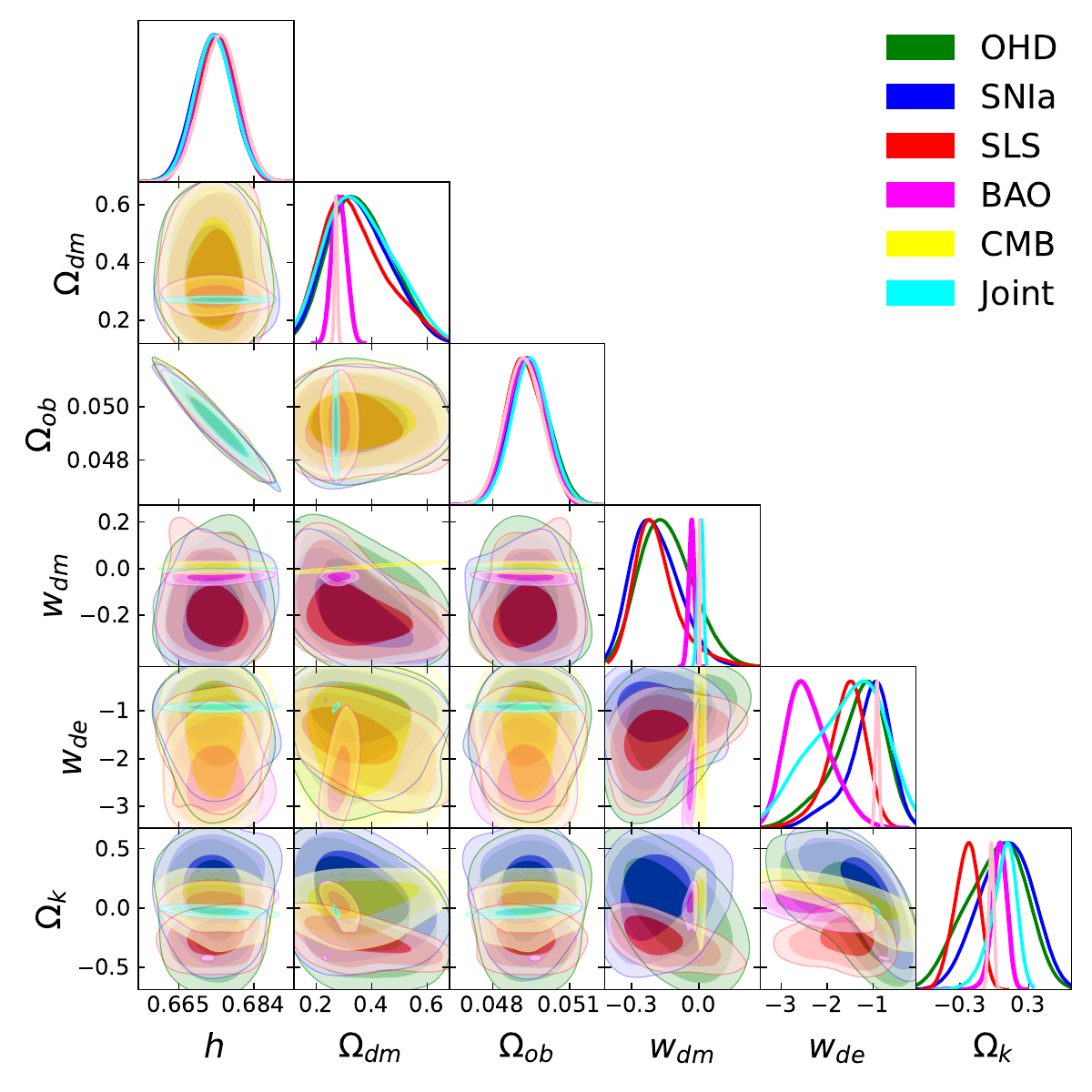}
    \hfill
    \includegraphics[width=0.48\textwidth]{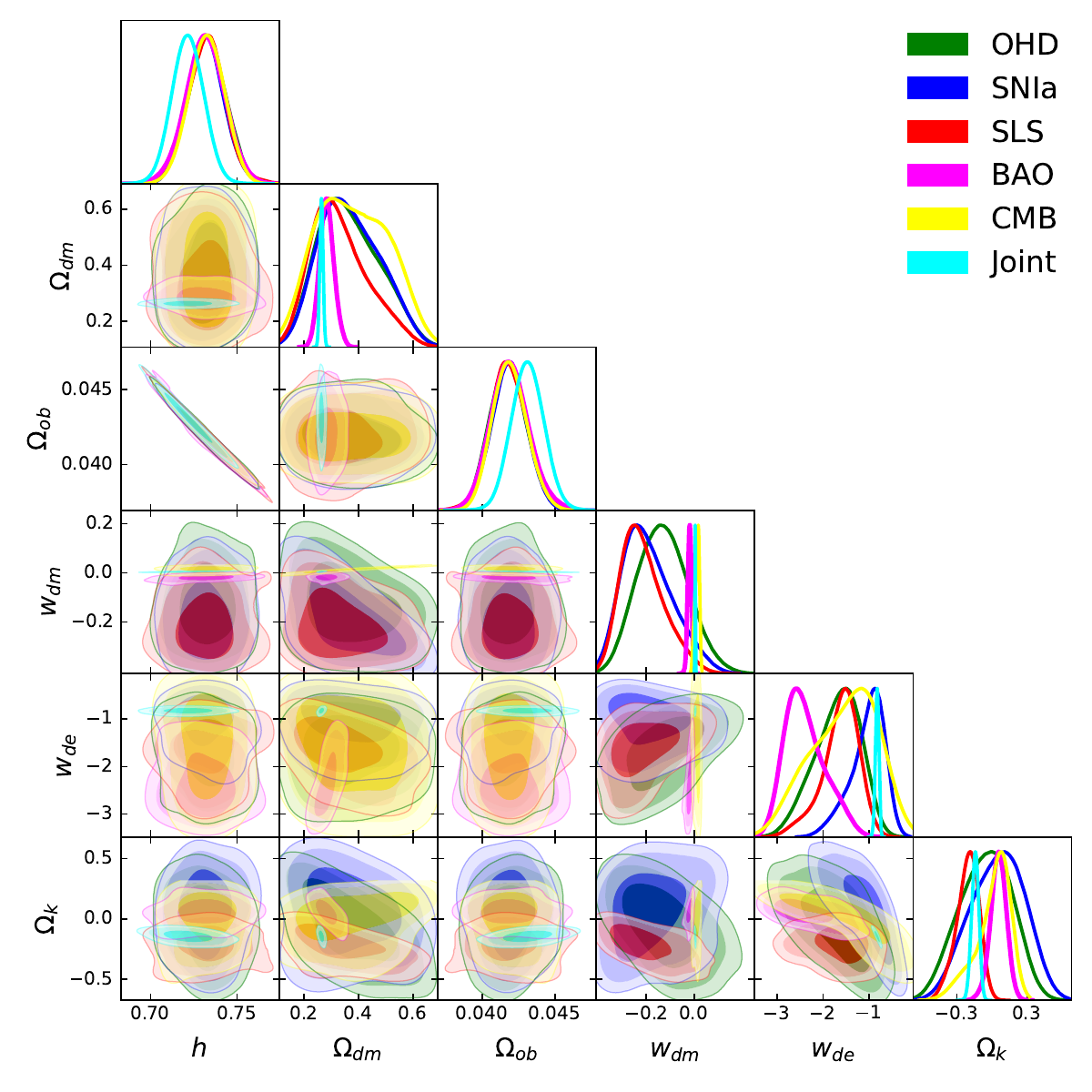}

    \vspace{0.5cm}

    \includegraphics[width=0.48\textwidth]{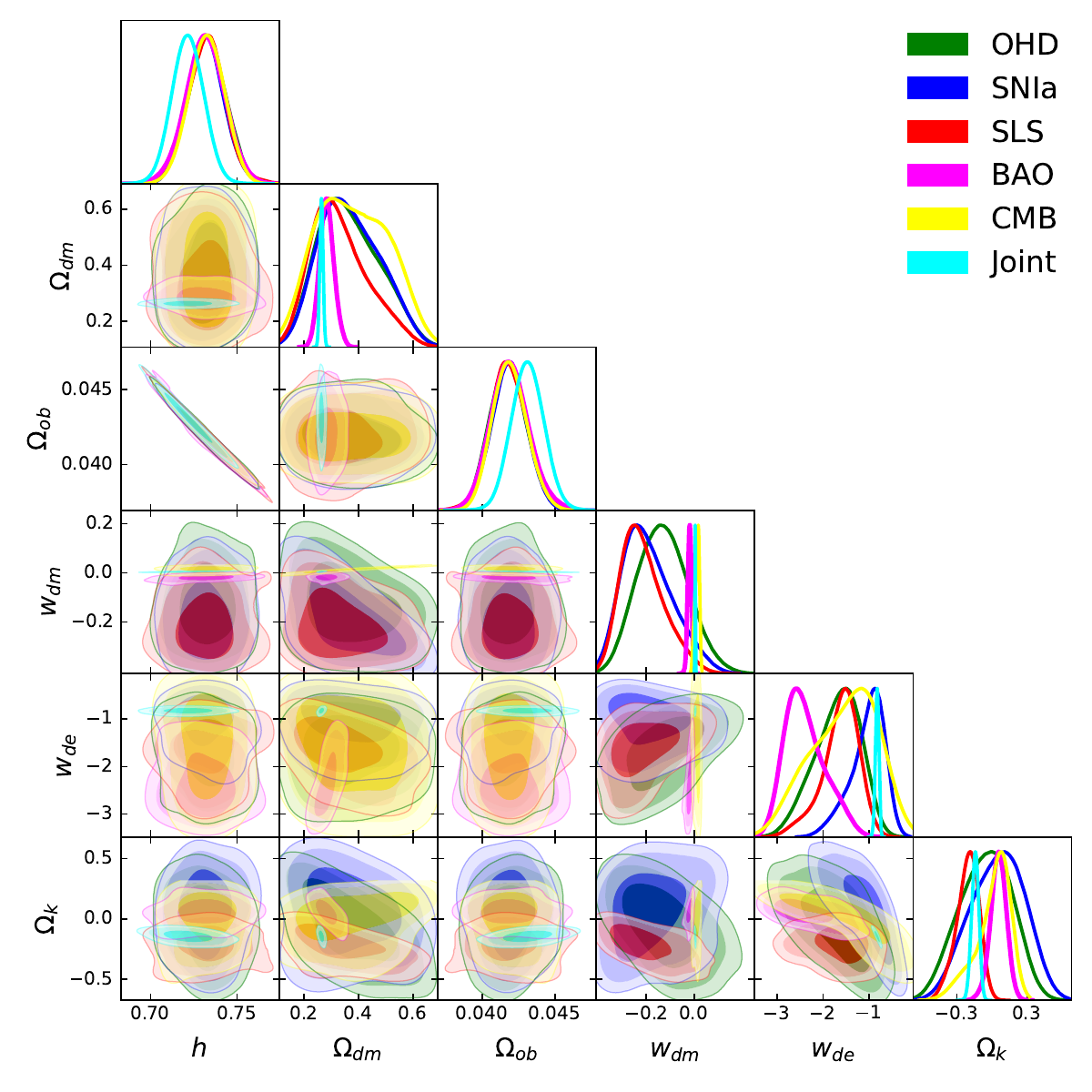}
    \hfill
    \includegraphics[width=0.48\textwidth]{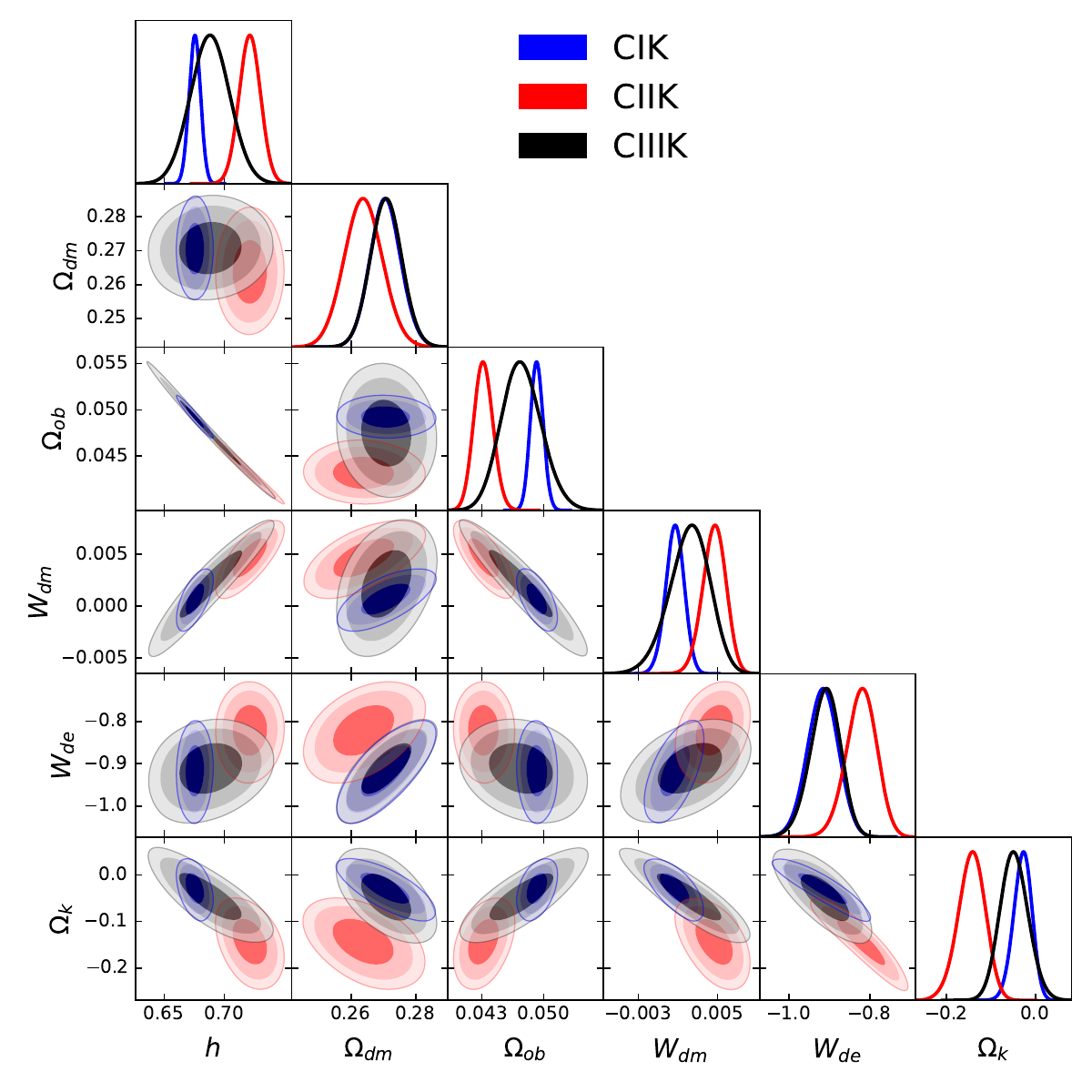}

    \caption{1D posterior distributions and 2D confidence contours for the Wdm-Wde model free parameters in a non-flat universe from different priors using OHD, SLS, SNIa, BAO, and their joint combinations. Top-left: CIK (Gaussian prior on $h$ from \cite{Planck:2020} and  $\Omega_{b0} h^2$ from \cite{DESI:2024mwx}); Top-right: CIIK (Gaussian prior on $h$ from \cite{Riess_2022} and on $\Omega_{b0} h^2$ from \cite{DESI:2024mwx}); Bottom-left: CIIIK (Gaussian prior on $\Omega_{b0} h^2$ from \cite{DESI:2024mwx}; and Bottom-right: comparison of the three cases using the joints constraints. Contours represent 1$\sigma$, 2$\sigma$, and 3$\sigma$ levels (from darker to lighter color bands).}
    \label{fig:wdmwde_curvature_all}
\end{figure}

\begin{figure}[H]
\centering
\includegraphics[width=0.4\textwidth]{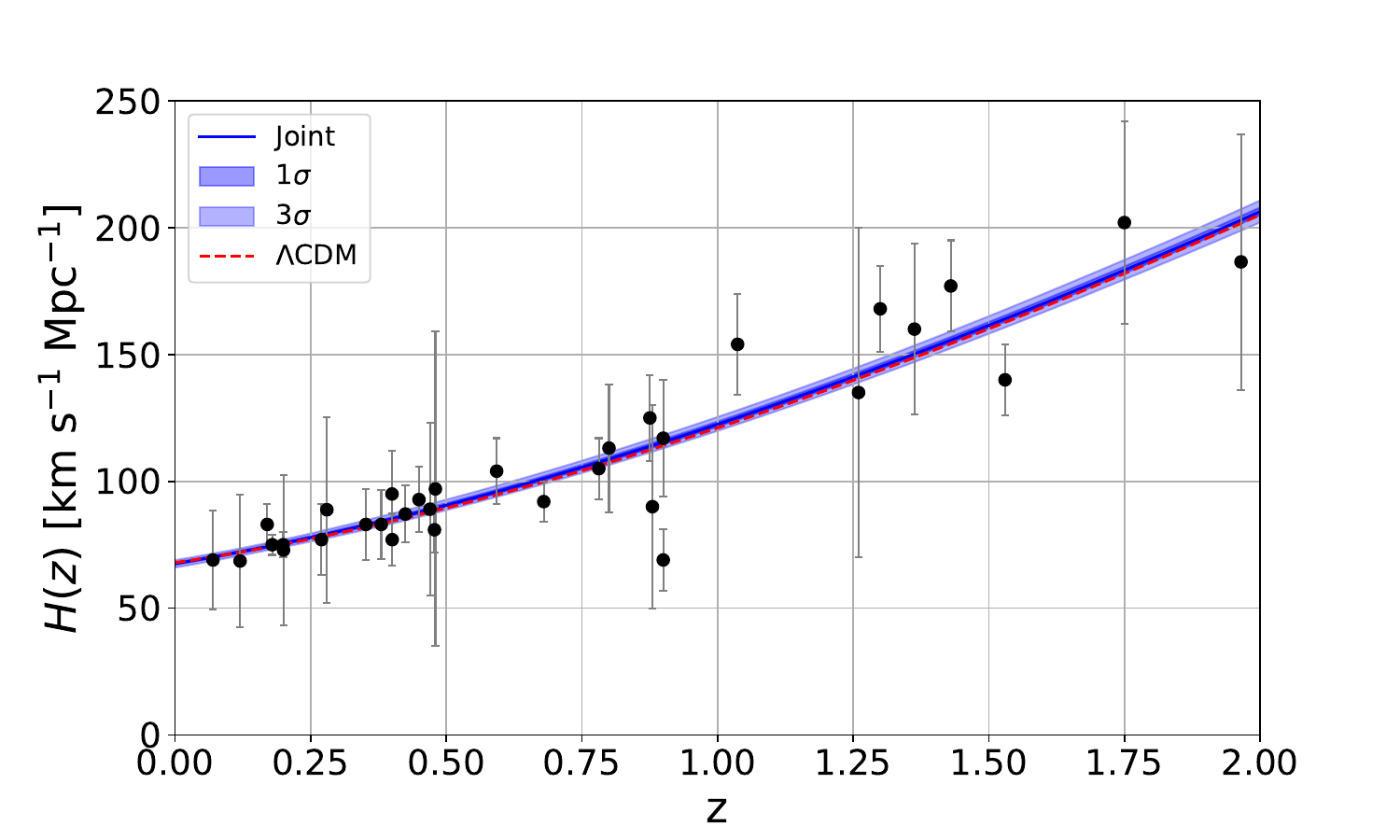}
\includegraphics[width=0.4\textwidth]{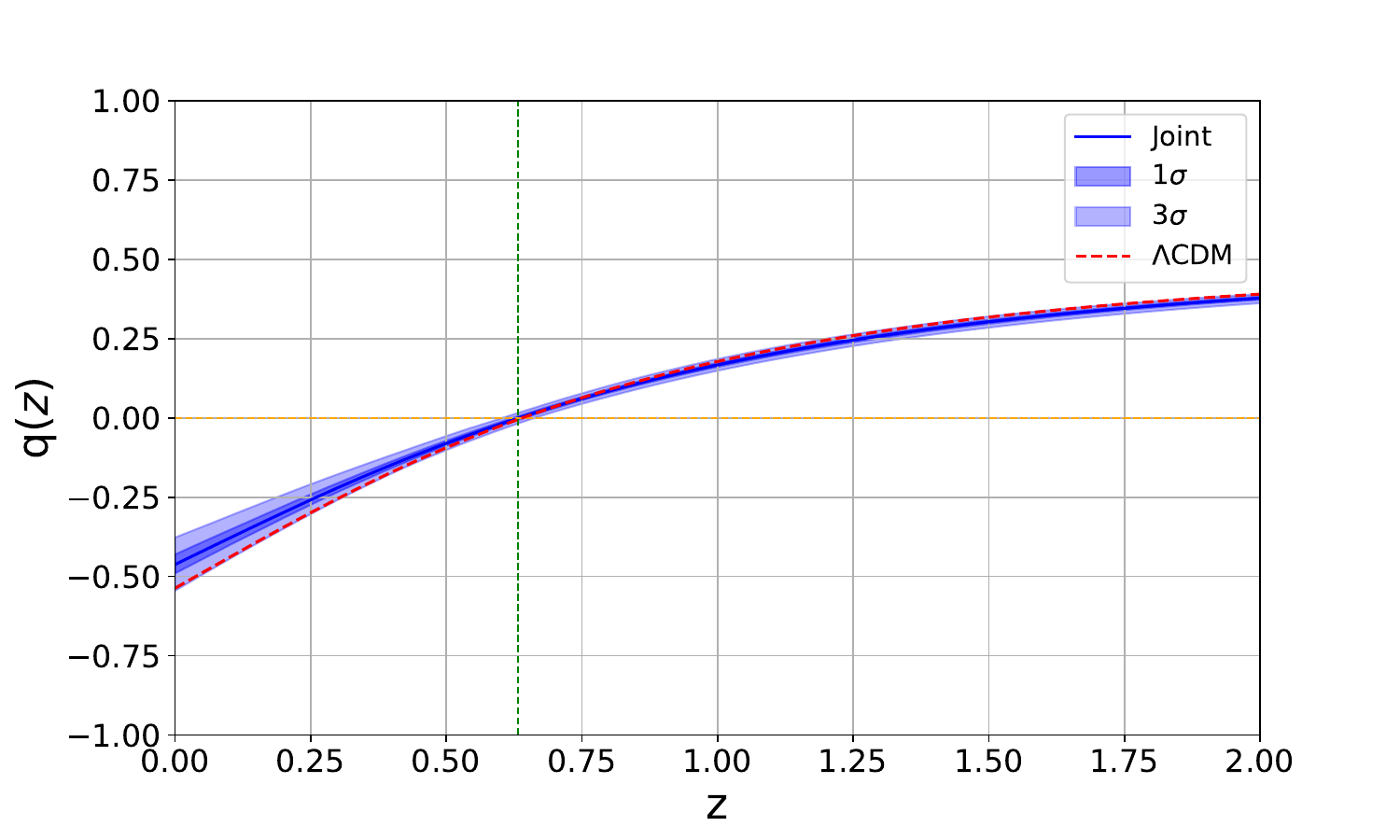} \\
\includegraphics[width=0.4\textwidth]{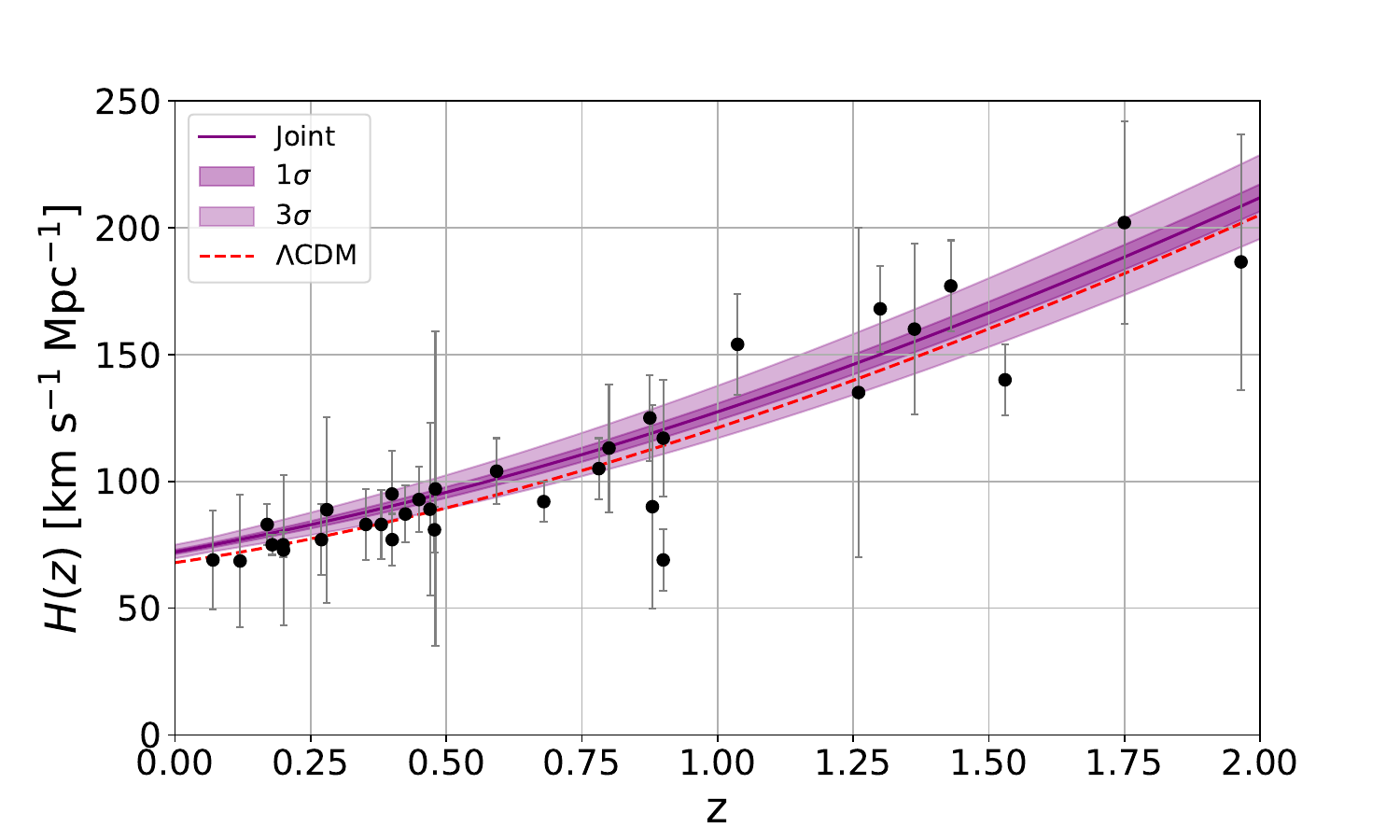}
\includegraphics[width=0.4\textwidth]{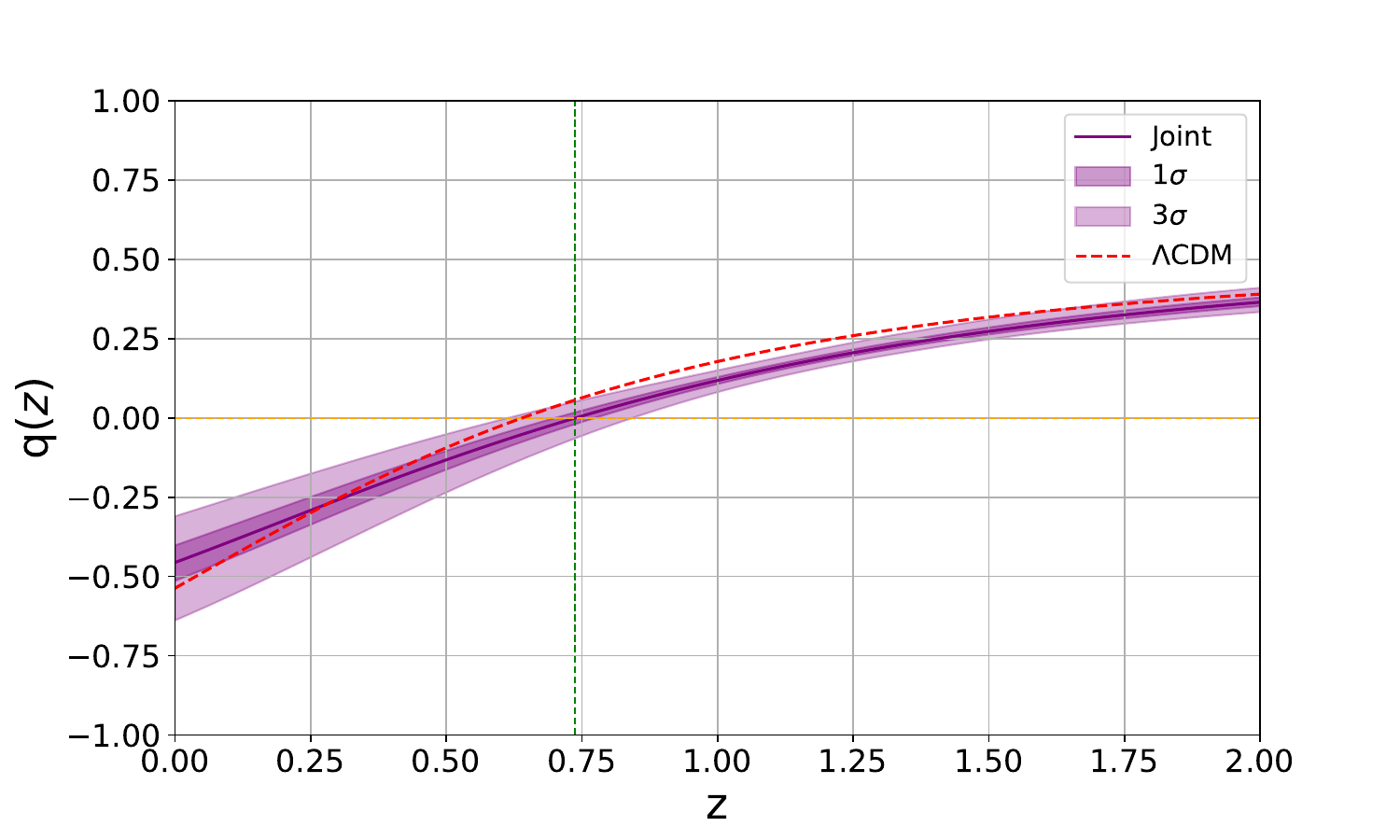}\\
\includegraphics[width=0.4\textwidth]{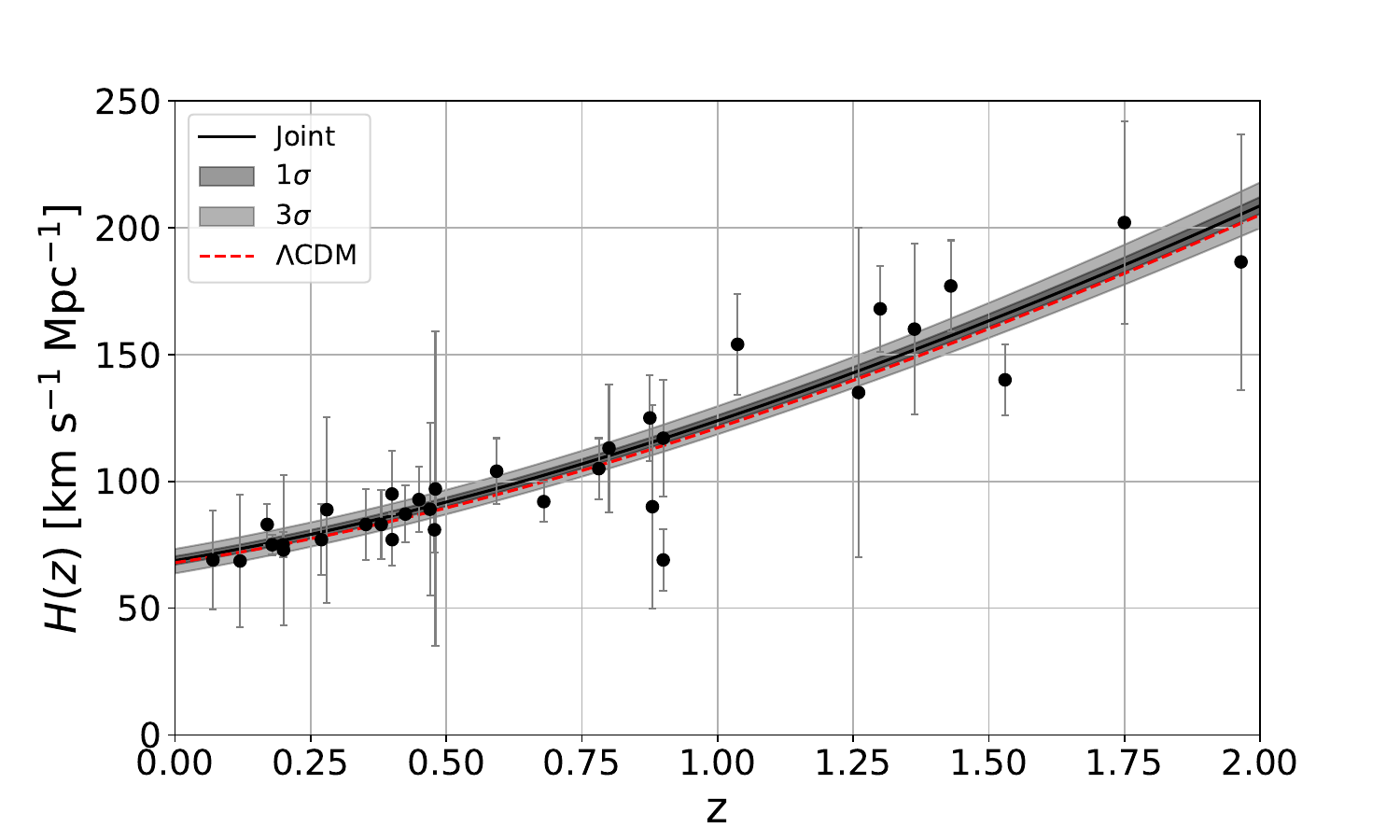}
\includegraphics[width=0.4\textwidth]{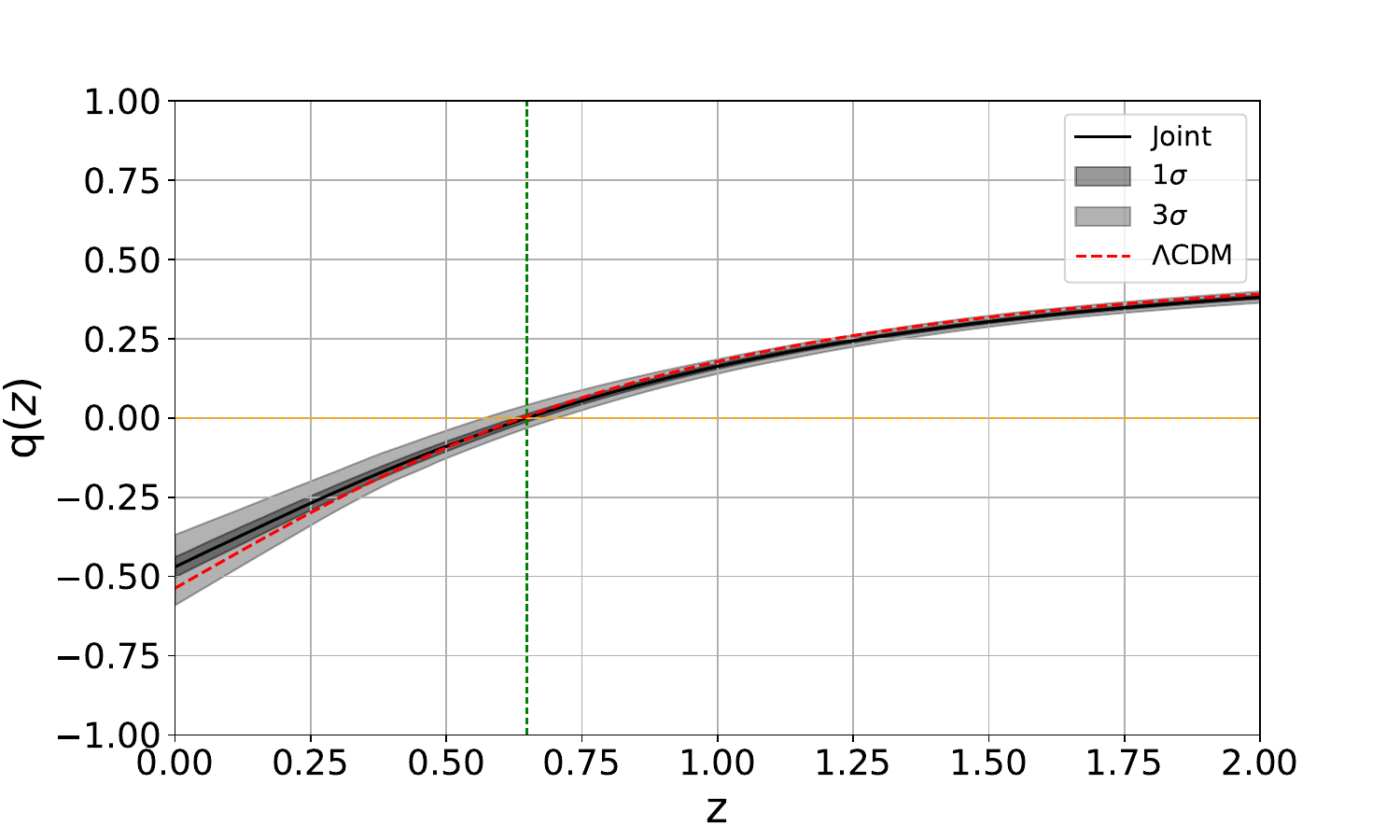}\\
\caption{Reconstruction of the Hubble parameter (left columns) and the deceleration parameter (right columns) for a non-flat universe using the mean values from the joint constraints for cases CI (top), CII (middle), and CIII (bottom). The shaded regions represent the $1\sigma$ (darker) and $3\sigma$ (lighter) error bands.}
    \label{hzcombined_curvature}
\end{figure}

In what follows, we analyze the consequences of the constraints obtained from the joint analysis of the $w_{dm}$–$w_{de}$ model. 

\subsection{Discussion} 

Figure \ref{fig:wdmwde_flat_grid} and Table \ref{tab:flatparameters} shows that there is a discrepancy in the value of $\Omega_{dm}$ when compared to the $\Lambda$CDM model reported by \cite{Planck:2020}, for the joint analysis of cases CI, CII, and CIII in a flat universe, the equations of state (EoS) parameters $w_{dm}$ and $w_{de}$ are consistent with the $\Lambda$CDM values at the $3\sigma$ confidence level, as reported in \cite{Planck:2020}, except for $w_{de}$ in CIII. The non-zero EoS of dark matter suggests an influence on the expansion history of the Universe with positive pressure; however, it does not drive acceleration, given that a component with $w < -1/3$ is required to produce an accelerated expansion. Conversely, the DE EoS constraints indicate a quintessence behavior, characterized by $w_{de} > -1$. It is worth mentioning that the choice of prior on $h$, whether from SH0ES or CMB data (see cases CI and CII), does not lead to a significant change in the values of $w_{dm}$ and $w_{de}$. Nevertheless, a slight variation in the both dark EoS parameters is observed for case CIII. It its denoted that for the three cases when the EoS of DM decreases the EoS of DE increases, this might indicate a interaction between both dark fluids. To alleviate the Hubble Tension, CIII shows that the value of the parameter $h$ is more consistent to $1\sigma$ with Planck et al.(2018) \cite{Planck:2020}.

Figure~\ref{hzcombined} shows that CI, CII, and CIII provide a good fit to the Hubble parameter across the full redshift range, remaining within $3\sigma$ of $\Lambda$CDM. Regarding the deceleration parameter $q(z)$, CI, CII and CIII match $\Lambda$CDM at the $3\sigma$ level across the full range. The transition redshift $z_t$ is consistent with the $\Lambda$CDM prediction, except for CI. For $H(z=0)$, models CI and CIII remain consistent with the $\Lambda$CDM value reported by~\cite{Planck:2020}.
A combination of $w_{dm}>0$ and $w_{de}>-1$ would, in principle, imply a lower expansion, such behavior is observed. This suggests that deviations from the standard EoS for DM and DE within this model are sufficient to substantially alter the expansion rate. Nevertheless, we found that the age of the universe is consistent with that estimated from the standard model.

In the flat scenario, CI, CII, and CIII constraints at 3$\sigma$ confidence level can accommodate either cold or warm dark matter, which may influence the expansion of the universe due to an EoS for $w_{dm} > 0$. When comparing these results with the previous work by \cite{avelino2012testing}, who found an EoS for DM of $w_{dm} \approx  0.006$ and for DE of $w_{de} \approx -1.11$, it is noted that our constraints on the dark matter EoS are slightly different. In contrast, our constraints on the dark energy EoS show a significant deviation from those reported by \cite{avelino2012testing} (see Fig. \ref{fig:wdmwde_flat_grid}). This discrepancy could be attributed to updated cosmological data samples or the Gaussian priors employed in the respective cases. 

Figure \ref{fig:wdmwde_curvature_all} and table \ref{tab:nonflat_parameters} show that, for the joint analysis of cases CIK, CIIK, and CIIIK in a non-flat universe, the parameter $w_{dm}$ is consistent with the $\Lambda$CDM values at the $3\sigma$ confidence level, as reported in \cite{Planck:2020}. The non-zero EoS of dark matter $w_{dm} > 0$ suggests an influence on the expansion history of the Universe with positive pressure; however, it does not drive acceleration, given that a fluid with $w < -1/3$ is required to produce an accelerated expansion. Conversely, the DE EoS constraints indicate a quintessence behavior, characterized by $w_{de} > -1$, but is not consistent with $\Lambda$CDM values at the $3\sigma$ confidence level, as reported in \cite{Planck:2020}. It is worth mentioning that the choice of prior on $h$, whether from SH0ES or CMB data (see cases CIK and CIIK), does lead to a significant change in the values of $w_{dm}$ and $w_{de}$. Nevertheless, a slight variation for both dark EoS parameters is observed for case CIIIK. The curvature parameter $\Omega_k$ for CI and CIII is consistent with $\Lambda$CDM to $3\sigma$, this mean value results in a open universe which is for values of $\Omega_k < 0$.

Figure \ref{hzcombined_curvature} illustrates that CIK, CIIK and CIIIK provide a good fit to the Hubble parameter across the full redshift range, remaining within $3\sigma$ of $\Lambda$CDM. For cases CI and CIII the parameter $H(z=0)$ is consistent at the $3\sigma$ level compared to $\Lambda$CDM and \cite{Planck:2020}.

Figure \ref{hzcombined_curvature} illustrates that $w_{dm}-w_{de}$ CIK, CIIK, and CIIIK models accurately describe the cosmic expansion at the $3\sigma$ level across at early times, nonetheless at values of $z\lesssim0.2$, becomes inconsistent compared to $\Lambda$CDM. The current deceleration parameter, $q(z=0)$, for all cases is consistent with $\Lambda$CDM model, while the transition redshift, $z_t$, shows significant deviation from the $\Lambda$CDM estimation. On the other hand, we found that the age of the universe is consistent with that estimated from the standard model.

Table \ref{tablaAICBIC} indicates that the evidence supporting the $w_{dm}-w_{de}$ model CI, CII consireably less and CIII is substantial. The evidence against the $w_{dm}-w_{de}$ model CI, CII and CIII is very strong. The evidence supporting the $w_{dm}-w_{de}$ model CIK is essentially none, for CIIK is substantial, for $\Lambda$CDM is considerably less and CIIK  is substantial. The evidence against the $w_{dm}-w_{de}$ model CI, CII and CII is very strong. The Gaussian priors may have influenced these results since none of the models utilized the same Gaussian priors across all parameters.

\begin{table*}
\centering
\resizebox{.95\linewidth}{!}{%
\begin{tabular}{|c|c|c|c|c|c|c|c|c|}
\hline
\multicolumn{9}{|c|}{Flat universe}\\
\hline
Data set & \multicolumn{2}{|c|}{$\Lambda$CDM} & \multicolumn{2}{|c|}{CI}& \multicolumn{2}{|c|}{CII} & \multicolumn{2}{|c|}{CIII}\\
\hline
& $\Delta$ AIC & $\Delta$ BIC & $\Delta$ AIC & $\Delta$BIC & $\Delta$AIC & $\Delta$BIC & $\Delta$AIC & $\Delta$BIC\\
\hline
Joint & 0.0 & 0.0 & 9.57 & 20.66 & 9.94 & 21.41 & 1.97 & 13.07 \\
\hline
\multicolumn{9}{|c|}{Non-flat universe}\\
\hline
Data set & \multicolumn{2}{|c|}{$\Lambda$CDM} & \multicolumn{2}{|c|}{CIK}& \multicolumn{2}{|c|}{CIIK} & \multicolumn{2}{|c|}{CIIIK}\\
\hline
& $\Delta$ AIC & $\Delta$ BIC & $\Delta$ AIC & $\Delta$BIC & $\Delta$AIC & $\Delta$BIC & $\Delta$AIC & $\Delta$BIC\\
\hline
Joint & 3.3161 & 0.0 & 12.7607 & 26.095  & 1.8346 & 15.1689 & 0.0 & 13.3343 \\
\hline
\end{tabular}}
\caption{Differences in AIC and BIC values with respect to the minimum among all models, computed using the chi-square value from the joint analysis.}
\label{tablaAICBIC}
\end{table*}

For the case of a flat universe, we find that the standard model, $\Lambda$CDM, yields the minimum values for both the AIC and BIC, and we compare the CI, CII, and CIII cases relative to it. On the other hand, for the non-flat universe, the minimum AIC value is obtained from the CIIIK case, while the minimum BIC value is obtained from the $\Lambda$CDM model. Therefore, a careful comparison is required to determine which model the data prefers. Notice that the higher AIC and BIC values obtained for the CI, CII, and CIII cases suggest that cosmological observations favor cold dark matter with $w_{dm} = 0$ and a cosmological constant ($w_{de} = -1$) as dark energy. On the other hand, although the AIC value for the CIIIK case indicates substantial support for a non-flat universe, its BIC value provides a very strong evidence against this model. However, the BIC value for the $\Lambda$CDM model suggests that there is no evidence against it; however, the corresponding AIC value shows considerably less support. Therefore, in a universe with curvature, there remains the possibility that dark matter could be warm ($w_{dm} = 0.0022^{+0.0052}_{-0.0069}$), dark energy could correspond to a quintessence field
($w_{de}=-0.9101^{+0.1025}_{-0.1281}$), and  
$\Omega_k=-0.0483^{+0.1011}_{-0.0880}$.

\section{Conclusion} \label{sec:conclusiones}

In this paper, we revisited a cosmological model in which both the dark matter and dark energy components are characterized by barotropic equations of state, allowing deviations from the standard $\Lambda$CDM assumptions i.e. DM could have an EoS different to zero $w_{dm}\neq 0$ and DE a value $w_{de} \neq -1$ under the assumption of a flat and non-flat curvature \cite{avelino2012testing}.

To study how these models evolve, we used a dynamical systems approach. This revealed key equilibrium points and critical values of $w_{dm}$ and $w_{de}$ that influence cosmic behavior. Autonomous systems were built to track radiation, dark matter, dark energy, and curvature. In flat universes ($k = 0$), the evolution resembles $\Lambda$CDM: early radiation dominates, matter slows expansion, and dark energy drives acceleration. In negatively curved universes ($k = -1$), dynamics depend more on parameter values, with dark energy helping stabilize expansion. Positively curved universes ($k = +1$) show complex behavior, including bounce scenarios and rare mixed scaling solutions, shaped by the normalized Hubble parameter $Q$ which compactifies the phase space and distinguishes between expansion and contraction branches.

Scaling solutions, where energy density ratios stay constant, are central to understanding long-term cosmic evolution. These occur when the system reaches equilibrium and each component contributes steadily. For acceleration to occur, $w_{de} < -1/3$ is required. We also examined how changes in $w_{de}$ affect stability and curvature growth. In unified models where $w_{dm} = w_{de} = w$, the system simplifies, and outcomes depend on whether $w$ is positive, negative, or zero.

Curvature whether flat, negative, or positive, plays a pivotal role in shaping cosmic dynamics. Bifurcation values mark transitions in behavior, such as the onset of acceleration, curvature dominance, or bounce scenarios. Scaling solutions emerge under specific conditions and may act as attractors or unstable regimes. Flat curvature provides a baseline scenario consistent with standard $\Lambda$CDM evolution, while curved models reveal richer dynamical structures and critical values for cosmic transitions.

We also performed a Bayesian analysis to constrain the EoS parameters $w_{dm}$ and $w_{de}$ using updated cosmological datasets, including observational Hubble data, Pantheon+ Type Ia Supernovae, strong lensing systems, baryon acoustic oscillations (BAO), Cosmic Microwave Background (CMB) and their joint analysis. The priors for this analysis were informed by the behavior of the dynamical system and the bifurcation values identified earlier. We also tested different priors on the Hubble constant $H_0$, and found that all cases consistently explain the universe's behavior at late and early times. Each case aligns with $\Lambda$CDM to 3$\sigma$ for all parameters for both, flat and non-flat curvature. However, none of the cases strictly reproduce the standard $\Lambda$CDM values of $w_{dm} = 0$ and $w_{de} = -1$. This implies that if the EoS of DM is not a dust fluid, the large-scale structure formation in the Universe could be affected, and DE will shift towards quintessence or phantom fluid.

Model selection criteria (AIC and BIC) indicate that the best-performing model without curvature is the standard $\Lambda$CDM, while for non-flat scenarios, the CIIIK model emerges as viable, with:
\begin{equation*}
w_{dm} = 0.0022^{+0.0052}_{-0.0069}, \quad
w_{de} = -0.9101^{+0.1025}_{-0.1281}, \quad
\Omega_k = -0.0483^{+0.1011}_{-0.0880}.
\end{equation*}

This model favors a warm dark matter component ($w_{dm} > 0$) and a quintessence-like dark energy component ($w_{de} > -1$), consistent for the parameter $w_{dm}$ with previous findings \cite{avelino2012testing} who reported $ w_{dm} \approx 0.006 $ and $w_{de} \approx -1.1$. Concurrently with the submission of this work, \cite{Non-zeroEOS_LI} independently explored a phenomenologically emergent dark energy scenario with a constant but non-cold dark matter equation of state, finding $w_{dm}<0$, in contrast to our results ($w_{dm}>0$). Additionally, the curvature parameter is 
$\Omega_k=-0.0281^{+0.0559}_{-0.0684}$ (CIK) \allowbreak
and $-0.0483^{+0.1011}_{-0.0883}$ (CIIIK), 
in agreement with Planck 2018 ($\Omega_k=0.0007\pm0.0019$) 
at the $3\sigma$ level, indicating that curvature inclusion does not significantly deviate from spatial flatness but may improve statistical performance in barotropic two-fluid models.

Ultimately, this research underscores that while $\Lambda$CDM remains a robust framework, alternative models incorporating non-standard equations of state for the dark sector are viable and essential for a deeper understanding of cosmic evolution. The inclusion of curvature and dynamical analysis reveals rich structures and critical values that may guide future observational and theoretical efforts. Further exploration of these phenomenological models could provide new insights into the nature of dark matter, dark energy, and the formation of large-scale structure in the Universe.

\section*{Acknowledgements}
We thank anonymous referees for thoughtful remarks and suggestions. The numerical analysis was also conducted on the GUINA cluster at the Central University of Chile, which was acquired through Fondequip EQM 200216. 
D.A., C.C., J.M., and M.A.G.-A. acknowledge partial support from the project Agencia Nacional de Investigación y Desarrollo (ANID) Vinculación Internacional FOVI220144. 
D. A., J.M., and G.L. acknowledge the financial support of ANID, Chile, through Proyecto Fondecyt Regular 2024, Folio 1240514, Etapa 2025. G.L. thanks  Vicerrectoría de Investigación y Desarrollo Tecnológico (VRIDT) at Universidad Católica del Norte (UCN)  through the scientific support of Núcleo de Investigación~Geometría~Diferencial y Aplicaciones, Resolución VRIDT N°096/2022. M.A.G.-A. acknowledges support from Catedra Marcos Moshinsky, Universidad Iberoamericana, for the support with the National Research System (SNI) grant and the project 0056 from Universidad Iberoamericana: Nuestro Universo en Aceleración, energía oscura o modificaciones a la relatividad general.

\bibliographystyle{elsarticle-num} 
\bibliography{LEGACY/references}

@article{Di_Valentino_2021,
   title={In the realm of the Hubble tension—a review of solutions*},
   volume={38},
   ISSN={1361-6382},
   url={http://dx.doi.org/10.1088/1361-6382/ac086d},
   DOI={10.1088/1361-6382/ac086d},
   number={15},
   journal={Classical and Quantum Gravity},
   publisher={IOP Publishing},
   author={Di Valentino, Eleonora and Mena, Olga and Pan, Supriya and Visinelli, Luca and Yang, Weiqiang and Melchiorri, Alessandro and Mota, David F and Riess, Adam G and Silk, Joseph},
   year={2021},
   month=jul, pages={153001} }

@article{Garcia-Aspeitia:2022uxz,
    author = "Garc{\'\i}a-Aspeitia, Miguel A. and Fernandez-Anaya, Guillermo and Hern{\'a}ndez-Almada, A. and Leon, Genly and Maga{\~n}a, Juan",
    title = "{Cosmology under the fractional calculus approach}",
    eprint = "2207.00878",
    archivePrefix = "arXiv",
    primaryClass = "gr-qc",
    doi = "10.1093/mnras/stac3006",
    journal = "Mon. Not. Roy. Astron. Soc.",
    volume = "517",
    number = "4",
    pages = "4813--4826",
    year = "2022"
}

@article{RoyChoudhury:2024wri,
    author = "Roy Choudhury, Shouvik and Okumura, Teppei",
    title = "{Updated Cosmological Constraints in Extended Parameter Space with Planck PR4, DESI Baryon Acoustic Oscillations, and Supernovae: Dynamical Dark Energy, Neutrino Masses, Lensing Anomaly, and the Hubble Tension}",
    eprint = "2409.13022",
    archivePrefix = "arXiv",
    primaryClass = "astro-ph.CO",
    doi = "10.3847/2041-8213/ad8c26",
    journal = "Astrophys. J. Lett.",
    volume = "976",
    number = "1",
    pages = "L11",
    year = "2024"
}

@article{Angulo:2021kes,
    author = "Angulo, Raul E. and Hahn, Oliver",
    title = "{Large-scale dark matter simulations}",
    eprint = "2112.05165",
    archivePrefix = "arXiv",
    primaryClass = "astro-ph.CO",
    doi = "10.1007/s41115-021-00013-z",
    month = "12",
    year = "2021"
}

@ARTICLE{Ghose:2021EPJC,
       author = {{Ghose}, Souvik and {Bhadra}, Arunava},
        title = "{Is non-particle dark matter equation of state parameter evolving with time?}",
      journal = {European Physical Journal C},
         year = 2021,
        month = aug,
       volume = {81},
       number = {8},
          eid = {683},
        pages = {683},
          doi = {10.1140/epjc/s10052-021-09490-1},
archivePrefix = {arXiv},
       eprint = {2108.04686},
 primaryClass = {physics.gen-ph},
       adsurl = {https://ui.adsabs.harvard.edu/abs/2021EPJC...81..683G}
}

@article{Barranco:2015,
    author = {Barranco, J. and Bernal, A. and Núñez, D.},
    title = {Dark matter equation of state from rotational curves of galaxies},
    journal = {Monthly Notices of the Royal Astronomical Society},
    volume = {449},
    number = {1},
    pages = {403-413},
    year = {2015},
    month = {03},
    issn = {0035-8711},
    doi = {10.1093/mnras/stv302},
    url = {https://doi.org/10.1093/mnras/stv302},
    eprint = {https://academic.oup.com/mnras/article-pdf/449/1/403/4142328/stv302.pdf},
}

@article{Perivolaropoulos:2021jda,
    author = "Perivolaropoulos, Leandros and Skara, Foteini",
    title = "{Challenges for \ensuremath{\Lambda}CDM: An update}",
    eprint = "2105.05208",
    archivePrefix = "arXiv",
    primaryClass = "astro-ph.CO",
    doi = "10.1016/j.newar.2022.101659",
    journal = "New Astron. Rev.",
    volume = "95",
    pages = "101659",
    year = "2022"
}

@article{Armendariz-Picon:2013jej,
    author = "Armendariz-Picon, Cristian and Neelakanta, Jayanth T.",
    title = "{How Cold is Cold Dark Matter?}",
    eprint = "1309.6971",
    archivePrefix = "arXiv",
    primaryClass = "astro-ph.CO",
    doi = "10.1088/1475-7516/2014/03/049",
    journal = "JCAP",
    volume = "03",
    pages = "049",
    year = "2014"
}

@article{Kopp:2018zxp,
    author = "Kopp, Michael and Skordis, Constantinos and Thomas, Daniel B. and Ili\'c, St\'ephane",
    title = "{Dark Matter Equation of State through Cosmic History}",
    eprint = "1802.09541",
    archivePrefix = "arXiv",
    primaryClass = "astro-ph.CO",
    doi = "10.1103/PhysRevLett.120.221102",
    journal = "Phys. Rev. Lett.",
    volume = "120",
    number = "22",
    pages = "221102",
    year = "2018"
}

@article{Weinberg:1988cp,
    author = "Weinberg, Steven",
    editor = "Hsu, Jong-Ping and Fine, D.",
    title = "{The Cosmological Constant Problem}",
    reportNumber = "UTTG-12-88",
    doi = "10.1103/RevModPhys.61.1",
    journal = "Rev. Mod. Phys.",
    volume = "61",
    pages = "1--23",
    year = "1989"
}

@article{Chevallier:2000qy,
    author = "Chevallier, Michel and Polarski, David",
    title = "{Accelerating universes with scaling dark matter}",
    eprint = "gr-qc/0009008",
    archivePrefix = "arXiv",
    doi = "10.1142/S0218271801000822",
    journal = "Int. J. Mod. Phys. D",
    volume = "10",
    pages = "213--224",
    year = "2001"
}

@article{Linder:2002et,
    author = "Linder, Eric V.",
    title = "{Exploring the expansion history of the universe}",
    eprint = "astro-ph/0208512",
    archivePrefix = "arXiv",
    doi = "10.1103/PhysRevLett.90.091301",
    journal = "Phys. Rev. Lett.",
    volume = "90",
    pages = "091301",
    year = "2003"
}

@article{Hinshaw_2013,
doi = {10.1088/0067-0049/208/2/19},
url = {https://dx.doi.org/10.1088/0067-0049/208/2/19},
year = {2013},
month = {sep},
publisher = {The American Astronomical Society},
volume = {208},
number = {2},
pages = {19},
author = {G. Hinshaw and others},
title = {NINE-YEAR WILKINSON MICROWAVE ANISOTROPY PROBE (WMAP) OBSERVATIONS: COSMOLOGICAL PARAMETER RESULTS},
journal = {The Astrophysical Journal Supplement Series},
}

@article{eBOSS:2020yzd,
    author = "Alam, Shadab and others",
    collaboration = "eBOSS",
    title = "{Completed SDSS-IV extended Baryon Oscillation Spectroscopic Survey: Cosmological implications from two decades of spectroscopic surveys at the Apache Point Observatory}",
    eprint = "2007.08991",
    archivePrefix = "arXiv",
    primaryClass = "astro-ph.CO",
    doi = "10.1103/PhysRevD.103.083533",
    journal = "Phys. Rev. D",
    volume = "103",
    number = "8",
    pages = "083533",
    year = "2021"
}

@article{DiValentino:2022fjm_challenges,
    author = "Di Valentino, Eleonora",
    title = "{Challenges of the Standard Cosmological Model}",
    doi = "10.3390/universe8080399",
    journal = "Universe",
    volume = "8",
    number = "8",
    pages = "399",
    year = "2022"
}

@misc{desi2025,
    author = "Abdul Karim, M. and others",
    collaboration = "DESI",
    title = "{DESI DR2 Results II: Measurements of Baryon Acoustic Oscillations and Cosmological Constraints}",
    eprint = "2503.14738",
    archivePrefix = "arXiv",
    primaryClass = "astro-ph.CO",
    reportNumber = "FERMILAB-PUB-25-0169-PPD",
    month = "3",
    year = "2025"
}

@article{Arbey:2021gdg,
    author = "Arbey, A. and Mahmoudi, F.",
    title = "{Dark matter and the early Universe: a review}",
    eprint = "2104.11488",
    archivePrefix = "arXiv",
    primaryClass = "hep-ph",
    reportNumber = "CERN-TH-2021-066",
    doi = "10.1016/j.ppnp.2021.103865",
    journal = "Prog. Part. Nucl. Phys.",
    volume = "119",
    pages = "103865",
    year = "2021"
}

@article{distancesumrule,
  title = {New Test of the Friedmann-Lema\^{\i}tre-Robertson-Walker Metric Using the Distance Sum Rule},
  author = {R\"as\"anen, Syksy and Bolejko, Krzysztof and Finoguenov, Alexis},
  journal = {Phys. Rev. Lett.},
  volume = {115},
  issue = {10},
  pages = {101301},
  numpages = {5},
  year = {2015},
  month = {Sep},
  publisher = {American Physical Society},
  doi = {10.1103/PhysRevLett.115.101301},
  url = {https://link.aps.org/doi/10.1103/PhysRevLett.115.101301}
}

@article{DILATIONSCALE2005,
   title={Detection of the Baryon Acoustic Peak in the Large‐Scale Correlation Function of SDSS Luminous Red Galaxies},
   volume={633},
   ISSN={1538-4357},
   url={http://dx.doi.org/10.1086/466512},
   DOI={10.1086/466512},
   number={2},
   journal={The Astrophysical Journal},
   publisher={American Astronomical Society},
   author={Eisenstein, D. J. and others},
   year={2005},
   month=nov, pages={560–574} }

@ARTICLE{biografiaSNIa,
       author = {{Raskin}, Cody and {Timmes}, F.~X. and {Scannapieco}, Evan and {Diehl}, Steven and {Fryer}, Chris},
        title = "{On Type Ia supernovae from the collisions of two white dwarfs}",
      journal = {\mnras},
         year = 2009,
        month = oct,
       volume = {399},
       number = {1},
        pages = {L156-L159},
          doi = {10.1111/j.1745-3933.2009.00743.x},
archivePrefix = {arXiv},
       eprint = {0907.3915},
 primaryClass = {astro-ph.SR},
       adsurl = {https://ui.adsabs.harvard.edu/abs/2009MNRAS.399L.156R}
}

@article{Copeland:2006wr,
    author = "Copeland, Edmund J. and Sami, M. and Tsujikawa, Shinji",
    title = "{Dynamics of dark energy}",
    eprint = "hep-th/0603057",
    archivePrefix = "arXiv",
    doi = "10.1142/S021827180600942X",
    journal = "Int. J. Mod. Phys. D",
    volume = "15",
    pages = "1753--1936",
    year = "2006"
}

@article{Copeland:1997et,
    author = "Copeland, Edmund J. and Liddle, Andrew R and Wands, David",
    title = "{Exponential potentials and cosmological scaling solutions}",
    eprint = "gr-qc/9711068",
    archivePrefix = "arXiv",
    reportNumber = "SUSX-TH-97-022, SUSSEX-AST-97-11-1, PU-RCG-97-20",
    doi = "10.1103/PhysRevD.57.4686",
    journal = "Phys. Rev. D",
    volume = "57",
    pages = "4686--4690",
    year = "1998"
}

@article{Barreiro:1999zs,
    author = "Barreiro, T. and Copeland, Edmund J. and Nunes, N. J.",
    title = "{Quintessence arising from exponential potentials}",
    eprint = "astro-ph/9910214",
    archivePrefix = "arXiv",
    reportNumber = "SUSX-TH-016",
    doi = "10.1103/PhysRevD.61.127301",
    journal = "Phys. Rev. D",
    volume = "61",
    pages = "127301",
    year = "2000"
}

@article{Pourtsidou:2013nha,
    author = "Pourtsidou, A. and Skordis, C. and Copeland, E. J.",
    title = "{Models of dark matter coupled to dark energy}",
    eprint = "1307.0458",
    archivePrefix = "arXiv",
    primaryClass = "astro-ph.CO",
    doi = "10.1103/PhysRevD.88.083505",
    journal = "Phys. Rev. D",
    volume = "88",
    number = "8",
    pages = "083505",
    year = "2013"
}

@book{wainwright_ellis_1997, 
editor={J. Wainwright and G. F. R. Ellis},
title={Dynamical Systems in Cosmology},
DOI={10.1017/CBO9780511524660}, 
publisher={Cambridge University Press}, 
place={Cambridge}, 
year={1997}}

@book{Coley:2003mj,
    author = "Coley, A. A.",
    title = "{Dynamical systems and cosmology}",
    doi = "10.1007/978-94-017-0327-7",
    publisher = "Kluwer",
    address = "Dordrecht, Netherlands",
    year = "2003"
}

@book{Leon:2012ccj,
    author = "Leon, Genly and Fadragas, Carlos R.",
    title = "{Cosmological dynamical systems}",
    eprint = "1412.5701",
    archivePrefix = "arXiv",
    primaryClass = "gr-qc",
    isbn = "978-3-8473-0233-9",
    publisher = "LAP Lambert Academic Publishing",
    year = "2012"
}

@article{Hernandez-Almada:2020uyr,
    author = "Hern\'andez-Almada, A. and Leon, Genly and Maga\~na, Juan and Garc\'\i{}a-Aspeitia, Miguel A. and Motta, V.",
    title = "{Generalized Emergent Dark Energy: observational Hubble data constraints and stability analysis}",
    eprint = "2002.12881",
    archivePrefix = "arXiv",
    primaryClass = "astro-ph.CO",
    doi = "10.1093/mnras/staa2052",
    journal = "Mon. Not. Roy. Astron. Soc.",
    volume = "497",
    number = "2",
    pages = "1590--1602",
    year = "2020"
}

@article{Leon:2021wyx,
    author = "Leon, Genly and Maga\~na, Juan and Hern\'andez-Almada, A. and Garc\'\i{}a-Aspeitia, Miguel A. and Verdugo, Tom\'as and Motta, V.",
    title = "{Barrow Entropy Cosmology: an observational approach with a hint of stability analysis}",
    eprint = "2108.10998",
    archivePrefix = "arXiv",
    primaryClass = "astro-ph.CO",
    doi = "10.1088/1475-7516/2021/12/032",
    journal = "JCAP",
    volume = "12",
    number = "12",
    pages = "032",
    year = "2021"
}

@article{Hernandez-Almada:2021aiw,
    author = "Hern\'andez-Almada, A. and Leon, Genly and Maga\~na, Juan and Garc\'\i{}a-Aspeitia, Miguel A. and Motta, V. and Saridakis, Emmanuel N. and Yesmakhanova, Kuralay",
    title = "{Kaniadakis-holographic dark energy: observational constraints and global dynamics}",
    eprint = "2111.00558",
    archivePrefix = "arXiv",
    primaryClass = "astro-ph.CO",
    doi = "10.1093/mnras/stac255",
    journal = "Mon. Not. Roy. Astron. Soc.",
    volume = "511",
    number = "3",
    pages = "4147--4158",
    year = "2022"
}

@article{Hernandez-Almada:2021rjs,
    author = "Hern\'andez-Almada, A. and Leon, Genly and Maga\~na, Juan and Garc\'\i{}a-Aspeitia, Miguel A. and Motta, V. and Saridakis, Emmanuel N. and Yesmakhanova, Kuralay and Millano, Alfredo D.",
    title = "{Observational constraints and dynamical analysis of Kaniadakis horizon-entropy cosmology}",
    eprint = "2112.04615",
    archivePrefix = "arXiv",
    primaryClass = "astro-ph.CO",
    doi = "10.1093/mnras/stac795",
    journal = "Mon. Not. Roy. Astron. Soc.",
    volume = "512",
    number = "4",
    pages = "5122--5134",
    year = "2022"
}

@article{Bahamonde:2017ize,
    author = {Bahamonde, Sebastian and B\"ohmer, Christian G. and Carloni, Sante and Copeland, Edmund J. and Fang, Wei and Tamanini, Nicola},
    title = "{Dynamical systems applied to cosmology: dark energy and modified gravity}",
    eprint = "1712.03107",
    archivePrefix = "arXiv",
    primaryClass = "gr-qc",
    doi = "10.1016/j.physrep.2018.09.001",
    journal = "Phys. Rept.",
    volume = "775-777",
    pages = "1--122",
    year = "2018"
}

@article{Sokal1996MonteCM,
  title={Monte Carlo Methods in Statistical Mechanics: Foundations and New Algorithms Note to the Reader},
  author={Alan D. Sokal},
  year={1996},
  url={https://api.semanticscholar.org/CorpusID:14817657}
}

@misc{abedin2025darkmatterheatsup,
      title={When Dark Matter Heats Up: A Model-Independent Search for Non-Cold Behavior}, 
      author={Mazaharul Abedin and Luis A. Escamilla and Supriya Pan and Eleonora Di Valentino and Weiqiang Yang},
      year={2025},
      eprint={2505.09470},
      archivePrefix={arXiv},
      primaryClass={astro-ph.CO},
      url={https://arxiv.org/abs/2505.09470}, 
}

@article{Perlmutter:1999,
    author = "Perlmutter, S. and others",
    collaboration = "Supernova Cosmology Project",
    title = "{Measurements of $\Omega$ and $\Lambda$ from 42 High Redshift Supernovae}",
    eprint = "astro-ph/9812133",
    archivePrefix = "arXiv",
    reportNumber = "LBNL-41801, LBL-41801",
    doi = "10.1086/307221",
    journal = "Astrophys. J.",
    volume = "517",
    pages = "565--586",
    year = "1999"
}

@ARTICLE{Planck:2020,
       author = {{Planck Collaboration}},
        title = "{Planck 2018 results. VI. Cosmological parameters}",
      journal = {\aap},
         year = 2020,
        month = sep,
       volume = {641},
          eid = {A6},
        pages = {A6},
          doi = {10.1051/0004-6361/201833910},
archivePrefix = {arXiv},
       eprint = {1807.06209},
 primaryClass = {astro-ph.CO},
       adsurl = {https://ui.adsabs.harvard.edu/abs/2020A&A...641A...6P}
}

@article{DATASN,
    author = "Scolnic, Dan and others",
    title = "{The Pantheon+ Analysis: The Full Data Set and Light-curve Release}",
    eprint = "2112.03863",
    archivePrefix = "arXiv",
    primaryClass = "astro-ph.CO",
    doi = "10.3847/1538-4357/ac8b7a",
    journal = "Astrophys. J.",
    volume = "938",
    number = "2",
    pages = "113",
    year = "2022"
}

@article{DATABAO,
    author = {C. Nunes,Rafael and K. Yada, Santosh  and Jesus, J.F. and Bernui,Armando},
    title = {Cosmological parameter analyses using transversal BAO data},
    journal = {Monthly Notices of the Royal Astronomical Society},
    volume= {497},
    pages = {2133–2141},
    year =  {2020},
    adsurl = {https://academic.oup.com/mnras/article/497/2/2133/5870123},
}

@article{DEMODELSSLS,
    author = {Amante, Mario H. and Motta, Ver\'onica and Garc\'ia-Aspeitia, Miguel A. and Maga\~na, Juan and Verdugo, Tom\'as},
    title = {Testing dark energy models with a new sample of
strong-lensing systems},
    journal =  {Monthly Notices of the Royal Astronomical Society},
    volume = {498},
    pages = {6013–6033},
    year = {2020},
    adsurl = {https://academic.oup.com/mnras/article/498/4/6013/5904771}
}

@article{DATAHZ,
       author = {{Moresco}, M. and others},
        title = "{Improved constraints on the expansion rate of the Universe up to z \raisebox{-0.5ex}\textasciitilde 1.1 from the spectroscopic evolution of cosmic chronometers}",
      journal = {Journal of Cosmology and Astroparticle Physics},
         year = 2012,
        month = aug,
       volume = {2012},
       number = {8},
          eid = {006},
        pages = {006},
          doi = {10.1088/1475-7516/2012/08/006},
archivePrefix = {arXiv},
       eprint = {1201.3609},
 primaryClass = {astro-ph.CO},
       adsurl = {https://ui.adsabs.harvard.edu/abs/2012JCAP...08..006M}
}

@article{BAOFORMULA,
    author = {Eisenstein,Daniel J. and Hu, Wayne},
    title = {Baryonic Features in the Matter Transfer Function},
    journal = {The Astrophysical Journal},
    year = {1998},
    volume = {496},
    pages = {L2},
    publisher = {IOP Puslishing},
    adsurl = {https://iopscience.iop.org/article/10.1086/305424}
}

@ARTICLE{1998RIESS,
    author = "Riess, Adam G. and others",
    collaboration = "Supernova Search Team",
    title = "{Observational evidence from supernovae for an accelerating universe and a cosmological constant}",
    eprint = "astro-ph/9805201",
    archivePrefix = "arXiv",
    doi = "10.1086/300499",
    journal = "Astron. J.",
    volume = "116",
    pages = "1009--1038",
    year = "1998"
}

@misc{avelino2012testing,
      title={Testing the EoS of dark matter with cosmological observations}, 
      author={Arturo Avelino and Norman Cruz and Ulises Nucamendi},
      year={2012},
      eprint={1211.4633},
      archivePrefix={arXiv},
      primaryClass={astro-ph.CO}
}

@ARTICLE{1933Zwicky,
       author = {{Zwicky}, F.},
        title = "{Die Rotverschiebung von extragalaktischen Nebeln}",
      journal = {Helvetica Physica Acta},
         year = 1933,
        month = jan,
       volume = {6},
        pages = {110-127},
       adsurl = {https://ui.adsabs.harvard.edu/abs/1933AcHPh...6..110Z}
}

@ARTICLE{1970ApJ...159..379R,
       author = {{Rubin}, Vera C. and {Ford}, W. Kent, Jr.},
        title = "{Rotation of the Andromeda Nebula from a Spectroscopic Survey of Emission Regions}",
      journal = {\apj},
         year = 1970,
        month = feb,
       volume = {159},
        pages = {379},
          doi = {10.1086/150317},
       adsurl = {https://ui.adsabs.harvard.edu/abs/1970ApJ...159..379R}
}

@ARTICLE{comparinphquitessce,
       author = {{Novosyadlyj}, B. and {Sergijenko}, O. and {Durrer}, R. and {Pelykh}, V.},
        title = "{Quintessence versus phantom dark energy: the arbitrating power of current and future observations}",
      journal = {\jcap},
         year = 2013,
        month = jun,
       volume = {2013},
       number = {6},
          eid = {042},
        pages = {042},
          doi = {10.1088/1475-7516/2013/06/042},
archivePrefix = {arXiv},
       eprint = {1212.5824},
 primaryClass = {astro-ph.CO},
       adsurl = {https://ui.adsabs.harvard.edu/abs/2013JCAP...06..042N}
}

@misc{bassett2009baryon,
      title={Baryon Acoustic Oscillations}, 
      author={Bruce A. Bassett and Renée Hlozek},
      year={2009},
      eprint={0910.5224},
      archivePrefix={arXiv},
      primaryClass={astro-ph.CO}
}

@article{Jimenez_2002,
doi = {10.1086/340549},
url = {https://dx.doi.org/10.1086/340549},
year = {2002},
month = {jul},
publisher = {},
volume = {573},
number = {1},
pages = {37},
author = {Raul Jimenez and Abraham Loeb},
title = {Constraining Cosmological Parameters Based on Relative Galaxy Ages},
journal = {The Astrophysical Journal}
}

@article{whitedwarcompo,
   title={The composition of massive white dwarfs and their dependence on C-burning modeling},
   volume={659},
   ISSN={1432-0746},
   url={http://dx.doi.org/10.1051/0004-6361/202142341},
   DOI={10.1051/0004-6361/202142341},
   journal={Astronomy \& Astrophysics},
   publisher={EDP Sciences},
   author={De Gerónimo, Francisco C. and Miller Bertolami, Marcelo M. and Plaza, Francisco and Catelan, Márcio},
   year={2022},
   month=mar, pages={A150} }

@ARTICLE{chandrasekharlimit,
       author = {{Chandrasekhar}, S.},
        title = "{The Maximum Mass of Ideal White Dwarfs}",
      journal = {\apj},
         year = 1931,
        month = jul,
       volume = {74},
        pages = {81},
          doi = {10.1086/143324},
       adsurl = {https://ui.adsabs.harvard.edu/abs/1931ApJ....74...81C}
}

@article{Escamilla:2023oce,
    author = "Escamilla, Luis A. and Giar\`e, William and Di Valentino, Eleonora and Nunes, Rafael C. and Vagnozzi, Sunny",
    title = "{The state of the dark energy equation of state circa 2023}",
    eprint = "2307.14802",
    archivePrefix = "arXiv",
    primaryClass = "astro-ph.CO",
    doi = "10.1088/1475-7516/2024/05/091",
    journal = "JCAP",
    volume = "05",
    pages = "091",
    year = "2024"
}

@article{Muller:2004yb,
    author = "Muller, Christian M.",
    title = "{Cosmological bounds on the equation of state of dark matter}",
    eprint = "astro-ph/0410621",
    archivePrefix = "arXiv",
    reportNumber = "HD-THEP-04-44",
    doi = "10.1103/PhysRevD.71.047302",
    journal = "Phys. Rev. D",
    volume = "71",
    pages = "047302",
    year = "2005"
}

@ARTICLE{Yang_IDE:2018,
       author = {{Yang}, Weiqiang and {Mukherjee}, Ankan and {Di Valentino}, Eleonora and {Pan}, Supriya},
        title = "{Interacting dark energy with time varying equation of state and the H$_{0}$ tension}",
      journal = {\prd},
         year = 2018,
        month = dec,
       volume = {98},
       number = {12},
          eid = {123527},
        pages = {123527},
          doi = {10.1103/PhysRevD.98.123527},
archivePrefix = {arXiv},
       eprint = {1809.06883},
 primaryClass = {astro-ph.CO},
       adsurl = {https://ui.adsabs.harvard.edu/abs/2018PhRvD..98l3527Y}
}

@article{Riess_2022,
 author = "Riess, Adam G. and others",
    title = "{A Comprehensive Measurement of the Local Value of the Hubble Constant with 1 km s$^{-1}$ Mpc$^{-1}$ Uncertainty from the Hubble Space Telescope and the SH0ES Team}",
    eprint = "2112.04510",
    archivePrefix = "arXiv",
    primaryClass = "astro-ph.CO",
    doi = "10.3847/2041-8213/ac5c5b",
    journal = "Astrophys. J. Lett.",
    volume = "934",
    number = "1",
    pages = "L7",
    year = "2022"
}

@ARTICLE{Navarro1997,
       author = {{Navarro}, Julio F. and {Frenk}, Carlos S. and {White}, Simon D.~M.},
        title = "{A Universal Density Profile from Hierarchical Clustering}",
      journal = {\apj},
         year = 1997,
        month = dec,
       volume = {490},
       number = {2},
        pages = {493-508},
          doi = {10.1086/304888},
archivePrefix = {arXiv},
       eprint = {astro-ph/9611107},
 primaryClass = {astro-ph},
       adsurl = {https://ui.adsabs.harvard.edu/abs/1997ApJ...490..493N}
}

@article{dwarfcuspycoreAmorisco_2011,
   title={Dark matter cores and cusps: the case of multiple stellar populations in dwarf spheroidals: Multiple stellar population in dSphs},
   volume={419},
   ISSN={0035-8711},
   url={http://dx.doi.org/10.1111/j.1365-2966.2011.19684.x},
   DOI={10.1111/j.1365-2966.2011.19684.x},
   number={1},
   journal={Monthly Notices of the Royal Astronomical Society},
   publisher={Oxford University Press (OUP)},
   author={Amorisco, N. C. and Evans, N. W.},
   year={2011},
   month=oct, pages={184–196} }

@misc{quasarhubbleproblem1,
      title={Cosmological constraints from the Hubble diagram of quasars at high redshifts}, 
      author={Guido Risaliti and Elisabeta Lusso},
      year={2018},
      eprint={1811.02590},
      archivePrefix={arXiv},
      primaryClass={astro-ph.CO}
}

@article{Nojiri_EDE:2021,
   title={Modeling and testing the equation of state for (Early) dark energy},
   volume={32},
   ISSN={2212-6864},
   url={http://dx.doi.org/10.1016/j.dark.2021.100837},
   DOI={10.1016/j.dark.2021.100837},
   journal={Physics of the Dark Universe},
   publisher={Elsevier BV},
   author={Nojiri, Shin’ichi and Odintsov, Sergei D. and Sáez-Chillón Gómez, Diego and Sharov, German S.},
   year={2021},
   month=may, pages={100837} }

@article{diValentino:2020,
   title={Soundness of dark energy properties},
   volume={2020},
   ISSN={1475-7516},
   url={http://dx.doi.org/10.1088/1475-7516/2020/07/045},
   DOI={10.1088/1475-7516/2020/07/045},
   number={07},
   journal={Journal of Cosmology and Astroparticle Physics},
   publisher={IOP Publishing},
   author={Valentino, Eleonora Di and Gariazzo, Stefano and Mena, Olga and Vagnozzi, Sunny},
   year={2020},
   month=jul, pages={045–045} }

@article{mcmchammer,
       author = {{Foreman-Mackey}, Daniel and {Hogg}, David W. and {Lang}, Dustin and {Goodman}, Jonathan},
        title = "{emcee: The MCMC Hammer}",
      journal = {\pasp},
         year = 2013,
        month = mar,
       volume = {125},
       number = {925},
        pages = {306},
          doi = {10.1086/670067},
archivePrefix = {arXiv},
       eprint = {1202.3665},
 primaryClass = {astro-ph.IM},
       adsurl = {https://ui.adsabs.harvard.edu/abs/2013PASP..125..306F}
}

@ARTICLE{1974ITAC...19..716A,
       author = {{Akaike}, H.},
        title = "{A New Look at the Statistical Model Identification}",
      journal = {IEEE Transactions on Automatic Control},
         year = 1974,
        month = jan,
       volume = {19},
        pages = {716-723},
       adsurl = {https://ui.adsabs.harvard.edu/abs/1974ITAC...19..716A}
}

@article{BIC1974,
author = {Gideon Schwarz},
title = {{Estimating the Dimension of a Model}},
volume = {6},
journal = {The Annals of Statistics},
number = {2},
publisher = {Institute of Mathematical Statistics},
pages = {461 -- 464},
year = {1978},
doi = {10.1214/aos/1176344136},
URL = {https://doi.org/10.1214/aos/1176344136}
}

@misc{DESI:2024mwx,
    author = "Adame, A. G. and others",
    collaboration = "DESI",
    title = "{DESI 2024 VI: Cosmological Constraints from the Measurements of Baryon Acoustic Oscillations}",
    eprint = "2404.03002",
    archivePrefix = "arXiv",
    primaryClass = "astro-ph.CO",
    reportNumber = "FERMILAB-PUB-24-0154-PPD",
    month = "4",
    year = "2024"
}

@misc{HZnewpoint1,
      title={Addressing the Hubble tension with cosmic chronometers}, 
      author={Michele Moresco},
      year={2023},
      eprint={2307.09501},
      archivePrefix={arXiv},
      primaryClass={astro-ph.CO},
      url={https://arxiv.org/abs/2307.09501}, 
}

@article{hznewpoint2,
   title={A new measurement of the expansion history of the Universe at z = 1.26 with cosmic chronometers in VANDELS},
   volume={679},
   ISSN={1432-0746},
   url={http://dx.doi.org/10.1051/0004-6361/202346992},
   DOI={10.1051/0004-6361/202346992},
   journal={Astronomy \& Astrophysics},
   publisher={EDP Sciences},
   author={Tomasetti, E. and Moresco, M. and Borghi, N. and Jiao, K. and Cimatti, A. and Pozzetti, L. and Carnall, A. C. and McLure, R. J. and Pentericci, L.},
   year={2023},
   month=nov, pages={A96} }

@article{hznewpoint3,
   title={New Observational H(z) Data from Full-spectrum Fitting of Cosmic Chronometers in the LEGA-C Survey},
   volume={265},
   ISSN={1538-4365},
   url={http://dx.doi.org/10.3847/1538-4365/acbc77},
   DOI={10.3847/1538-4365/acbc77},
   number={2},
   journal={The Astrophysical Journal Supplement Series},
   publisher={American Astronomical Society},
   author={Jiao, Kang and Borghi, Nicola and Moresco, Michele and Zhang, Tong-Jie},
   year={2023},
   month=mar, pages={48} }

@ARTICLE{CMB_1969,
       author = {{Zel'dovich}, Ya. B. and {Kurt}, V.~G. and {Syunyaev}, R.~A.},
        title = "{Recombination of Hydrogen in the Hot Model of the Universe}",
      journal = {Soviet Journal of Experimental and Theoretical Physics},
         year = 1969,
        month = jan,
       volume = {28},
        pages = {146},
       adsurl = {https://ui.adsabs.harvard.edu/abs/1969JETP...28..146Z}
}

@ARTICLE{CMB_1968,
       author = {{Peebles}, P.~J.~E.},
        title = "{Recombination of the Primeval Plasma}",
      journal = {\apj},
         year = 1968,
        month = jul,
       volume = {153},
        pages = {1},
          doi = {10.1086/149628},
       adsurl = {https://ui.adsabs.harvard.edu/abs/1968ApJ...153....1P}
}

@ARTICLE{CMB_DATA2,
       author = {{Chen}, Lu and {Huang}, Qing-Guo and {Wang}, Ke},
        title = "{Distance priors from Planck final release}",
      journal = {\jcap},
         year = 2019,
        month = feb,
       volume = {2019},
       number = {2},
          eid = {028},
        pages = {028},
          doi = {10.1088/1475-7516/2019/02/028},
archivePrefix = {arXiv},
       eprint = {1808.05724},
 primaryClass = {astro-ph.CO},
       adsurl = {https://ui.adsabs.harvard.edu/abs/2019JCAP...02..028C}
}

@ARTICLE{SHIFT_PARAMETER,
       author = {{Bond}, J.~R. and {Efstathiou}, G. and {Tegmark}, M.},
        title = "{Forecasting cosmic parameter errors from microwave background anisotropy experiments}",
      journal = {\mnras},
         year = 1997,
        month = nov,
       volume = {291},
       number = {3},
        pages = {L33-L41},
          doi = {10.1093/mnras/291.1.L33},
archivePrefix = {arXiv},
       eprint = {astro-ph/9702100},
 primaryClass = {astro-ph},
       adsurl = {https://ui.adsabs.harvard.edu/abs/1997MNRAS.291L..33B}
}

@article{Hu_1996,
   title={Small‐Scale Cosmological Perturbations: An Analytic Approach},
   volume={471},
   ISSN={1538-4357},
   url={http://dx.doi.org/10.1086/177989},
   DOI={10.1086/177989},
   number={2},
   journal={The Astrophysical Journal},
   publisher={American Astronomical Society},
   author={Hu, Wayne and Sugiyama, Naoshi},
   year={1996},
   month=nov, pages={542–570} }

@misc{Non-zeroEOS_LI,
      title={Is non-zero equation of state of dark matter favored by DESI DR2?}, 
      author={Tian-Nuo Li and Yi-Min Zhang and Yan-Hong Yao and Peng-Ju Wu and Jing-Fei Zhang and Xin Zhang},
      year={2025},
      eprint={2506.09819},
      archivePrefix={arXiv},
      primaryClass={astro-ph.CO},
      url={https://arxiv.org/abs/2506.09819}, 
}

@article{Freedman_2019,
doi = {10.3847/1538-4357/ab2f73},
url = {https://dx.doi.org/10.3847/1538-4357/ab2f73},
year = {2019},
month = {aug},
publisher = {The American Astronomical Society},
volume = {882},
number = {1},
pages = {34},
author = {Freedman, Wendy L. and Madore, Barry F. and Hatt, Dylan and Hoyt, Taylor J. and Jang, In Sung and Beaton, Rachael L. and Burns, Christopher R. and Lee, Myung Gyoon and Monson, Andrew J. and Neeley, Jillian R. and Phillips, M. M. and Rich, Jeffrey A. and Seibert, Mark},
title = {The Carnegie-Chicago Hubble Program. VIII. An Independent Determination of the Hubble Constant Based on the Tip of the Red Giant Branch*},
journal = {The Astrophysical Journal}
}

@article{Freedman_2020,
doi = {10.3847/1538-4357/ab7339},
url = {https://dx.doi.org/10.3847/1538-4357/ab7339},
year = {2020},
month = {mar},
publisher = {The American Astronomical Society},
volume = {891},
number = {1},
pages = {57},
author = {Freedman, Wendy L. and Madore, Barry F. and Hoyt, Taylor and Jang, In Sung and Beaton, Rachael and Lee, Myung Gyoon and Monson, Andrew and Neeley, Jill and Rich, Jeffrey},
title = {Calibration of the Tip of the Red Giant Branch},
journal = {The Astrophysical Journal}
}

@article{H0_review_verde,
    author = {Verde, Licia and Sch{\"o}neberg, Nils and Gil-Mar{\'\i}n, H{\'e}ctor},
    title = "{A Tale of Many H0}",
    eprint = "2311.13305",
    archivePrefix = "arXiv",
    primaryClass = "astro-ph.CO",
    doi = "10.1146/annurev-astro-052622-033813",
    journal = "Ann. Rev. Astron. Astrophys.",
    volume = "62",
    number = "1",
    pages = "287--331",
    year = "2024"
}

@Inbook{H0_review_Tully2024,
author="Tully, R. Brent",
editor="Di Valentino, Eleonora
and Brout, Dillon",
title="The Hubble Constant: A Historical Review",
bookTitle="The Hubble Constant Tension",
year="2024",
publisher="Springer Nature Singapore",
address="Singapore",
pages="7--26",
isbn="978-981-99-0177-7",
doi="10.1007/978-981-99-0177-7_2",
url="https://doi.org/10.1007/978-981-99-0177-7_2"
}

@article{H0_review_Shah_2021,
   title={A buyer’s guide to the Hubble constant},
   volume={29},
   ISSN={1432-0754},
   url={http://dx.doi.org/10.1007/s00159-021-00137-4},
   DOI={10.1007/s00159-021-00137-4},
   number={1},
   journal={The Astronomy and Astrophysics Review},
   publisher={Springer Science and Business Media LLC},
   author={Shah, Paul and Lemos, Pablo and Lahav, Ofer},
   year={2021},
   month=dec }

@article{H0_review_Khalife_2024,
doi = {10.1088/1475-7516/2024/04/059},
url = {https://dx.doi.org/10.1088/1475-7516/2024/04/059},
year = {2024},
month = {apr},
publisher = {IOP Publishing},
volume = {2024},
number = {04},
pages = {059},
author = {Khalife, Ali Rida and Zanjani, Maryam Bahrami and Galli, Silvia and Günther, Sven and Lesgourgues, Julien and Benabed, Karim},
title = {Review of Hubble tension solutions with new SH0ES and SPT-3G data},
journal = {Journal of Cosmology and Astroparticle Physics}
}
\end{document}